\documentclass[useAMS,usenatbib]{mn2e}
\usepackage{epsfig}
\usepackage{graphicx}
\usepackage{times}
\usepackage{color}

\def\cm3{cm$^{-3}$}
\def\kms{km~s$^{-1}$}

\def\lsun{L$_{\odot}$}

\def\msun{M$_{\odot}$}

\def\one{\ts {\,\sc i}}
\def\two{\ts {\,\sc ii}}
\def\three{\ts {\,\sc iii}}
\def\four{\ts {\,\sc iv}}

\def\six{\ts {\sc vi}}
\def\beq{\begin{equation}}
\def\eeq{\end{equation}}

\def\lesssim{\mathrel{\hbox{\rlap{\hbox{\lower4pt\hbox{$\sim$}}}\hbox{$<$}}}}
\def\gtrsim{\mathrel{\hbox{\rlap{\hbox{\lower4pt\hbox{$\sim$}}}\hbox{$>$}}}}

\def\lesssim{\mathrel{\hbox{\rlap{\hbox{\lower4pt\hbox{$\sim$}}}\hbox{$<$}}}}
\def\gtrsim{\mathrel{\hbox{\rlap{\hbox{\lower4pt\hbox{$\sim$}}}\hbox{$>$}}}}

\def\isoni{$^{56}{\rm Ni}$}

\def\one{{\,\sc i}}
\def\two{{\,\sc ii}}
\def\three{{\,\sc iii}}
\def\four{{\,\sc iv}}

\def\six{{\sc vi}}
\def\sev{{\sc vii}}

\newcommand{\opar}{\hbox{$^{\scriptsize \rm  o}$}}
\newcommand{\oparsub}[1]{\hbox{$^{\scriptsize \rm  o}_{#1}$}}

\def\cmfgen{{\sc cmfgen}}

\def\ergs{erg\,s$^{-1}$}

\newcommand{\iso}[2]{\ensuremath{^{#1}\rm{#2}}}

\def\aj{AJ}
\def\pasp{PASP}
\def\apj{ApJ}
\def\apjs{ApJS}
\def\apjl{ApJL}
\def\aap{A\&A}

\def\aaps{A\&AS}
\def\mnras{MNRAS}
\def\nat{Nature}

\def\solphys{Sol.~Phys.}

\title[Radiative-transfer modelling of supernovae Ia]
{Critical ingredients of supernova Ia radiative-transfer modelling}

\author[Luc Dessart, D.J. Hillier, St\'ephane Blondin, and Alexei Khokhlov]
{Luc Dessart,$^{1,2}$ D. John Hillier,$^{3}$ St\'ephane Blondin,$^{1}$ and Alexei Khokhlov$^{4}$ \\ \\
$^{1}$ Aix Marseille Universit\'e, CNRS, LAM (Laboratoire d'Astrophysique
de Marseille) UMR 7326, 13388, Marseille, France.\\
$^{2}$ Laboratoire Lagrange, UMR7293, Universit\'e Nice Sophia-Antipolis, CNRS,
Observatoire de la C\^{o}te d'Azur, 06300 Nice, France. \\
$^3$ Department of Physics and Astronomy \& Pittsburgh Particle physics, Astrophysics,
and Cosmology Center (PITT PACC), \\
University of Pittsburgh,  Pittsburgh, PA 15260, USA \\
$^4$ Department of Astronomy \& Astrophysics, the Enrico Fermi Institute, and the Computational Institute,
The University of Chicago, \\
Chicago, IL 60637, USA}

\begin{document}

\date{Accepted 2014 April 17.  Received 2014 April 16; in original form 2013 August 28}

\pagerange{\pageref{firstpage}--\pageref{lastpage}} \pubyear{2014}

\maketitle

\label{firstpage}

\begin{abstract}
We explore the physics of SN Ia light curves and spectra using
the 1-D non-LTE time-dependent radiative-transfer code \cmfgen.  Rather than
adjusting ejecta properties to match observations, we select
as input one ``standard''  1-D Chandrasekhar-mass delayed-detonation
hydrodynamical model, and then explore the sensitivity of radiation and gas
properties of the ejecta on radiative-transfer modelling assumptions.
The correct computation of SN Ia radiation is not exclusively a solution to
an ``opacity problem", characterized by the treatment of a large number of lines.
We demonstrate that the key is to identify and treat important atomic processes consistently.
This is not limited to treating line blanketing in non-LTE.
We show that including forbidden line transitions of metals, and in particular Co, is increasingly
important for the temperature and ionization of the gas beyond maximum light.
Non-thermal ionization and excitation are also critical since they affect 
 the color evolution and the $\Delta M_{15}$ decline rate of our model.
While impacting little the bolometric luminosity, a more complete
treatment of decay routes leads to enhanced line blanketing, e.g., associated with \iso{48}Ti in the $U$
and $B$ bands. Overall, we find that SN Ia radiation properties are influenced in a complicated
way by the atomic data we employ, so that obtaining converged results is a real challenge.
Nonetheless, with our fully-fledged \cmfgen\ model, we obtain good
agreement with the golden standard type Ia SN 2005cf in the optical and near-IR, from 5 to 60\,d after 
explosion, suggesting that assuming spherical symmetry is not detrimental to SN Ia 
radiative-transfer modeling at these times.
Multi-D effects no doubt matter, but they are perhaps less important than accurately treating
the non-LTE processes that are crucial to obtain reliable temperature and ionization structures.
\end{abstract}


\def\nifs{\iso{56}Ni}

\section{Introduction}

Over the last two decades type Ia supernovae (SNe), have become important tools
for measuring basic cosmological parameters and the energy content of
the Universe \citep{riess_etal_98,perlmutter_etal_99}.
SNe Ia are likely the explosions of carbon-oxygen degenerate stars in binary systems
\citep{hoyle_fowler_60}.
However the evolutionary
channels leading to SN Ia events are only crudely understood, and the
physical state of the progenitor star and details of the explosion
mechanism(s) are still debated.

In a single-degenerate scenario the white dwarf (WD) evolves towards explosion by
accreting hydrogen or helium from a non-degenerate stellar companion
\citep{whelan_iben_73,nomoto_82}, whereas in a double-degenerate
scenario the explosion is caused by a merger of two degenerate stars
\citep{iben_tutukov_84,webbink_84}.
At present the delayed detonation models
with varying deflagration-to-detonation transition density
 in a Chandrasekhar-mass white dwarf offer a good
agreement with SN Ia observations, e.g., for the range of
luminosities and the stratification of chemical elements
\citep{khokhlov_etal_93,hoeflich_etal_96}.

The study of the photometric and spectroscopic properties of SNe Ia requires numerical 
radiative-transfer tools. The early work
of \cite{arnett_82} and \cite{pinto_eastman_00a} used analytic modeling
to extract a basic understanding of their bolometric light curve
and to estimate the ejecta kinetic energy and $^{56}$Ni mass.
Unfortunately, this approach does not yield constraints on important
ejecta properties, such as chemical composition and stratification,
and lacks information on color evolution.

There are a number of approaches for doing the radiative transfer of SNe Ia more accurately.
One approach is to treat  the photospheric layers exclusively and assume steady state.
Radiation transport is undertaken using the Monte Carlo technique \citep[e.g.,][]{mazzali_etal_93}
or by solving the transfer equation
\citep[e.g.][]{nugent_etal_95,pauldrach_etal_96,baron_etal_96,blondin_etal_06,branch_etal_06b,pauldrach_etal_13}.
The advantages of this approach are computational speed, and the ease with which
model parameters and abundances can be altered to  fit models to observations.
The drawback is that model adjustments may be used erroneously
to overcome missing physics (either in the atomic data or in the model).
One major disadvantage of the method is that  the infrared spectral range is optically thin even before the
$B$-band maximum, and the concept of a well defined photosphere becomes meaningless.

Another approach is to
perform time-dependent radiation transport and model the entire SN ejecta.
This has been done using gray or multi-group radiative transfer
(see, e.g., \citealt{hoeflich_etal_93,hoeflich_khokhlov_96,blinnikov_etal_98,blinnikov_etal_06}).
The gas is treated in Local-Thermodynamic-Equilibrium (LTE) and the opacities
are approximated using the formalism of
\citet{karp_etal_77} and \cite{pinto_eastman_00b}. Continuum and line opacity
contributions are summed over an energy bin and the transfer is solved for each bin.
One alternative uses the Monte Carlo technique
\citep{lucy_05,kasen_etal_06,sim_07,kromer_sim_09,sim_etal_13}.
The benefit is the possibility of extension to 3D but the drawback is again the approximate
treatment of the thermodynamic state of the gas.

Whether we consider SNe or stellar atmospheres, the leakage of radiative energy through
the photosphere is known to drive the material out of LTE \citep{mihalas_78}. However,
because of the fast expansion and small mass of SNe Ia, the low ejecta density
prevents LTE conditions even at depth as early as the peak of the bolometric light curve.
There is thus much interest in designing radiative-transfer tools that explicitly treat the
non-LTE aspects of the problem, i.e., by solving the statistical equilibrium equations directly
\citep{baron_etal_96}, while solving simultaneously for the radiation transport transport problem
time dependently \citep{hoeflich_etal_02,jack_etal_11,HD12}. 

In this work, we discuss our own efforts, started in 2008, to model SNe Ia with \cmfgen.
In its present form, the code contains a number of important improvements implemented
in recent years, primarily for the modelling of core-collapse SNe.
The line blanketed aspects are discussed in \citet{HM98_lb} in the context of hot star winds;
the modifications to treat SN atmospheres are discussed in \citet{DH05a,DH05b};
the extension for the time-dependent treatment of the statistical-equilibrium equations
is presented in \citet{DH08};
the philosophy of the full time-dependent approach for both the gas and the radiation is
given in \cite{DH10}, with details given in \citet{HD12}.
Our approach is non-LTE, time dependent, and solves for the gas and radiation properties at all depths,
from the innermost to the outermost ejecta mass shells.

The non-LTE treatment applies to the full radiative-transfer problem, hence allows the same level
of sophistication for the computation of the light curves and the spectra.
The multi-band light curves are computed by direct integration of the emergent wavelength-dependent
flux computed by the non-LTE time-dependent solver along the time sequence.
Our non-LTE approach conserves energy and provides
a physical solution to multi-band light curves and spectral evolution {\it simultaneously}.
The interaction between radiation and matter
is solved exactly, i.e.,  without any ad-hoc prescription for the nature of opacity and emissivity sources.

In \citet{blondin_etal_13}, we presented the results for a set of delayed-detonation models
and compared their radiative properties to observed SNe Ia at bolometric maximum.
Here, we discuss the technical aspects of SN Ia radiative transfer modelling using
the hydrodynamical delayed detonation model DDC10. In a future paper we will
cover in greater depth the properties of the radiative transfer in SN Ia ejecta
and photospheres, discussing the departures from LTE, the thermalization/scattering character
of spectral lines, as well as spectrum formation. We delay to subsequent papers
the discussion of dependencies of the SN Ia radiation on ejecta properties, in particular
on the abundance of \iso{56}Ni synthesized in the explosion.

   An important message from our work on SNe Ia is that with detailed non-LTE radiative
transfer, we can reproduce the fundamental SN Ia light curve and spectral properties
with the basic delayed-detonation scenario, even with the assumption of {\it spherical
symmetry}. Although this represents a very important result, it has limited value
if we do not understand why or how it works. Until recently  we were unable
to reproduce  the fundamental radiative properties of type Ia SNe. The reasons for this
failure were related to assumptions in the modeling, rather than issues with the
properties of the progenitor and the explosion model. So, rather than only presenting
the properties of the radiative-transfer
model that works, we also present the original models we ran and describe how they failed
(Section~\ref{sect_atom}). We describe the numerous attempts to solve the discrepancies,
roughly in a chronological order, and present some ingredients that solve these problems
(Section~\ref{sect_sol}).
We then discuss the impact of non-local energy deposition and $\gamma$-ray escape
on SN Ia properties (Section~\ref{sect_loc_noloc}), as well as the influence of non-thermal
processes (Section~\ref{sect_nonte}).
Having covered the various ingredients controlling SN Ia radiation, we study the origin
of the secondary maximum observed in near-IR SN Ia light curves (Section~\ref{sect_nearir}).
We present our conclusions in Section~\ref{sect_conc}.

\begin{figure}
\begin{minipage}{0.45\linewidth}
\epsfig{file=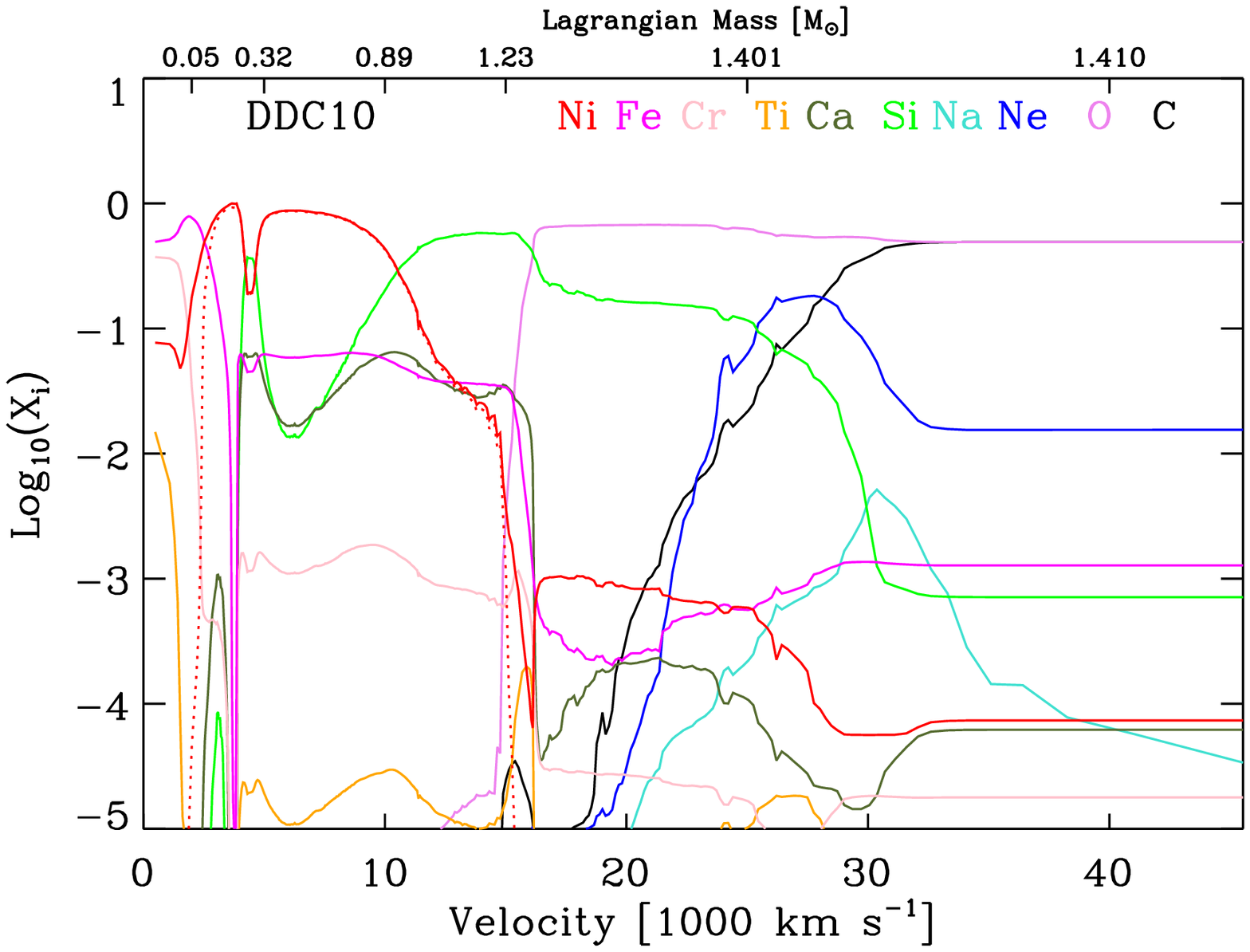,width=8.cm}\\
\epsfig{file=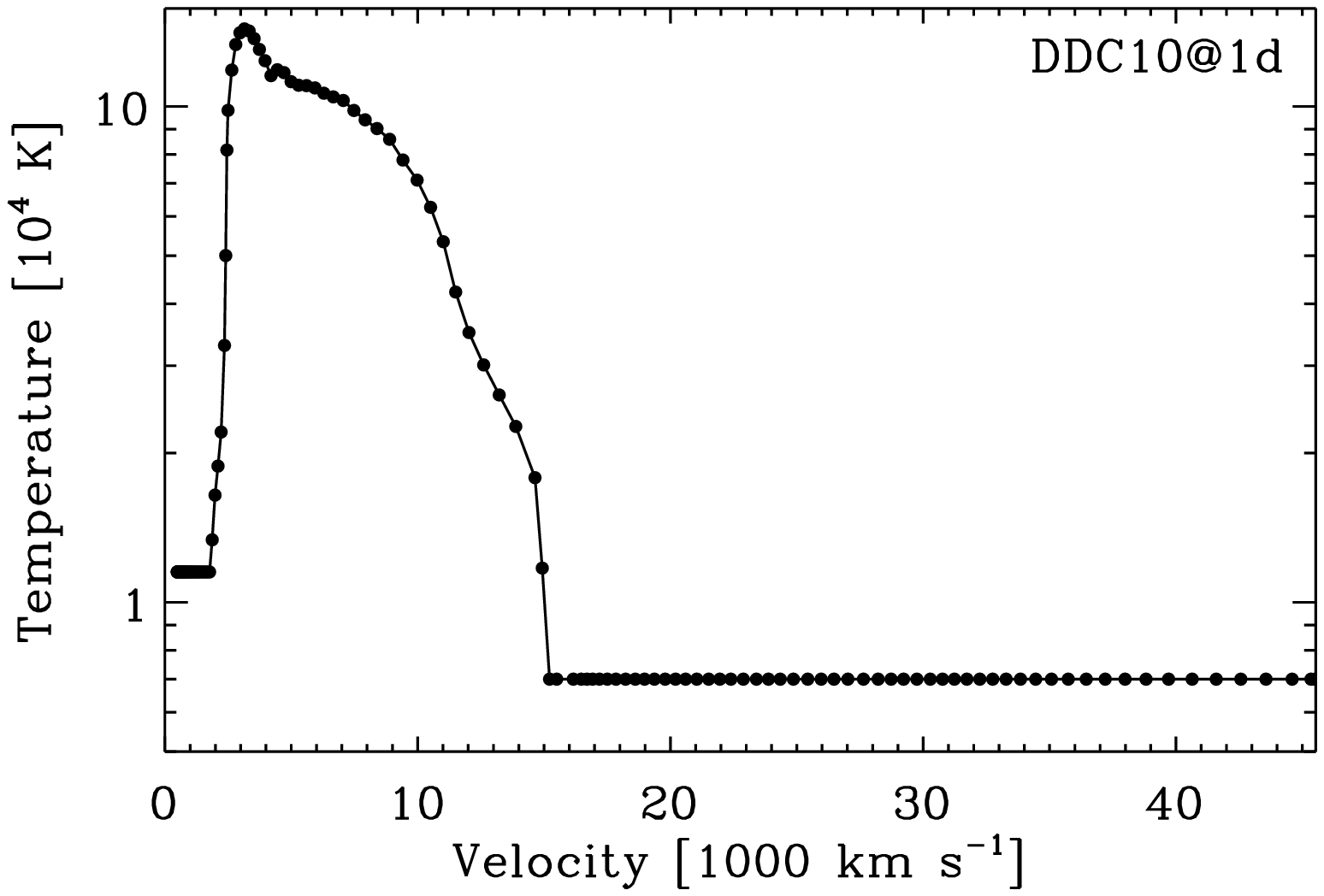,width=8cm}\\
\epsfig{file=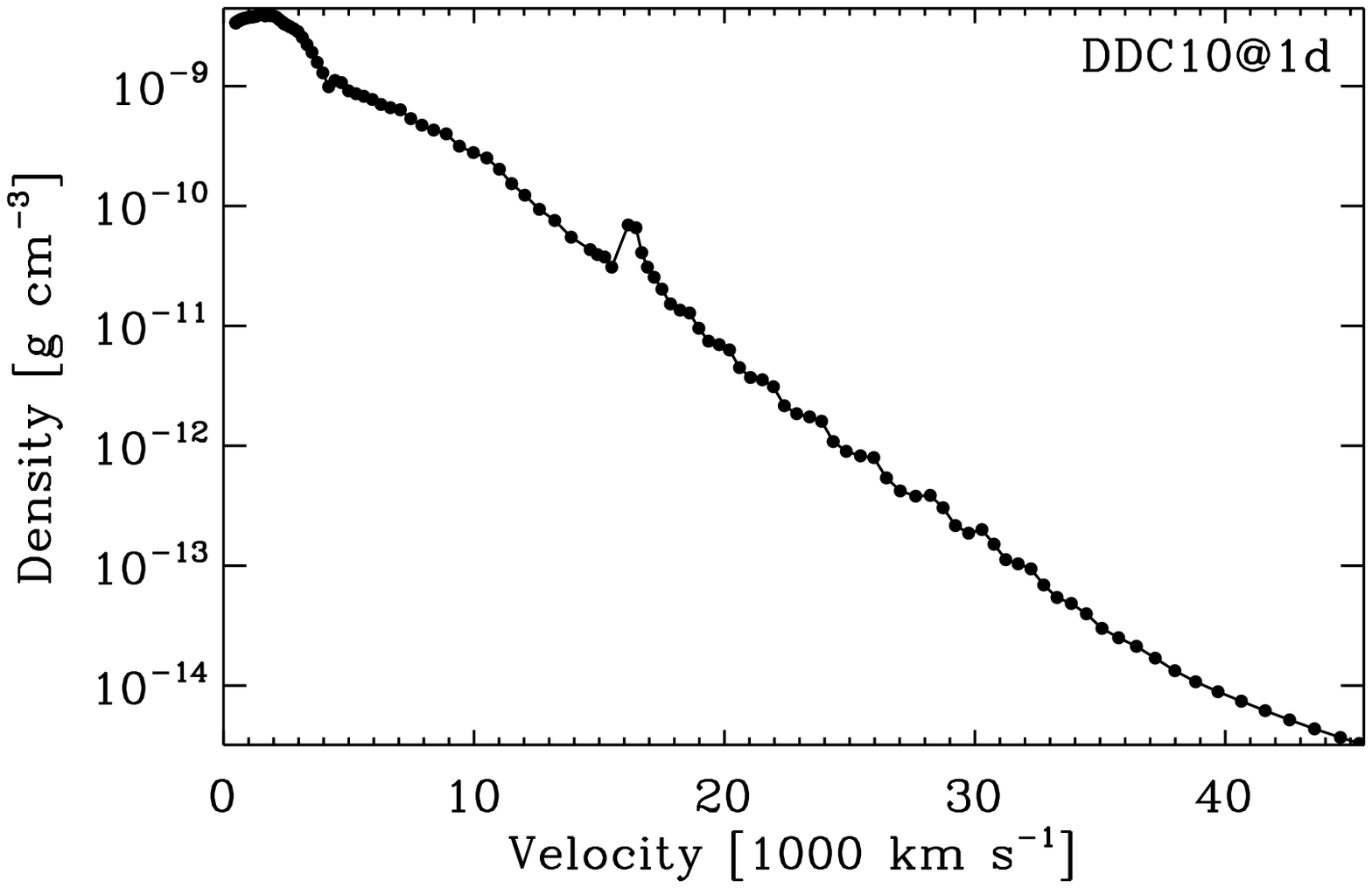,width=8.cm}\\
\end{minipage}
\caption{
{\it Top:} Illustration of the ejecta chemical stratification in velocity and mass space (top axis) for model DDC10
and for representative species including C/O/Ne, and representative IMEs and IGEs.
We also overplot the \iso{56}Ni distribution as a dotted line. Note the absence of \isoni\  in the inner ejecta layers.
The time is 29\,sec after explosion.
{\it Middle:} Ejecta gas temperature at 1\,d after explosion. Note the enforced floor temperature
in the outer ejecta, which is necessary for the initial relaxation of the first model in any time sequence.
{\it Bottom:} Same as middle, but now for the mass density.
\label{fig_hydro}
  }
\end{figure}

\section{Numerics}
\label{sect_numerics}

   All simulations presented in this work start from the same delayed-detonation
model DDC10. We refer the reader to \citet{blondin_etal_13} for a description
of the hydrodynamical model and the basic set up for the radiative transfer calculations.

   \cmfgen\ works in the same standard
form for any SN ejecta. Hence, the simulations we present here are all carried out
with the standard line-blanketed non-LTE time-dependent radiative-transfer technique
presented in \citet{HD12}. What differentiates one SN simulation from another is the
hydrodynamical input (composition etc.) and the model atoms employed.
Because the SN Ia ejecta are thin early on,  we allow for non-local energy deposition
in all models beyond 10\,d after explosion, unless stated otherwise.
As discussed in the appendix of \citet{HD12} we use a Monte Carlo approach for the $\gamma$-ray transport.
Similarly, given the large abundance of unstable nuclei, we treat non-thermal
processes, with the method presented in \citet{li_etal_12}.
In its original form, the non-thermal solver included excitation rates for all ions but ionization
rates only for the ions specified in an input file. In the course of this work, we realized that not
all intermediate-mass elements (IMEs) and
iron-group elements (IGEs) were included in this file. We had C, O, Ne, Na, Mg, Al, Si, S, Ar, Ca, Fe, and Ni,
but by mistake did not have entries for Ti, Cr, and Co. This was corrected in model DDC10\_A4D1
using additional cross sections from \citet{mazzotta_etal_98}, to complement those provided
by \citet{AR85}.

The flexibility in \cmfgen\ permits the testing of numerous effects, in particular the
influence of model atoms whose characteristics can be easily adjusted
(e.g., number of levels, number of transitions, super-level assignments, source of atomic data; 
see Appendix~\ref{appendix_atom}).
In our simulations, we include C\one--\four, O\one--\four, Ne\one--\three,
Na\one, Mg\two--\three, Al\two--\three, Si\two--\four, S\two--\four, Ar\one--\three,
Ca\two--\four, Sc\two--\three, Ti\two--\three, V\one, Cl\four, K\three, Cr\two--\four,
Mn\two--\three, Fe\one--\sev, Co\two--\sev, and Ni\two--\sev.
When a given ion becomes sub-dominant at all depths because of recombination,
its impact on the radiative transfer becomes negligible and  it is conveniently excluded from the
computation at all subsequent times.

Species V, Cl, K are only included if/because they belong to decay routes
that we wanted to incorporate in the calculation (see below). However, because we are not attempting
to describe accurately their potential impact on the transfer (and also because we do not have
a satisfactory set of model atoms for the corresponding ions), we only include the ground state
of a single ionization state for each of these species (i.e., V\one, Cl\four,
and K\three).

   We discuss the properties
of SN Ia radiation at times that encompass the initial brightening of the object (from
day one to the light curve peak), the maximum light properties (see also \citealt{blondin_etal_13})
and the transition to the nebular phase until 60\,d after explosion.
This bridges very diverse conditions --- from optically thick to optically thin and with different processes
dominant in different density regimes. The code handles this evolution in a smooth fashion.
In practice, the inner boundary for the radiative transfer is the same as the inner boundary of
the ejecta, i.e., we do time-dependent simulations for {\it the full ejecta} at all times. In model DDC10,
this inner boundary is at a velocity $V_0$ of 490\,\kms. Because there is no dynamics in our simulations,
the velocity distribution in mass space is fixed throughout a \cmfgen\ sequence.
When the ejecta is optical thick, we impose a zero-flux condition at $V_0$, i.e., $H_{\nu}=$\,0 at
all frequencies $\nu$, and the incoming intensity $I_{\nu}^{+}$ is set equal to
the local Planck function at $\nu$.
When the ejecta turns thin, we impose a nebular condition at $V_0$, i.e.,
$I_{\nu}^{+}=I_{\nu'}^{-}$, where $\nu$\ and $\nu'$ are shifted to account for the Doppler
shift along the ray (see \citealt{HD12} for details). It is important to realize that, in this approach,
the entire ejecta is modeled at all times by \cmfgen, i.e., that the radiative transfer is solved at
all depths with merely a change in inner boundary condition at $V_0$ when the ejecta turns
nebular. This transition occurs earlier in the near-IR than in the UV, so we tend to switch to
the nebular condition when the Rosseland-mean optical depth is still well above unity.

\begin{table}
\caption{Summary of model assumptions for our SN Ia radiative-transfer calculations with \cmfgen.
All radiative-transfer simulations are based on the  delayed-detonation model named DDC10, which
is characterized by an initial  \iso{56}Ni mass of 0.65\,\msun\  (see \citealt{blondin_etal_13} for details),
and start at one day after explosion.
Simulations include either one (\iso{56}Ni) or all 2-step decay chains presented
in Table~\ref{tab_nuc1}--\ref{tab_nuc2}.
$\gamma$-ray energy deposition is treated as local (``L") or solved for using a
Monte Carlo transport approach (``NL").
Atom refers to the characteristics of the model atoms used for the \cmfgen\ calculations.
\label{tab_modset}}
\begin{tabular}{lc@{\hspace{2mm}}c@{\hspace{2mm}}c@{\hspace{2mm}}
c@{\hspace{2mm}}c@{\hspace{2mm}}}
\hline
\hline
Model &   Decays   & E$_{\rm dep}$ & Non-thermal   & Atom   \\
\hline
DDC10\_A0              & \iso{56}Ni    & NL & Yes &     Small  \\
DDC10\_A1              & \iso{56}Ni   & NL & Yes  &     Big   \\
DDC10\_A2$^a$     & \iso{56}Ni   & NL & Yes  &     Huge   \\
DDC10\_A1D1         & 2-step      & NL & Yes &    Big    \\
DDC10\_A3               & \iso{56}Ni   & NL & Yes  &   Big + [Co\three]   \\
DDC10\_A3D1          &  2-step      & NL & Yes  &    Big + [Co\three]    \\
DDC10\_A3L             & \iso{56}Ni    &  L & Yes  &     Big  + [Co\three]  \\
DDC10\_A3T             & \iso{56}Ni   &   NL & No  &     Big + [Co\three]   \\
DDC10\_A4D1$^b$ &  2-step      & NL & Yes  &    Huge + [IME,IGE]  \\
\hline
\hline
\end{tabular}
\begin{flushleft}
$^a$: For model DDC10\_A2, we only perform a few calculations at selected
post-explosion times, rather than computing a full sequence (Section~\ref{sect_opac}).
$^b$: The model atom for DDC10\_A4D1 is the same as for model DDC10\_A2,
but also includes forbidden line transitions of all metals.
Furthermore, unlike previous simulations, non-thermal ionization is included for all IMEs and IGEs.
\end{flushleft}
\end{table}

  To begin a sequence we map the explosion structure into \cmfgen\ although
  we needed to impose a floor temperature of $\sim$\,6000\,K.\footnote{Newer
  models can use a substantially lower floor temperature.} This artificially stores energy in the
  corresponding layers.
  Hence, in all time sequences, the ejecta needs to first relax by radiating away this excess energy.
  This usually takes only a few time steps since these layers have a relatively low optical depth.
  We generally exclude such results from the presentation and focus on times sufficiently advanced
  that this initial tinkering bears no impact on the ejecta and radiation properties.
  The general lack of SN Ia observations prior to $\lesssim$\,1\,d and computational
  tractability motivates a start time at about 1\,d after explosion, which is our standard choice here.
  This time is early enough so that the initial evolution until
  1\,d after explosion can be done assuming no diffusion (we do this with a separate program; see
  \citealt{dessart_etal_11}).
  The standard procedure in \cmfgen\ is to adopt a time step equal to 10\% of the current time.

  We show various ejecta properties in Fig.~\ref{fig_hydro}. In the top panel, we plot the chemical stratification
versus velocity and Lagrangian mass at the end of the hydrodynamical simulation --- the time is then 29\,s
after the start of the combustion in the WD. We only show representative species, namely
C (unburnt), O (unburnt or produced by C burning), the most abundant IMEs, and the IGEs Fe and Ni.
Particularly striking is the low \nifs\ abundance at velocities less than 2000 km/s (termed the nickel hole).
This ``hole" is a signature of 1-D Chandrasekhar-mass SN Ia explosions and it stems from the relatively high central
density of such massive WDs.\footnote{This feature, however, does not seem to persist in 3-D simulations of delayed 
detonations \citep{seitenzahl_etal_13}.}
The lower two panels describe the temperature and the mass density at 1\,d after explosion.
The gridding of the \cmfgen\ calculation, which is approximately equally spaced on a logarithmic
optical-depth scale (but with constraints on the change in velocity across grid points), is shown
with symbols. We typically use 110 depth points in our radiative-transfer simulations,
but this number can increase or decrease by $\sim$\,10\% depending on ejecta conditions (e.g., formation
of steep ionization fronts). As time proceeds, the spectrum formation region recedes to deeper layers
so we tend to reduce the maximum radius (or velocity) with time, while keeping the same number
of depth points; the resolution thus improves as we progress along a time sequence.

In the following sections, we describe the various models we have computed
(see also the summary given in Table~\ref{tab_modset}).
For model DDC10\_A2, which treats nearly two million
lines in non-LTE, we only do a few simulations at selected times.

\subsection*{Variations on adopted model atoms}

   We have used 5 different sets, including a small
(suffix A0), a big  (A1,A3), and a huge model atom (A2, A4). We have also run sets
of sequences in which the Co\three\ model atom was modified to include
forbidden-line transitions (A3).
The most complete model atom is A4, which we use in one sequence after bolometric maximum.
It includes a huge model atom of the same size as A2 (with emphasis on
Co\two\ and Co\three), but forbidden line transitions are also included for all ions
associated with IMEs and IGEs. In model A4, we also updated several atomic models.
A description of the model
atoms is provided  in Appendix~\ref{appendix_atom}. The specific model
names are DDC10\_A0, DDC10\_A1, DDC10\_A2, DDC10\_A3 and DDC10\_A4.

   Extensive testing has shown that employing a large model atom for iron
is necessary, even when that species is not very abundant, as in Type II SNe.
So, all sets of model atoms include a large model atom for Fe\two\ to
Fe\four\ \citep{DH10,li_etal_12,dessart_etal_13,dessart_etal_13b}--- other ionization stages
of Fe are given a modest-size model atom because they do not dominate
and are only present at times when and locations where radiative diffusion is
very inefficient due to the small photon mean-free-path in the corresponding
ejecta regions.

\subsection*{Local versus non-local energy deposition}

   Explosions of Chandrasekhar-mass WDs yielding ejecta with a kinetic
energy on the order of 1\,B become thin to $\gamma$-rays as early as two weeks
after explosion (see, e.g., \citealt{hoeflich_etal_92}). To document the implications
on both SN Ia spectra and light curves, we have computed a sequence where local energy
deposition is assumed (model DDC10\_A3L) --- all other simulations are performed
with allowance for non-local energy deposition past ten days after explosion,
using the $\gamma$-ray Monte-Carlo transport code described in the appendix of \citet{HD12}.

\subsection*{Influence of decay chains included}

Because of the prevalent role of \iso{56}Ni and \iso{56}Co in controlling SN Ia
radiative properties, the general custom is to include only that decay chain.
In reality, these explosions produce a variety of unstable nuclei, either IMEs or IGEs,
that take part in 2-step or 1-step decay chains.
These nuclei have a range of life times, from less than a day to years, and can
thus influence SN Ia ejecta on very different time scales.

In this paper, we thus explore the effect associated with the treatment of additional
2-step decay chains (Table~\ref{tab_nuc1}--\ref{tab_nuc2}; additional 1-step decay chains,
with an especially important influence at  early times, is discussed in \citealt{dessart_etal_14a}).
Besides their impact on the internal energy of the gas, these decays modify the composition
and hence can alter the predicted spectra through changes in line-blanketing.

Some variants of model DDC10 are thus run with different assumptions regarding
nuclear decay -- some isotopes may be treated as stable even if unstable
by nature. In \cmfgen, physically-unstable isotopes are treated as unstable
only if the associated decay route is considered in the calculation.
By default,  simulations include only the \iso{56}Ni 2-step decay chain.
Simulations that include all 2-step decay chains described in Appendix~\ref{app_decay}
have suffix ``D1''.

\subsection*{Influence of non-thermal processes}

   In \citet{li_etal_12} and \citet{dessart_etal_12}, we have presented our treatment of
non-thermal processes arising from $\gamma$-ray emission of unstable nuclei.
We test their importance by running the model sequence DDC10\_A3T in which these
non-thermal processes are ignored --- all other model sequences include non--thermal
processes.

   In the course of this work, we realized that all models and their variants up to DDC10\_A3
   (see Table~\ref{tab_modset}) did not include
   non-thermal ionization for Ti, Cr, and Co, but treated non-thermal excitation as
   expected.
   We thus run the new model DDC10\_A4 with non-thermal ionization accounted for for all species/ions.
   Non-thermal processes are consequently stronger in model DDC10\_A4
   than in DDC10\_A3, implying that the differences we discuss with model DDC10\_A3T are
   in reality even larger. We do not show that specific comparison because model DDC10\_A4
   differs with model DDC10\_A3T in more ways than the non-thermal treatment alone.
   For historical reasons, the sequence using model atom A4 also included all 2-step decay chains,
   so this model is called DDC10\_A4D1.

\begin{figure}
\epsfig{file=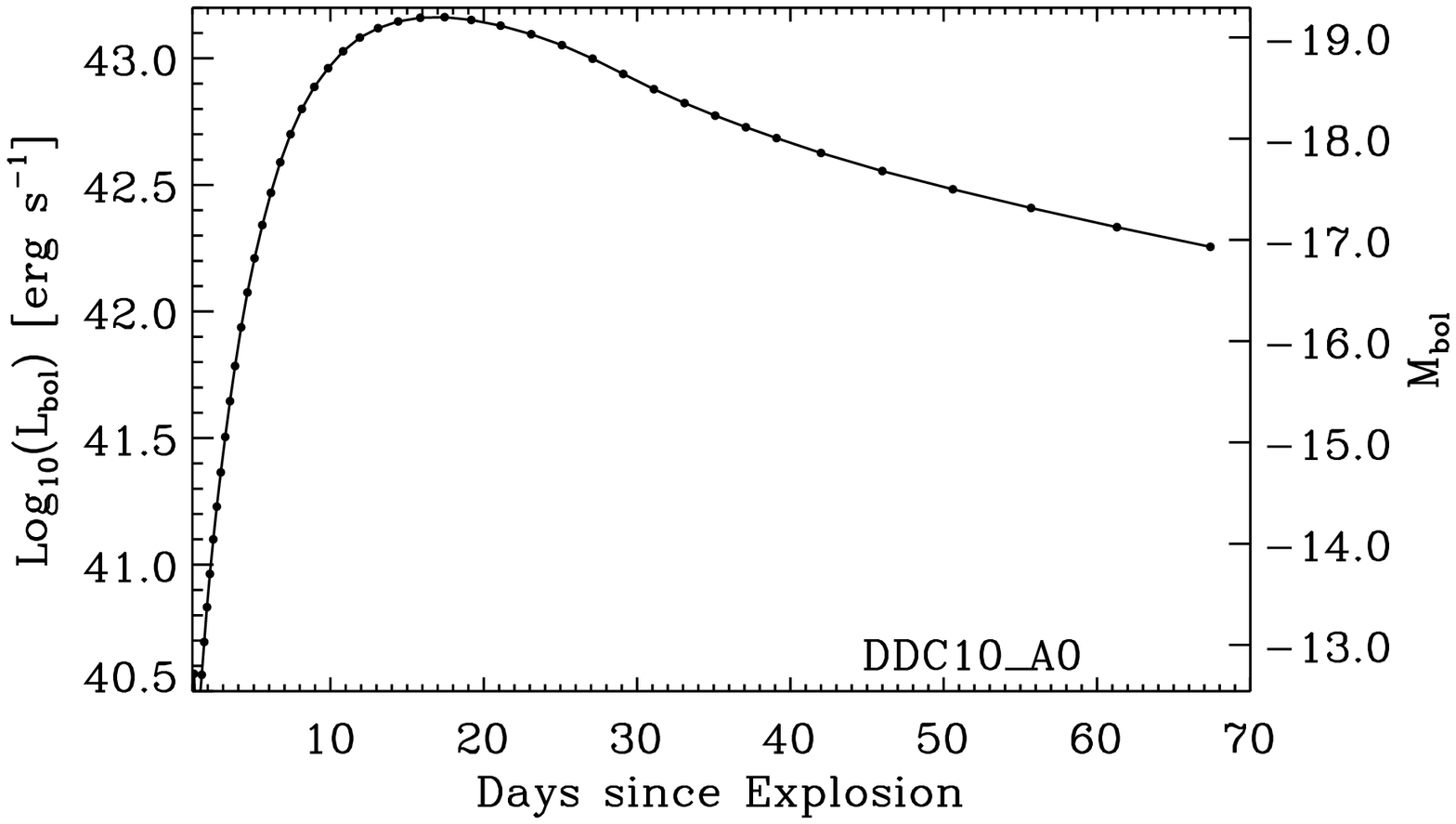,width=8.cm}
\epsfig{file=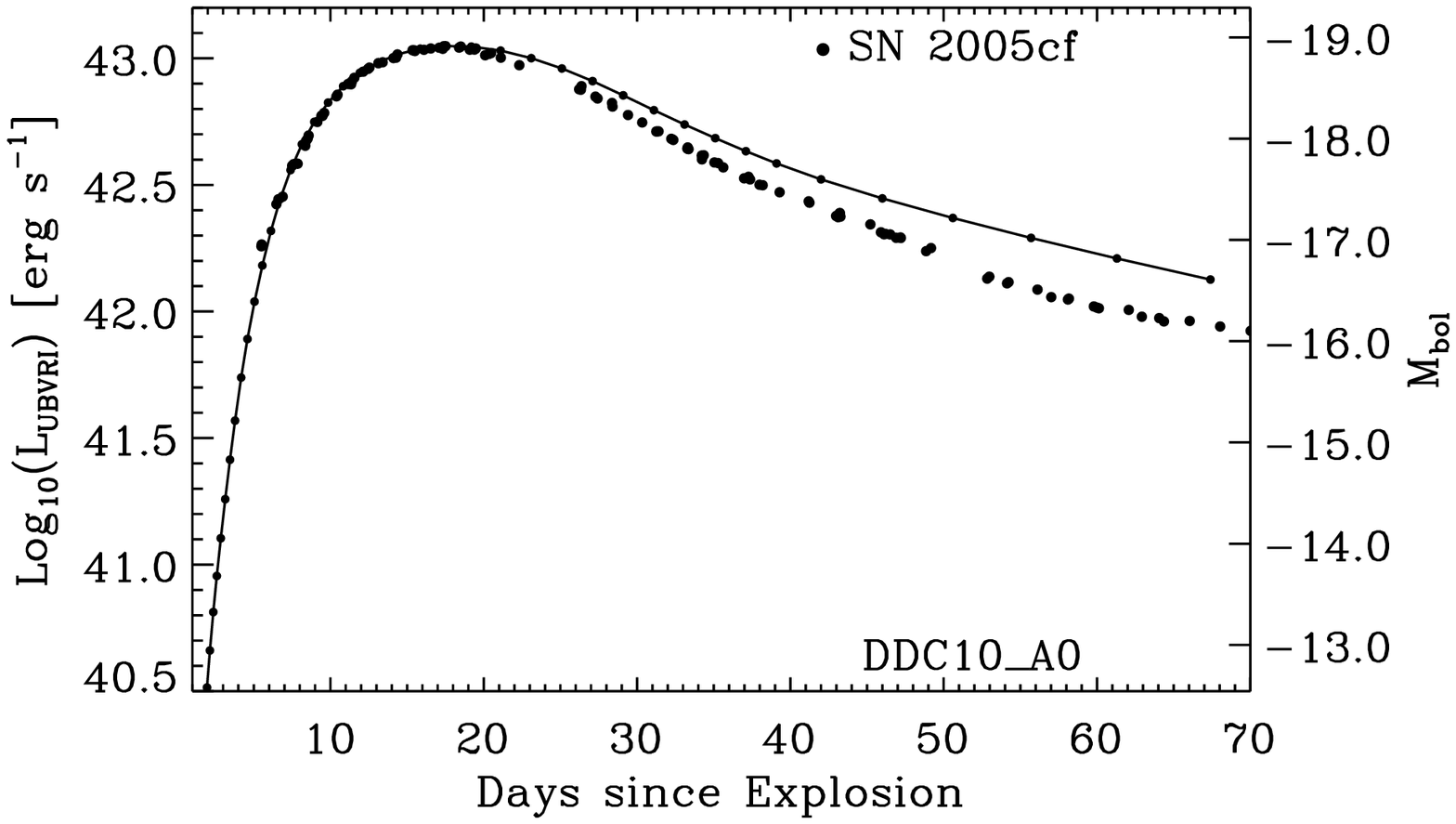,width=8.cm}
\epsfig{file=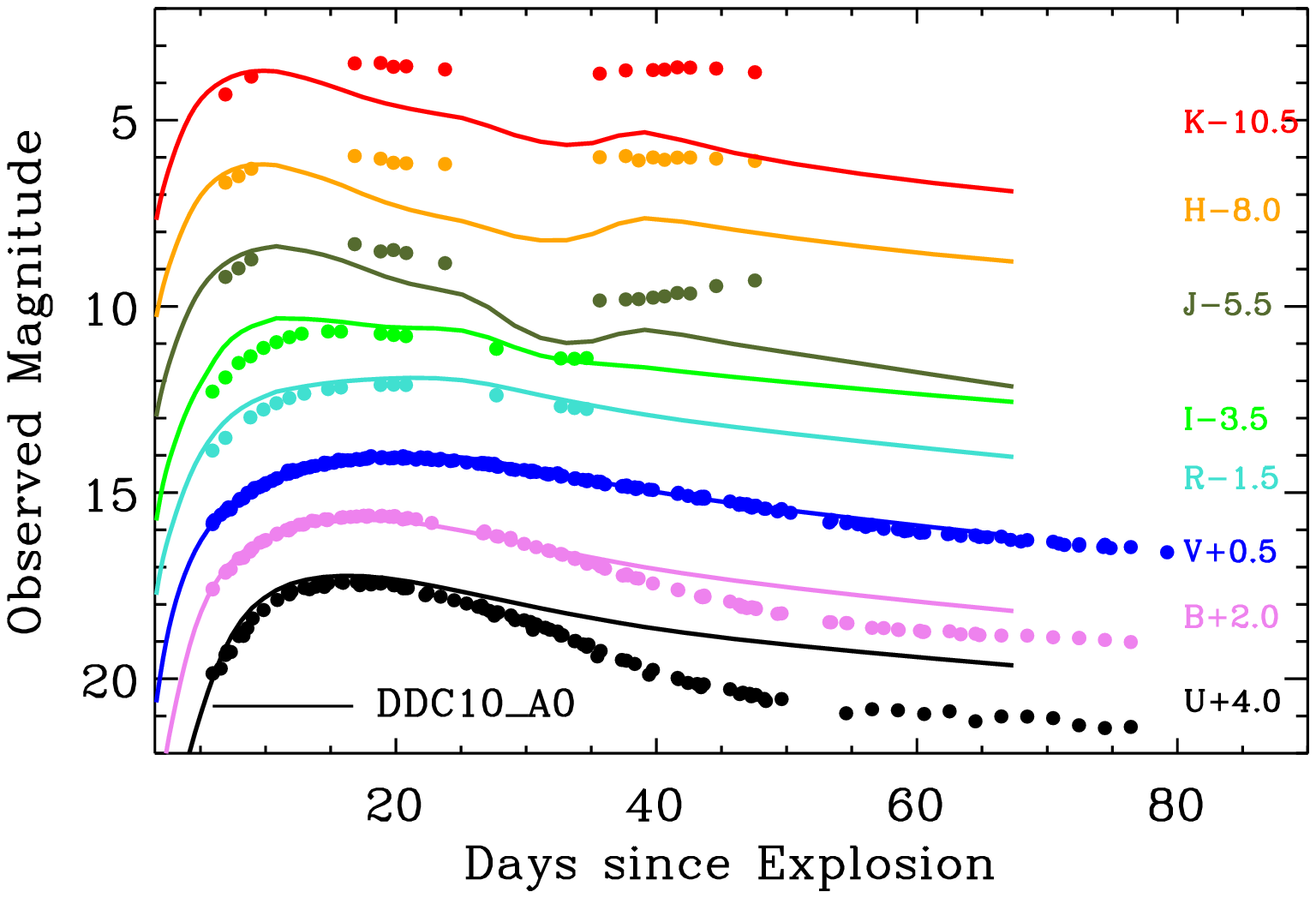,width=8.cm}
\caption{
{\it Top:} Bolometric light curve for model DDC10\_A0.
{\it Middle:} Same as top, but now showing the evolution of the fractional luminosity
falling over the photometric passbands $UBVRI$. For comparison, we show the
corresponding luminosity for SN\,2005cf inferred from its $UBVRI$ photometry.
{\it Bottom:} Same as in middle panel, but now showing a comparison of
model DDC10\_A0 photometry and counterparts for SN\,2005cf
--- the size of filled-dots is $\sim$\,0.25\,mag.
While the V band is matched at all epochs, there is very poor agreement beyond maximum
for the $UBJHK$ bands.
We use a distance modulus of 32.17\,mag, an $E(B-V)$ of 0.22\,mag, and an $R_V$ of 3.1.
\label{fig_lbol_A0}}
\end{figure}

\subsection*{Comparison to observations}

   Throughout this paper, we test the compatibility of our synthetic spectra and multi-band light curves
against the well observed SN\,2005cf.
We use the optical spectra published in \citet{garavini_etal_07} and \citet{bufano_etal_09}.
We use near-IR  spectra published in \citet{gall_etal_12}, from which we also adopt
the $B$-band maximum time of JD\,2453534.0.
We also use photometry from \citet{pastorello_etal_07}. As in \citet{blondin_etal_13},
we adopt a distance modulus of 32.17\,mag and a total reddening (Milky Way and host galaxy)
$E(B-V)$ of 0.22\,mag \citep{wang_etal_09}. We use $R_V=$\,3.1 and
the extinction law of \citet{cardelli_etal_89}.

  We use SN\,2005cf primarily to test how our model DDC10, which has
  a \nifs\ mass similar to that inferred for SN\,2005cf \citep{blondin_etal_13},
  compares with its multi-epoch multi-wavelength
  observations. For each comparison, we take the model that is the closest to the
  observation time, implying an offset of $\lesssim$\,1\,d around bolometric maximum
  and $\sim$\,2\,d at late times.
  A detailed discussion of the match to specific line features, line
  identifications, line widths, colors, decline rates etc. is left to a future paper on the specific
  modeling of SN\,2005cf.

  Throughout this paper, we use a single ejecta model, i.e., DDC10, without altering any of its
  properties. In this sense, it would be fortuitous if our synthetic spectra matched every spectral
  feature and if our multi-band light curves matched all the photometric properties of this SN.
  A corollary is that model DDC10, with its 0.65\,\msun\ of \nifs, cannot be used to compare with
  any sub-luminous or super-luminous SN Ia, for obvious reasons.

\begin{figure}
\epsfig{file=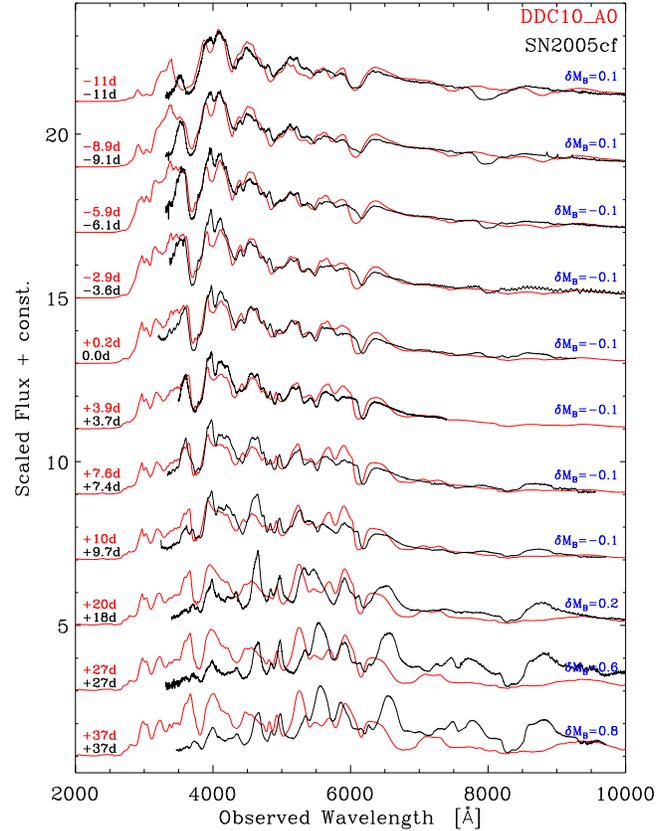,width=8.5cm}
\caption{
Comparison between the spectral evolution of model DDC10\_A0 (red) and the observations
of SN\,2005cf (black) --- we show $F_{\lambda}$ on a linear scale.
Times are given with respect to $B$-band maximum.
Synthetic spectra  have been reddened, redshifted, and scaled to match the distance
to SN\,2005cf.
Spectra are also scaled vertically for better visibility; the label
on the right gives the true $B$-band magnitude offset between model and
observations at each date.
After a good agreement up to bolometric maximum, model DDC10\_A0 and SN\,2005cf eventually
disagree, the model retaining a blue color that becomes more and more discrepant.
\label{fig_spec_A0}}
\end{figure}

\section{Reproducing the fundamental properties of  a SN I\lowercase{a}}
\label{sect_atom}

In this section, we present in a chronological way the work we have done.
For a number of years, we failed to reproduce the most fundamental
color properties of SNe Ia beyond the peak of the light curve, for reasons that became clear
only recently.

\subsection{Statement of the problem}

Previous studies have emphasized the difficulty of modelling SNe Ia -- the ejecta are
rich in metals and the problem is time dependent \citep{pinto_eastman_00a,pinto_eastman_00b}.
While some invoke the need for millions to billions of lines to model the transport
adequately \citep{kasen_etal_08}, some attempts in the mid-90s, which obtained a satisfactory
match to observations, employed not even a million lines \citep{hoeflich_95}.
Although \citet{baron_etal_96} emphasized the importance of non-LTE effects, many
SN Ia simulations assume either full LTE for the gas state (populations, ionization) or
use a nebular approximation for the ionization together with LTE for the level populations.
Since reasonable fits have been obtained with a variety of techniques it is still unclear what are the
critical ingredients for modeling type Ia SNe.

\begin{figure}
\epsfig{file=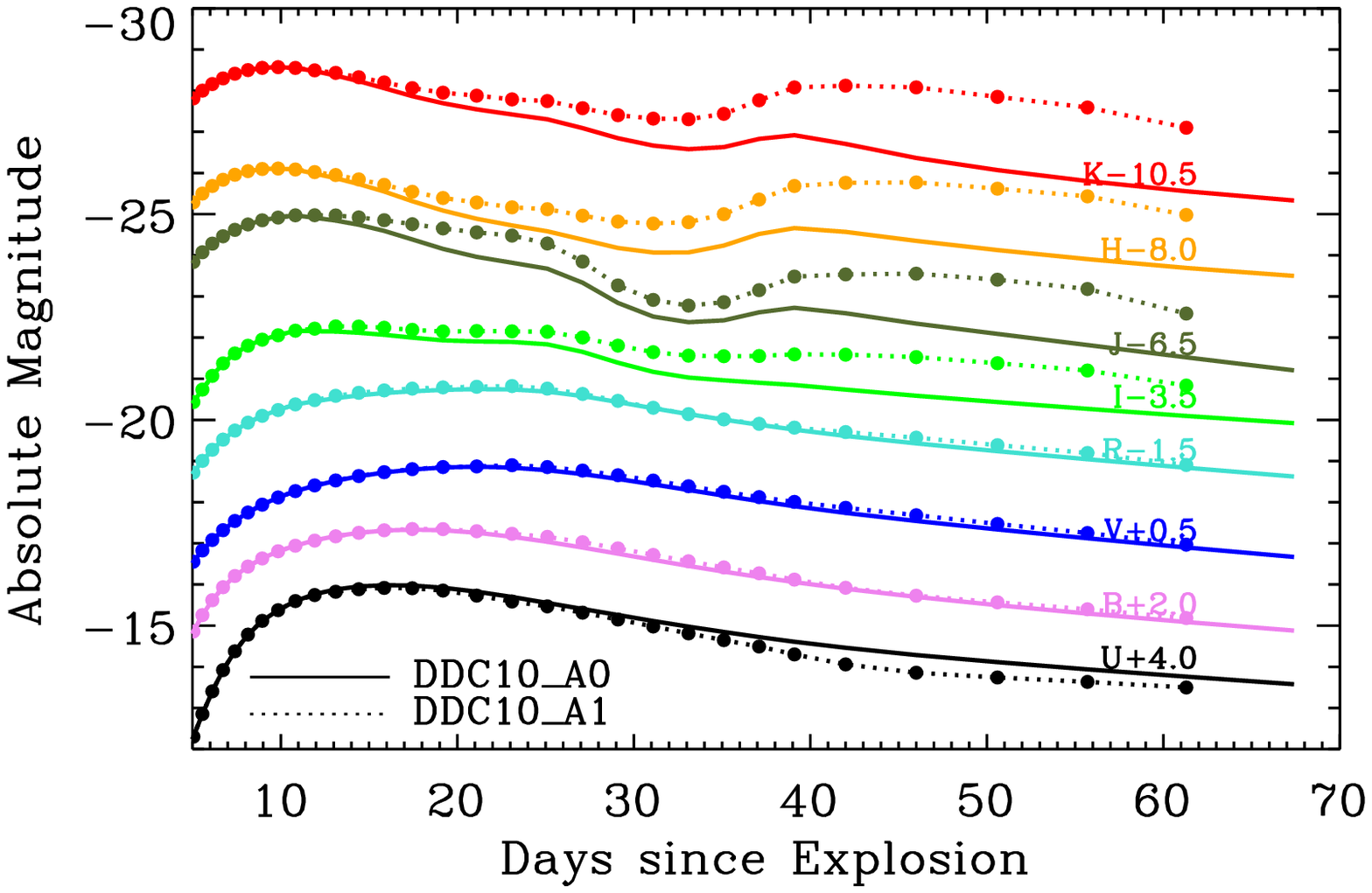,width=8.5cm}
\epsfig{file=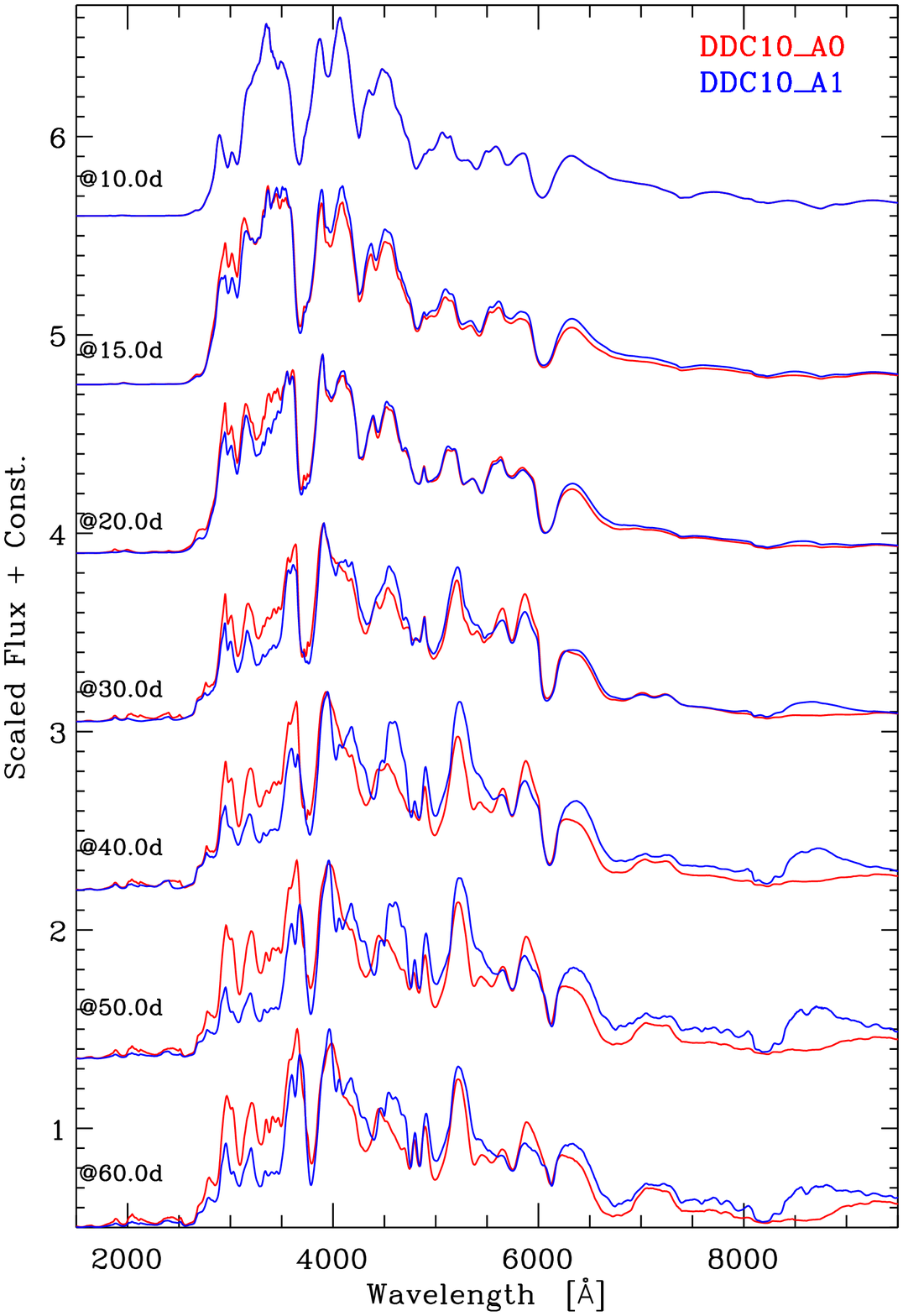,width=8.5cm}
\caption{Comparison between models DDC10\_A0 and DDC10\_A1
for the multi-band light curves (top) and spectral evolution (bottom).
Labels at left give the post-explosion time in days, which increases from top to bottom.
\label{fig_comp_A0_A1}}
\end{figure}

\begin{figure}
\epsfig{file=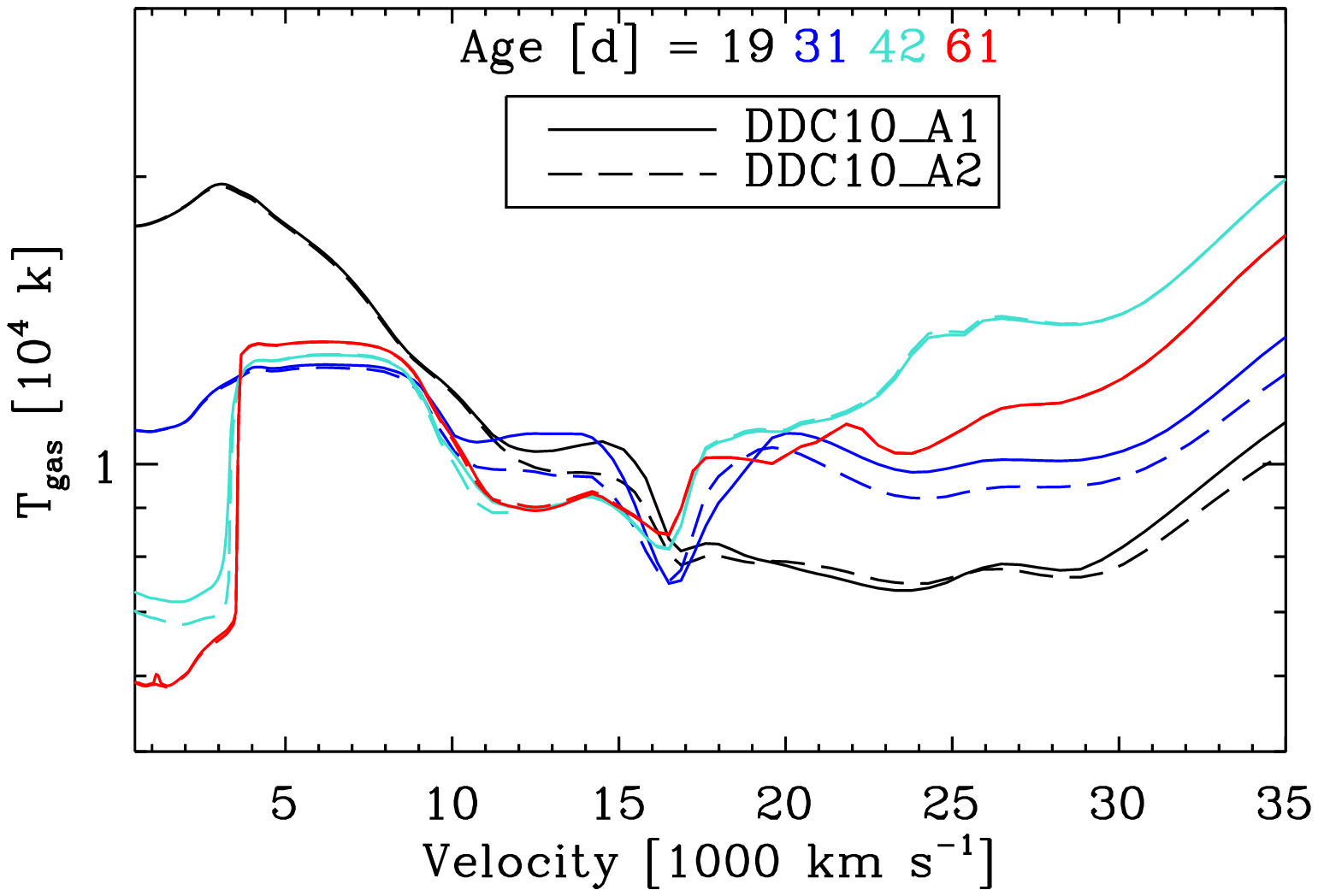,width=8.5cm}
\epsfig{file=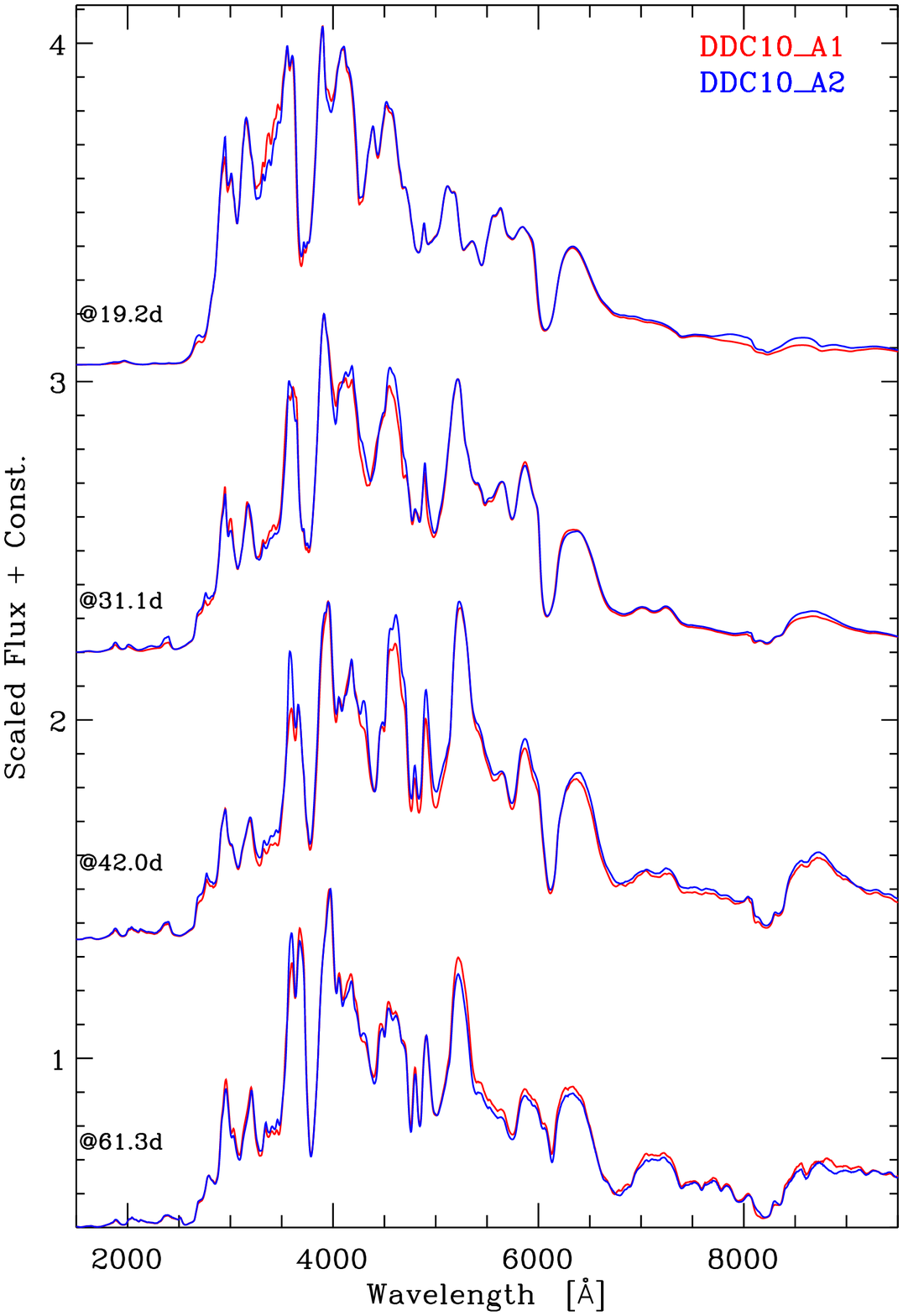,width=8.5cm}
\caption{Comparison at selected post-explosion times of the gas temperature (top)
and synthetic spectra (bottom) between model  DDC10\_A1 and DDC10\_A2.
The increase in the total number of transitions from 629\,396 to 1\,738\,088
has essentially no impact on the spectra while the effect on the temperature is weak.
\label{fig_comp_A1_A2}}
\end{figure}

   Over the years, we have explored the influence of model atoms on our results
   in the context of core-collapse SNe \citep{DH10,DH11,dessart_etal_11,li_etal_12,
   dessart_etal_13}. This revealed the critical importance of including large model atoms
   for iron, primarily Fe\one\ and Fe\two\ in the low ionization conditions of Type II/Ib/Ic SNe.
   The ionization conditions in SN Ia ejecta are typically much larger, so we started with large
   model atoms for Fe\two-\three-\four. The complete model atom for model DDC10\_A0 is
   given in Table~\ref{tab_atom_A0}. Starting at day one, we evolved the ejecta up until 60\,d
   after explosion and show the resulting bolometric light curve in the top panel of Fig.~\ref{fig_lbol_A0}.
   We obtain a rise time of 17.43\,d to a maximum bolometric luminosity of  1.45$\times$10$^{43}$\,\ergs.

   Since it is not possible to directly compare the bolometric luminosity to observations, we confront
   the fractional luminosity falling over the bands $UBVRI$ in model DDC10\_A0 and infer
   the corresponding quantity for SN\,2005cf.
   The agreement is satisfactory around the peak of the light curve, but the agreement
   becomes poorer as time progresses into the nebular phase (middle panel of Fig.~\ref{fig_lbol_A0}).
   This disagreement could arise in two ways --- we have too much late time energy deposition
   in the model, or alternatively, there is too much model flux coming out in the $UBVRI$ bands.
   In model DDC10\_A0, the bulk of the flux comes out in the range 3000\,\AA--1\,$\mu$m.
   This fraction rises steadily from 70\,\% at 1\,d to 90\,\% at the peak of the light
   curve, and it stays at 90\,\% until the end of the simulation at 60\,d. The flux falling
   shortward of 3000\,\AA\ is typically a few percent, and that falling in the near-IR is typically
   half that.

To understand the color evolution, we compare the multi-band light curves of
model DDC10\_A0 with those observed for SN\,2005cf (bottom panel of Fig.~\ref{fig_lbol_A0}).
While we obtain a good match to the $V$-band light curve at all times, the match to other bands
is satisfactory only up to the peak. Beyond the peak, the model is obviously too blue, showing
excess flux in the $U$ and $B$ bands, and a flux deficit in the near-IR.

 Spectroscopically, the mismatch between model DDC10\_A0 and the observations of SN\,2005cf
 is striking (Fig.~\ref{fig_spec_A0}). Up to the peak, the model shows the very standard SN Ia
 signatures, although the
 spectral-energy distribution (SED) is somewhat too red initially. The basic
 morphology of line profiles is also well matched. However, as time proceeds beyond bolometric
 maximum, our synthetic spectra are systematically too blue.
 This occurs in spite of the strong fading of the SN (well reproduced by our model),
 which is due to both the decreasing radioactive
 decay energy that is released and the increasing fraction of $\gamma$-rays that escape the ejecta.

  So, the problem with model DDC10\_A0 is not with the bolometric luminosity, or the rate at which
  radiant energy leaks out of the ejecta, but instead with the computed color evolution. Although a very basic
  property, SN color is one of the hardest  property to get right because it is sensitive
  to temperature, ionization, opacity etc. For example, increasing the size of the Fe\one\
  model atom in a SN II-P simulation yielded a fading of 2\,mag of our synthetic $U$ band magnitude
  \citep{DH11,dessart_etal_13}. Such a sensitivity is very problematic
  for the convergence of radiative-transfer results.

  The present problem is not limited to model DDC10\_A0. Earlier calculations
with \cmfgen, using the hydrodynamical inputs of \citet{kasen_etal_07},
yielded the same discrepancy. We also explored whether other
delayed-detonation models showed the same discrepancy. We tried a
new version of model N32 of \cite{hoeflich_khokhlov_96}, as well as the full series of models
of \citet{blondin_etal_13} which cover a factor of $>$5 in \iso{56}Ni mass, and
found that all resulting time sequences eventually develop this color problem
after the peak. The issue is probably related to the extraordinary conditions
existing in SN Ia ejecta. In the following sections, we explore various
routes to solve the problem.

\subsection{An opacity problem?}
\label{sect_opac}

A fundamental property of SN Ia ejecta that sets them aside from other SN ejecta is their unique
composition, split between IMEs and IGEs.
For once ionized atoms, the number of free electrons per nucleon goes from 1 for hydrogen, to 1/4 for helium,
to only 1/56 for \iso{56}Fe. Thus the electron scattering opacity per unit mass is much lower, by typically
more than an order of magnitude, than in Type II SN ejecta.

SNe Ia also have a representative ejecta kinetic energy of 1\,B for about a tenth of the ejecta mass of Type II
SNe. Compared to type II SNe, SN Ia ejecta have faster expansion rates and are characterized by lower
densities early on. Consequently, because of the weakening of the electron scattering opacity per unit mass
and the low ejecta density, the continuum mass absorption coefficient is reduced in Type Ia
compared to Type II SNe. However, in SNe Ia, metals, with their large mass fraction, are a strong source of additional opacity. The complex
atomic structures of metals, with their unfilled 3d/4s shell, leads to the presence of millions of lines, with those from iron and cobalt (e.g., Fe\two, Fe\three, Co\two, and Co\three) being of greater  importance.

 In model DDC10\_A0, we include all metal line transitions with a $gf$ value greater than 0.002 (but also
 limited by our adopted model atom).\footnote{There is no cut in the $gf$ value when we compute the final spectrum.}
 This cut only applies to elements whose atomic-mass number is greater
 than 20 (i.e., Ne), does not apply to the lowest $n$ levels
($n$ is typically 9), and a transition is omitted only when there are at least $m$ ($m$ is typically 9) stronger
downward transitions from the level. Thus, this procedure does not cut important transitions to ground levels,
and forbidden and semi-forbidden transitions among low-lying states. With the model atoms employed
in model DDC10\_A0 (Table~\ref{tab_atom_A0}), we include a total of 8370 (1773) full (super) levels
(see \citealt{HM98_lb} for a description of super-levels in \cmfgen), which corresponds to 174 674
bound-bound transitions.

We have a color problem with model DDC10\_A0, but in spite of the order of 10$^5$ lines included
in the simulation, which is rather small compared to the millions or billions of lines often invoked
\citep{kasen_etal_08}, the total $UBVRI$ flux and the $V$ band light curves are nonetheless
well matched. This suggests that the bulk of the energy diffusing out of the ejecta and producing the
SN Ia bolometric luminosity is not critically sensitive to the opacity, i.e., a reasonable description as in
model DDC10\_A0 is sufficient to capture the bolometric evolution. Paradoxically, the discrepancy with
observations occurs at later times when the ejecta turns optically thin, and thus when one would naively
think the opacity should matter less.

\begin{figure}
\epsfig{file=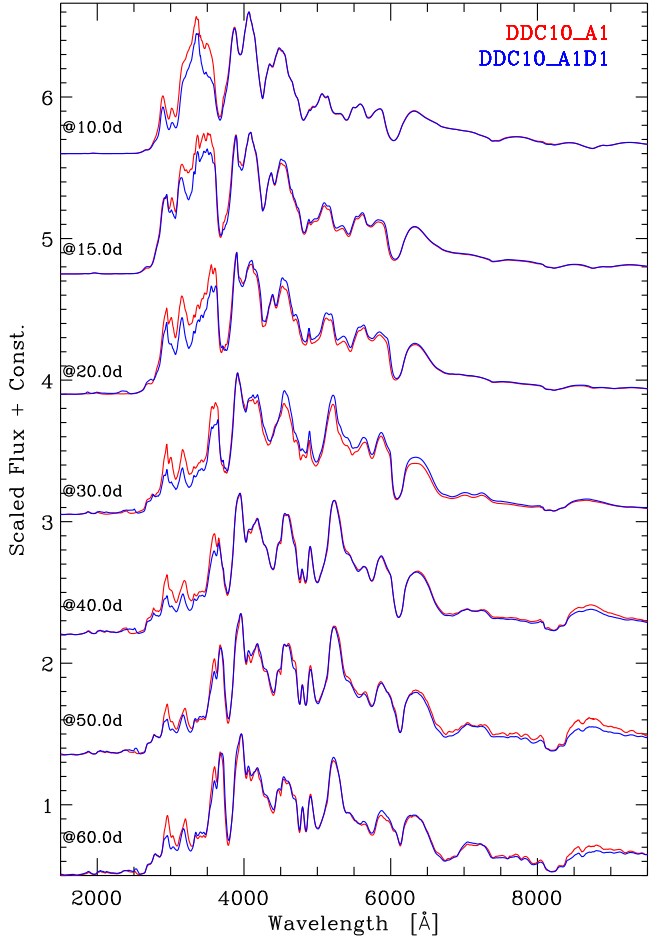,width=8.5cm}
\caption{Illustration of the impact on synthetic spectra of treating only one decay chain (associated
with \iso{56}Ni; model DDC10\_A1, red), or allowing for all 2-step decay chains presented in
Appendix~\ref{app_decay} (model DDC10\_A1D1, blue).
Note the influence of additional Ti\two\ opacity at $\sim$\,20\,d in the $U$ band, but the weak influence
of these decayed species throughout the DDC10\_A1D1 spectrum at nebular times.
\label{fig_comp_A1_A1D1}
}
\end{figure}

\begin{figure}
\epsfig{file=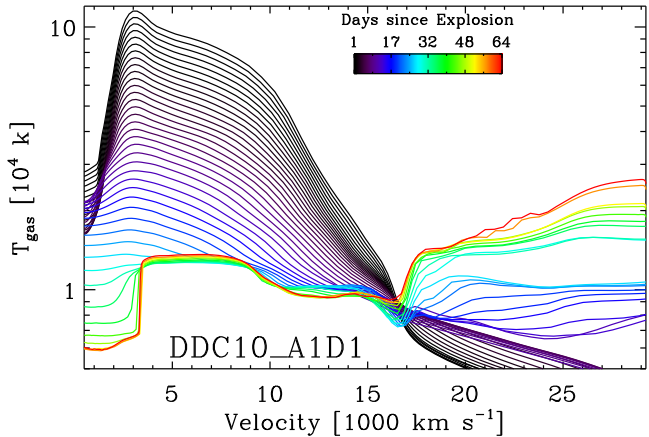,width=8.5cm}
\caption{Evolution of the gas temperature in model DDC10\_A1D1 from 1 to 64\,d after explosion.
Notice the development of a strong temperature jump at the inner edge of the \iso{56}Ni-rich region,
while the temperature in that region, between 3000 and 15000\,\kms\ retains a near constant
value at all times after bolometric maximum.
\label{fig_temp_A1D1}}
\end{figure}

We  investigate the effect of increasing the size of model atoms on model colors.
With model DDC10\_A1, we employ a larger model atom for Co\two-\four\ as well as Ni\two-\four\
(Table~\ref{tab_atom_A1}) for the modelling of near-peak and post-peak epochs.
We also lower the $gf$ cut from 0.002 to 0.0001. Model A1 includes 13 959 (2149) full (super) levels,
which corresponds to 629 396 transitions, which come primarily from Fe, Co, and Ni.
Despite such improvements, the radiative properties of models DDC10\_A0
and DDC10\_A1 remain very similar (Fig.~\ref{fig_comp_A0_A1}). The bolometric luminosity of each model agrees to within
a few percent at all times (shown further below), confirming that employing huge model atoms to solve
for the SN Ia bolometric luminosity is not critical. Enhanced opacity leads to enhanced blanketing in model
DDC10\_A1, which leads to a mild reddening of the colors, -- the brightness decreases in the blue
and augments in the red, in particular in the near-IR (top panel of Fig.~\ref{fig_comp_A0_A1}).
The impact on optical synthetic spectra remains small (bottom panel of Fig.~\ref{fig_comp_A0_A1}).
While 90\% of the flux falls within the range 3000\,\AA--1$\mu$m after bolometric maximum,
as in model DDC10\_A0, the UV and near-IR contributions are now at the same level ($\sim$\,5\,\%).

With model DDC10\_A2, we increase further the model atoms for Co\two\ and Co\three, with, respectively,
2747 (136) and 3917 (315) full (super) levels. The total number of full (super) levels
is 17533 (2338) and the total number of bound-bound transitions is 1\,738\,088.
Comparing these models at a few epochs,
we find that the difference is small (Fig.~\ref{fig_comp_A1_A2}),
comparable to or weaker than the change obtained between models A0 and A1.
This again is too small a change to resolve the color discrepancy.

\begin{figure*}
\epsfig{file=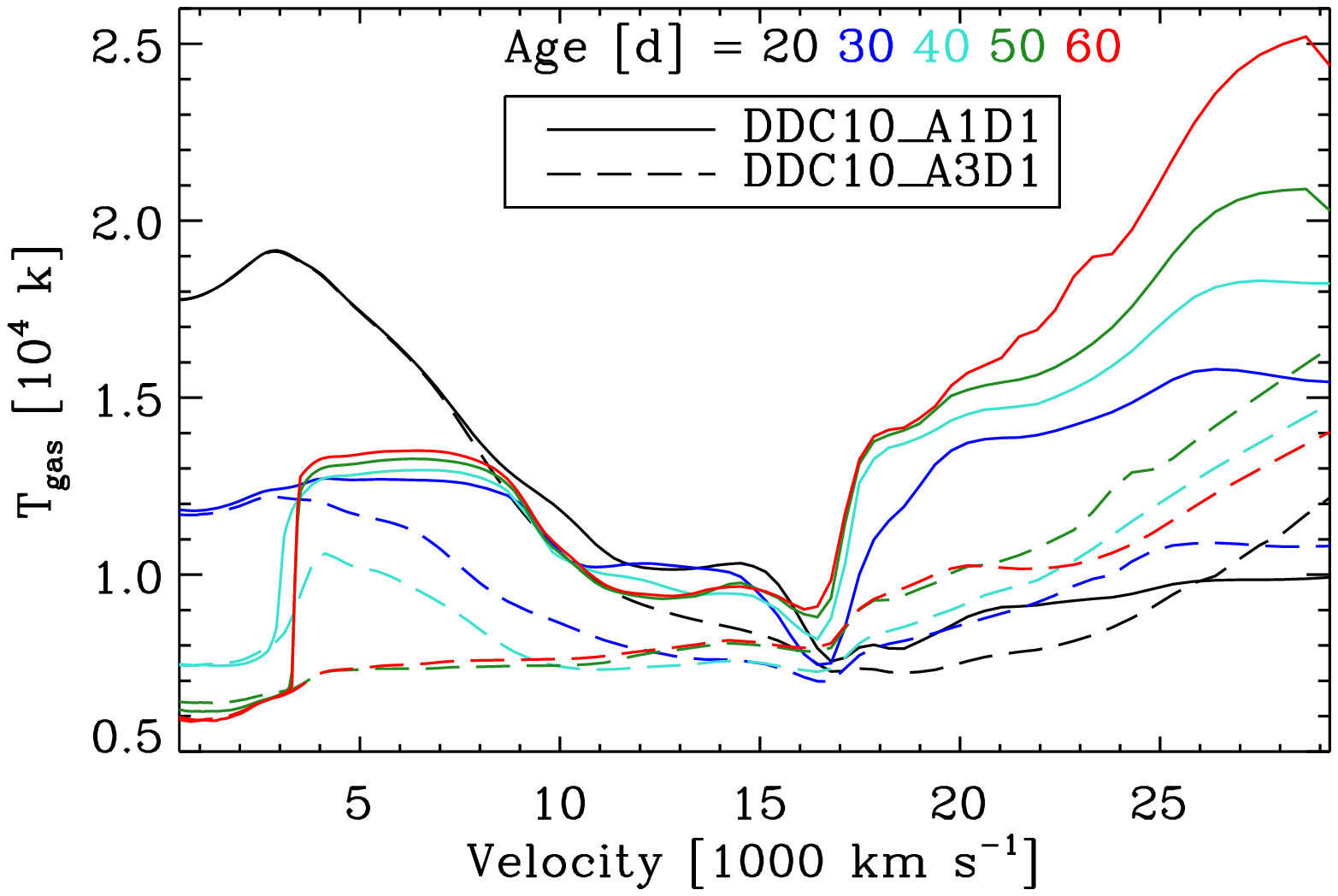,width=8.5cm}
\epsfig{file=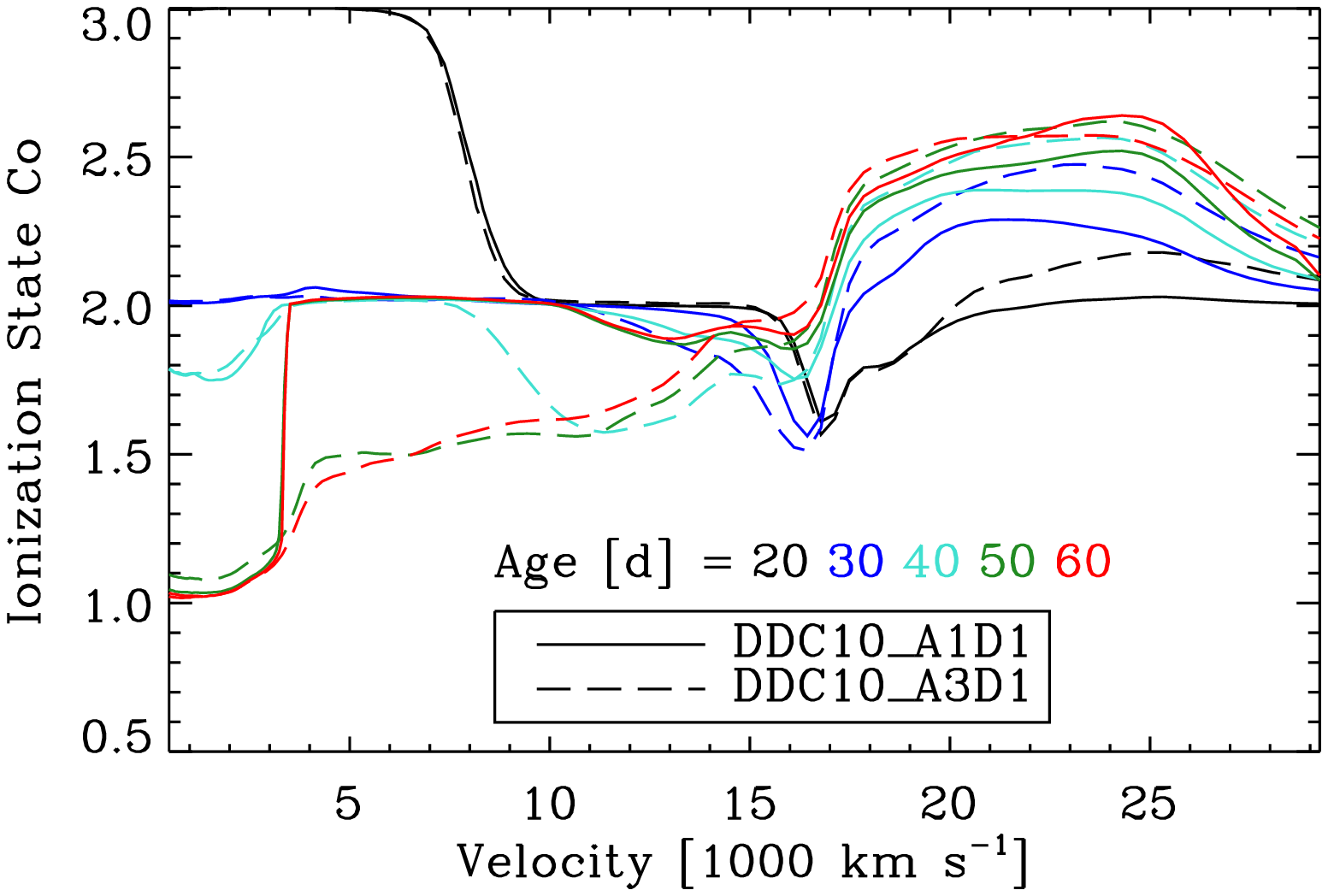,width=8.5cm}
\caption{{\it Left:} Evolution of the ejecta gas temperature with velocity (depth) and post-explosion
time for SN Ia models DDC10\_A1D1(solid)  and DDC10\_A3D1 (dashed). The only difference between
the two simulations is the treatment of forbidden-line transitions in model DDC10\_A3D1, while they
are ignored in model DDC10\_A1D1.
{\it Right:} Same as left, but now showing the ionization state for Co. Note the progressive recombination
from Co\three\ at light curve peak to Co\two--Co\three\ in the spectrum formation region at 60\,d
after explosion.
\label{fig_temp_ion}
}
\end{figure*}

The color problem we face concerns primarily the $U$ and $B$ bands. Rather than increasing the
opacity in these spectral regions by making the model atoms of IGEs more complete,
we finally investigated if a change in composition could help.
An obvious source of opacity in the $U$ and $B$ bands is Ti\two. Interestingly, \iso{48}Cr is an unstable
isotope at the origin of the chain \iso{48}Cr $\rightarrow$ \iso{48}V $\rightarrow$  \iso{48}Ti.
The first step has a half life of 0.89833\,d, and the second  step has a half life of 15.9735\,d,
hence comparable to the typical rise time of SNe Ia. Although the total mass of synthesized  \iso{48}Cr
is only on the order 0.001\,\msun, its mass fraction in the region 5,000--10,000\,\kms\ is 10$^{-3}$
while the mass fraction of Ti is  10$^{-5}$ (dominated by  \iso{44}Ti). Hence, by allowing for this decay chain,
the Ti mass fraction will rise by two orders of magnitude in this velocity range by the time the SN
gets to bolometric maximum.
In model DDC10\_A1D1, we thus repeat the model sequence DDC10\_A1
from scratch but now include all 2-step
decay chains compiled in Table~\ref{tab_nuc1}--\ref{tab_nuc2}. As for other time sequences, this simulation
took another 2-3 months. We show a spectral comparison between models DDC10\_A1 and
DDC10\_A1D1 in Fig.~\ref{fig_comp_A1_A1D1}. In the latter, the additional decay energy leads to
an increase in luminosity at the few percent level (shown further down). More importantly, the spectral differences
remain confined to the 3000-4000\,\AA\ region where the Ti\,\two\ opacity is the strongest 
\citep{filippenko_etal_92,nugent_etal_95,blondin_etal_13}.
At the light-curve peak, Ti is however three times ionized in those regions, interior to
10000\,\kms, where its abundance was most enhanced through \iso{48}Cr/\iso{48}V decay.

\subsection{Accounting for critical coolants}
\label{sect_sol}

  Part of the ambiguity with the color problem diagnosed above is that various processes
  can lead to a change of color. Historically, much of the color evolution of  SNe Ia has
  been associated with the redistribution of flux from the UV and blue part of the optical
  where the opacity is large to
  the near-IR where the opacity is low. This fluorescence process has been associated
  with an opacity issue -- the more complete the treatment of line opacity, the larger the number of
  transitions, the stronger the redistribution and the redder the SED
  \citep{hoeflich_etal_93,pinto_eastman_00b}.

  However, the color of the emergent radiation is also related to the temperature and
  ionization state of the gas. The hotter the gas, the bluer the SED.
  The thermodynamic state of the gas also controls
  what ions contribute to the opacity -- this  matters since higher ionization stages
  tend to have their opacity at shorter wavelengths.
  In Fig.~\ref{fig_temp_A1D1}, we show the evolution of the gas temperature in model
DDC10\_A1D1.  Crudely, the ejecta can be broken into three distinct regions,
each showing its own temperature evolution. In the outer region (above 17000\,\kms)
the gas is initially cool due to the rapid expansion of the ejecta. However, it gets hotter with time
due to the non-local energy deposition. These fast expanding layers have a very low density,
hence cool very inefficiently.

In the intermediate region (4000 to 17000\,\kms), where \iso{56}Ni is the most abundant,
there is initially a strong temperature gradient.
With time the gas cools,  the gradient  decreases, and the temperature levels off to a value of
$\gtrsim$\,10000\,K. From day 30 to 60 the temperature remains almost constant.

In the nickel hole, the temperature initially is much lower than that in the intermediate region.
Due to diffusion, and later on the deposition of energy by $\gamma$-rays, its cooling is slower so that
by day 30 the temperature across the hole, and out to 10000\,\kms, is roughly constant.
However, unlike the intermediate region, the ejecta continues to cool producing a large temperature
jump at $\sim$\,3000\,\kms.

  It now becomes clear that the color problem we face comes from the overestimated temperature
  in the spectrum forming region after the light curve peak. The material stays hot, the ionization high,
  and the SED appears blue. Models DDC10\_A0, DDC10\_A1, former attempts with model n32n,
  as well as other delayed-detonation models eventually form that temperature plateau and no longer
  cool. This does not affect their bolometric luminosity, which follows the decay energy deposition
  rate. What seems fundamentally discrepant is the cooling rate.

What held us away from the solution for a long time is that the ejecta, while turning nebular,
has high densities.  In the region 5000--10000\,\kms, the free-electron density is on the order
of 10$^9$\,cm$^{-3}$ at 30\,d, and hence we would not expect cooling by forbidden lines to be important
since the typical critical density for [Co\,\two] and [Co\three] lines  is 2 to 3 orders of magnitude lower.
  Consequently (and especially since many atoms were developed for stellar atmosphere calculations)
  most model atoms for IMEs and IGEs did not originally include forbidden-line transitions
  (but with the exception of Fe\two, Co\two, Si\two, Si\three, S\two, and S\three).
  High temperature and ionization conditions being generally met at high
  density and optical depth, model atoms for ions like Fe\three, Fe\four, Co\three, or Co\four\
  did not include forbidden-line transitions in models DDC10\_A0, DDC10\_A1, and DDC10\_A2.
  We experimented their potential role by doing a new model sequence named
  DDC10\_A3D1, which is identical to DDC10\_A1D1 apart from the treatment of [Co\three] lines.

  The impact on the evolution of the gas temperature and ionization is drastic (Fig.~\ref{fig_temp_ion}).
Prior to day 20, the difference in temperature between models   DDC10\_A1D1 and DDC10\_A3D1
remains small, and limited to the ejecta regions above 10000\,\kms. As time progresses, the ejecta expands
and thins out, and the temperature contrast between the two models grows. Instead of forming a temperature
plateau, the new model DDC10\_A3D1 continuously cools.
At 20--30\,d after explosion, the main coolants at $\sim$\,20000\,\kms\ are
[S\three]\,9533\,\AA\ (3p$^2$\,$^3$P--3p$^2$\,$^1$D transition),
[Si\two]\,2334--2335\,\AA\ (3p$^2$\,$^2$P--3p$^2$\,$^4$P transitions),
Si\three\,1206\,\AA\ (transition 3s-3p),
and [S\two]\,4069--4077\,\AA\ (3p$^3$\,$^4$S--3p$^3$\,$^2$P transitions). As
we progress deeper, e.g., at 11000\,\kms, the main coolants are
[S\two]\,4069--4077\,\AA,
[Co\three]\,5888\,\AA\ (3d$^7$\,$^4$F--3d$^7$\,$^2$G),
Si\three\,1206\,\AA,
[S\three]\,9533\,\AA.
Deeper still, e.g., at 3800\,\kms, the main coolants are
[Co\three]\,5888\,\AA, [S\three]\,9533\,\AA, Si\three\,1206\,\AA,
Fe\three\,1914\,\AA\ (3d$^5$\,4s\,$^7$S--3d$^5$\,4p\,$^7$P).
In each case, the quoted line on its own represents 10-50\% of the
total line-cooling rate for the corresponding ion, and whatever the depth, it
is generally the same line that dominates.
It is in stark contrast with the notion that millions of lines would be needed to
model SN Ia radiation accurately.
At 60d after explosion and in regions $\lesssim$\,10000\,\kms\ where the
spectrum forms, the main coolants are [Co\three]\,5888\,\AA,
Co\two\,2286\,\AA\ (not a forbidden line, but instead a strong 4s--4p transition),
[Fe\two]\,12570\,\AA, and [S\three]\,9533\,\AA.
A description of forbidden-line transitions is given in \citet{hansen_etal_84}
and \citet{quinet_98}.

\begin{figure}
\epsfig{file=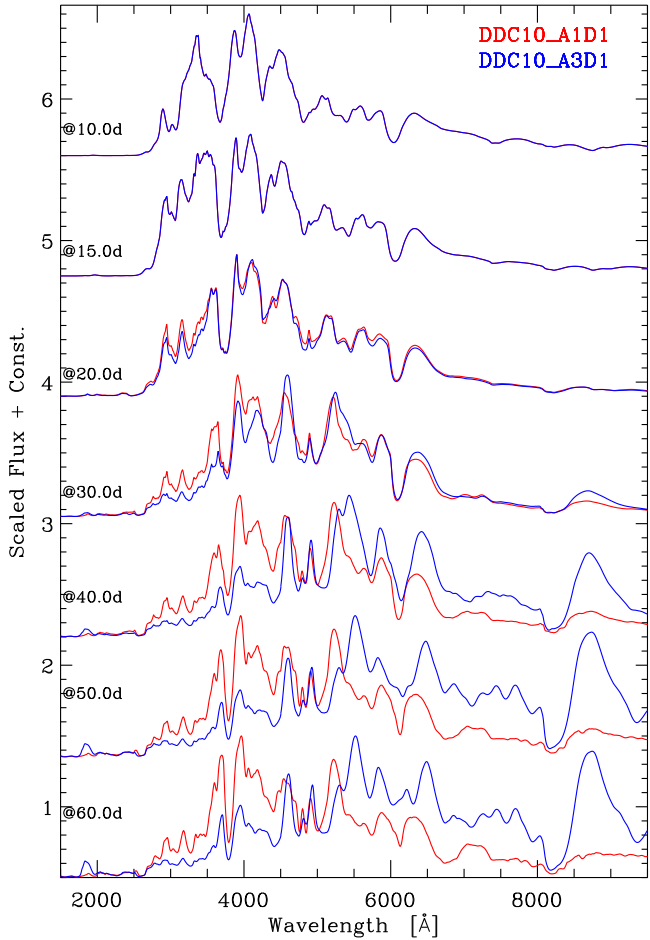,width=8.5cm}
\caption{Illustration of the impact on synthetic spectra of introducing forbidden-line transitions
in the Co\three\ model atom (model DDC10\_A3D1, blue), all else being the same as in model
DDC10\_A1D1 (red).
While the bolometric luminosity is identical between the two, the extra cooling reduces the temperature
and the ionization in the spectrum formation region, leading to large differences in color and spectral
morphology.
\label{fig_comp_ab}
}
\end{figure}

This cooling leads to a progressive recombination of the ejecta, in particular from Co\three\ to
Co\two\ in model DDC10\_A3D1 (right panel of Fig.~\ref{fig_temp_ion}).
Cobalt represents 70\% of the total mass fraction around 5000-10000\,\kms,
and so this ionization change eventually makes Co\two\ the primary source of line blanketing, in particular
in the blue part of the spectrum. The combined effects
of enhanced cooling (i.e., cooler photosphere) and enhanced blanketing (i.e., strengthening of Co\two\
opacity) leads to a significant reddening of the emergent radiation (Fig.~\ref{fig_comp_ab}).
The process is a runaway since more cooling induces more recombination, stronger optically-thin
line emission by Co\two\ in the near-IR, which induces further cooling and recombination etc.

  So, our color problem was fundamentally associated with the inadequate handling of coolants
  rather than insufficient opacity. Even well above the critical density, forbidden-line transitions
  can act as efficient coolants
  as soon as the time of bolometric maximum in SN Ia ejecta. In hindsight this is not surprising.
  Below the critical density, the cooling due to collisionally excited lines scales as the
  density squared (assuming the ion is
  dominant) while above the critical density the cooling scales only linearly with the density. As the
  electron density increases above the critical density the relative importance of the line falls
  relative to other processes which are still scaling with the square of the density. However, in
  SNe Ia, Co and Fe are not impurity species, thus enhancing their importance for the energy balance,
  and allowing forbidden lines to be important coolants well above their critical density.

 \begin{figure}
\epsfig{file=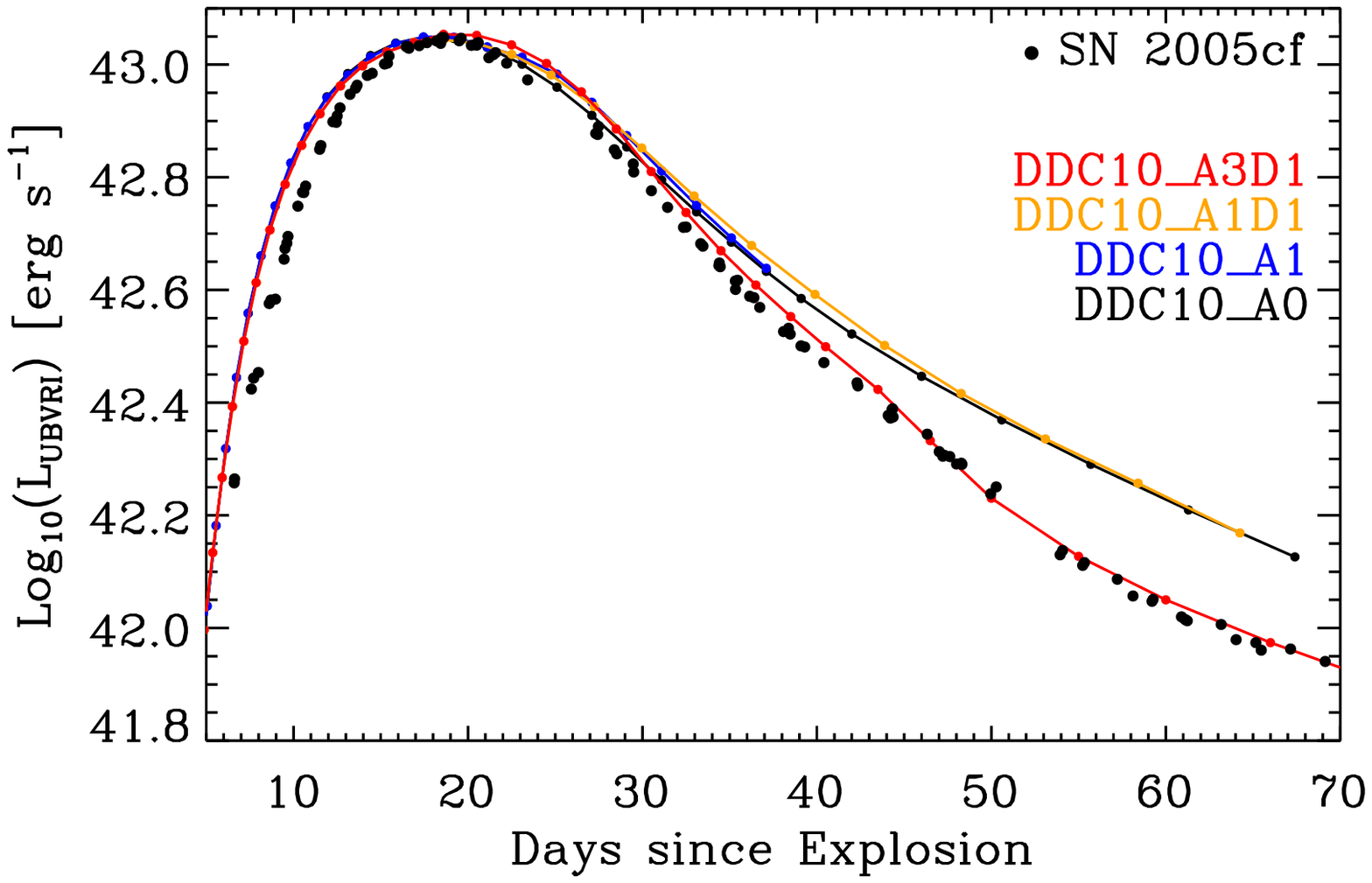,width=8.75cm}
\epsfig{file=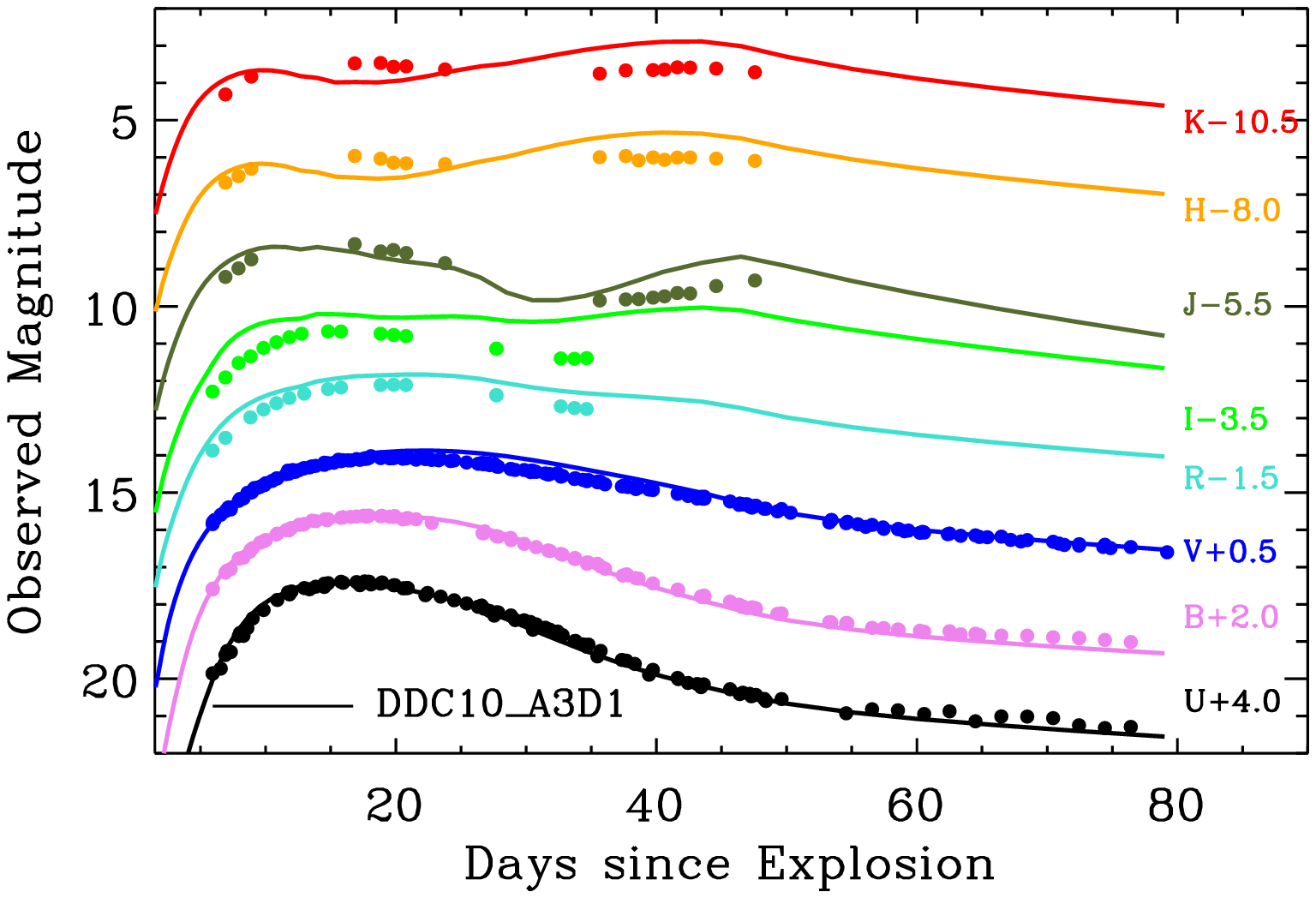,width=8.75cm}
\caption{{\it Left:} $UBVRI$ luminosity for models A0, A1, A1D1, A3, together with the corresponding
inferred luminosity for SN\,2005cf.
{\it Right:} Comparison between multi-band light curves of SN\,2005cf and model DDC10\_A3D1.
The synthetic photometry has been corrected for extinction, redshift, and distance dilution.
\label{fig_lubvri_ddc10}}
\end{figure}

\begin{figure}
\epsfig{file=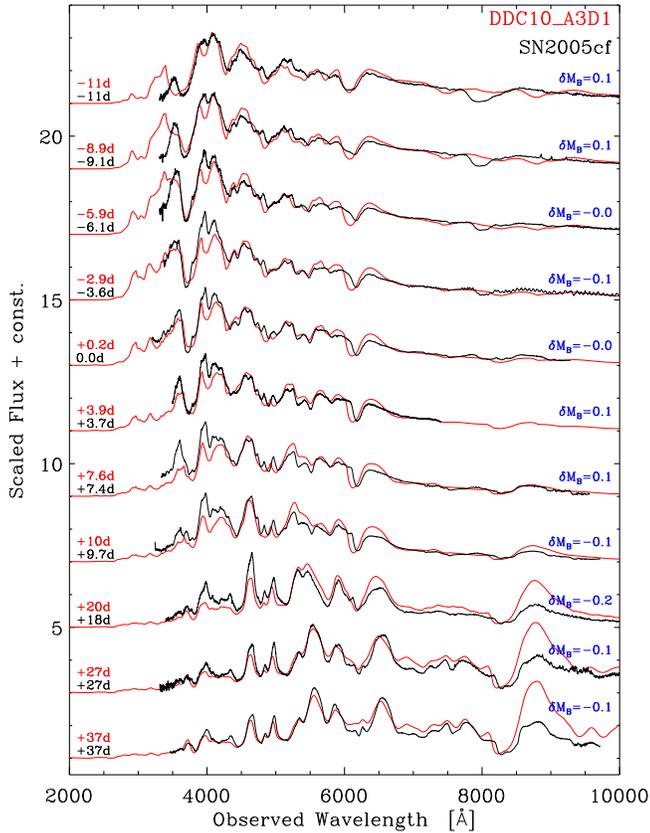,width=8.5cm}
\caption{
Same as Fig.~\ref{fig_spec_A0}, but now showing a comparison between model DDC10\_A3D1 (red)
and the observations of SN\,2005cf (black).
Contrary to model DDC10\_A0, model DDC10\_A3D1 includes both
big model atoms, [Co\three] lines, and all 2-step decay chains presented in
Tables~\ref{tab_nuc1}--\ref{tab_nuc2}.
Apart from the 8500\,\AA\ region after bolometric maximum, the agreement with observations
is very satisfactory.
\label{fig_spec_A3D1}
}
\end{figure}

Figure~\ref{fig_lubvri_ddc10} clearly shows that the new model DDC10\_A3D1 does a very good job
at reproducing the $UBVRI$ luminosity of SN\,2005cf, where other models DDC10\_A0, DDC10\_A1,
and DDC10\_A1D1 yielded essentially the same discrepancy. Comparing observed to synthetic
light curves, we find
model DDC10\_A3D1 yields a very satisfactory match to the SN\,2005cf color evolution. Prior to peak, the
mismatch appears to be related to a global flux offset (at the few tens of percent level) since
the colors agree well between the two. This offset may stem from the overestimated kinetic energy
of our delayed-detonation models \citep{blondin_etal_13}.

More spectacular is the spectroscopic match to the observations
of SN\,2005cf (Fig.~\ref{fig_spec_A3D1}). The model matches the
color evolution, line-profile morphology,
in particular as we proceed through bolometric maximum and progressively evolve from
thick to thin conditions. The spectrum is well matched throughout the optical.
A noticeable exception is the 8500\,\AA\
region where the Ca\two\ triplet is overestimated in our simulations. We find that the Ca\two\ line strength
is quite sensitive to overlapping Co\two\ line emission. The calcium lines form outside of the cobalt emitting
region, so that the strength of the Ca\two\ triplet depends sensitively on this background flux controlled
by Co line emission. This discrepancy is also at the origin of the
excess $I$-band flux at $\gtrsim$\,30\,d after explosion (bottom panel of
Fig.~\ref{fig_lubvri_ddc10}).

   After much exploration to identify the origin of this last problem, we realized that the non-thermal
ionization routine did not include all species (the non-thermal excitation routine was OK).
Namely, while non-thermal excitation rates were
computed for all ions and levels, we did not have any entry for Ti, Cr, and Co
for the non-thermal ionization cross sections. Using data from
\citet{mazzotta_etal_98}, we updated these rates and reran our SN Ia model DDC10.
Non-thermal processes have little impact before the peak so we restarted the
sequence DDC10\_A3D1 at bolometric maximum using this updated non-thermal
solver. Unfortunately, we also found that using a huge model atom for Co\two\
and Co\three\ produced a redder SED, so we needed to redo the sequence to address this.

\begin{figure*}
\epsfig{file=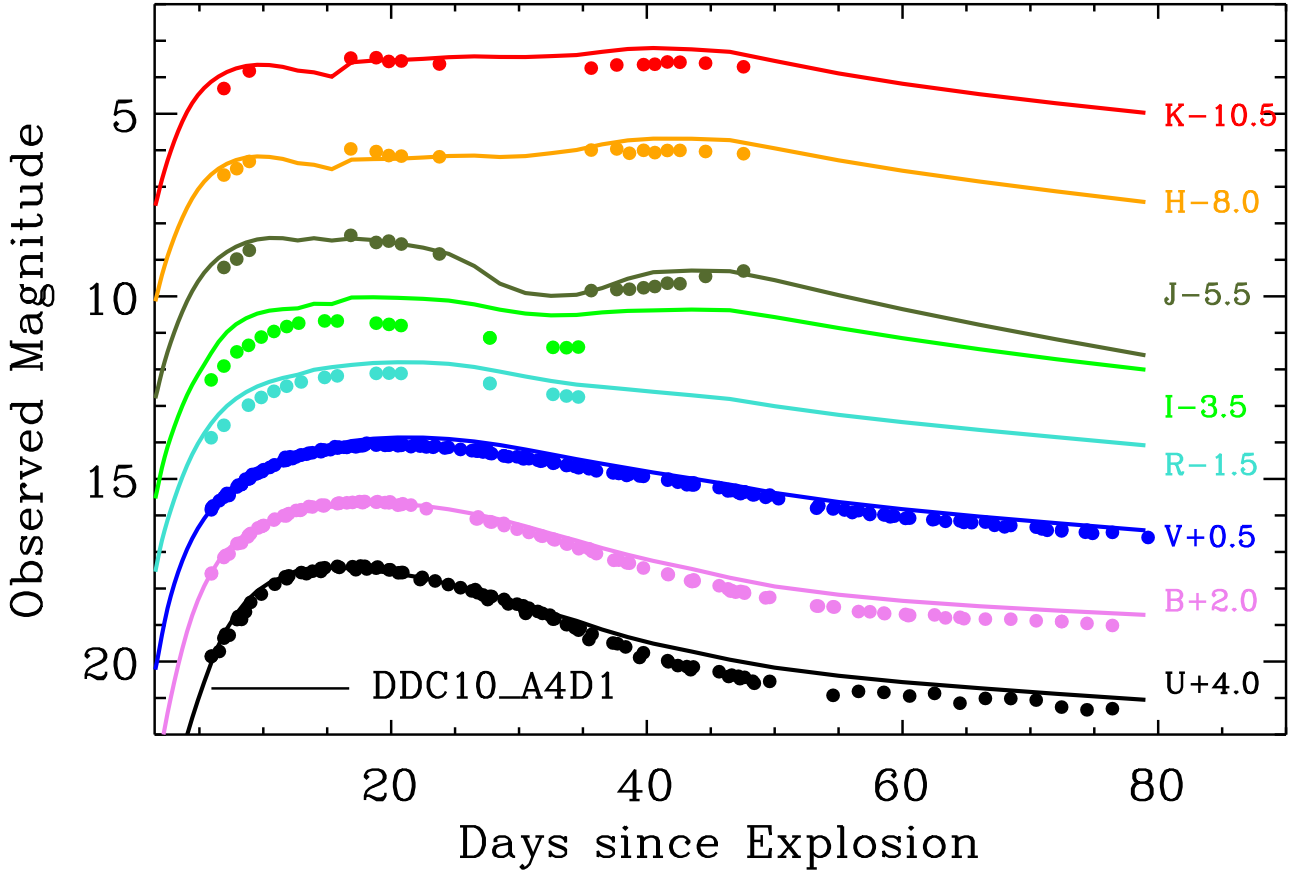,width=17cm}
\caption{
Comparison between multi-band light curves of SN\,2005cf and model DDC10\_A4D1.
The synthetic photometry has been corrected for extinction, redshift, and distance dilution.
The size of the dots in the figure is approximately 0.3 magnitudes.
\label{fig_mag_a4d1}}
\end{figure*}

\begin{figure*}
\epsfig{file=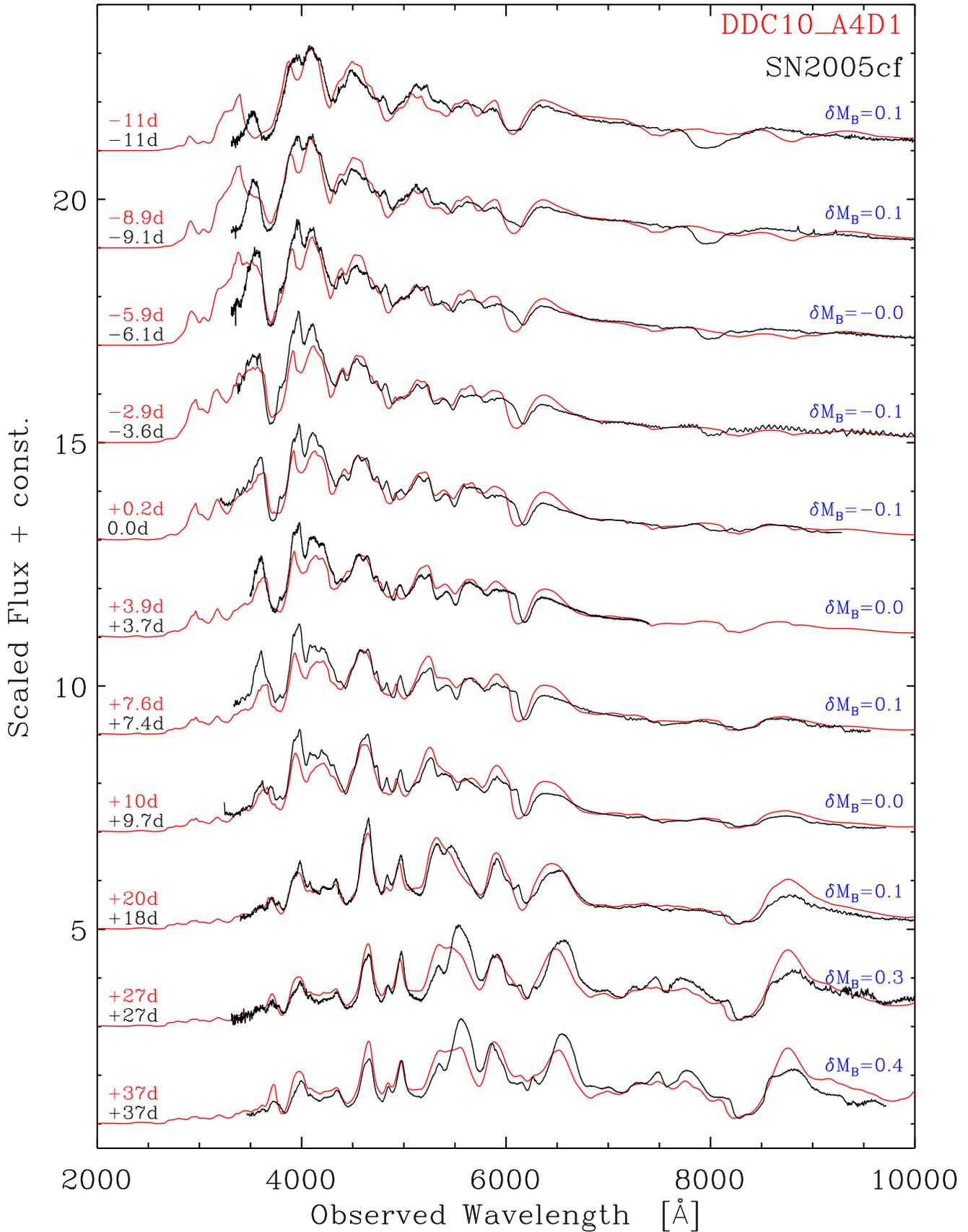,width=17.5cm}
\caption{
Comparison between model DDC10\_A4D1 and the
observed spectra of SN\,2005cf. We correct the synthetic flux to account for the
distance, redshift, and extinction of 05cf.
Spectra are scaled vertically for convenience, although the label
on the right gives the true $B$-band magnitude offset between model and
observations at each date --- this offset is typically small.
\label{fig_spec_A4D1}
}
\end{figure*}

\begin{figure*}
\epsfig{file=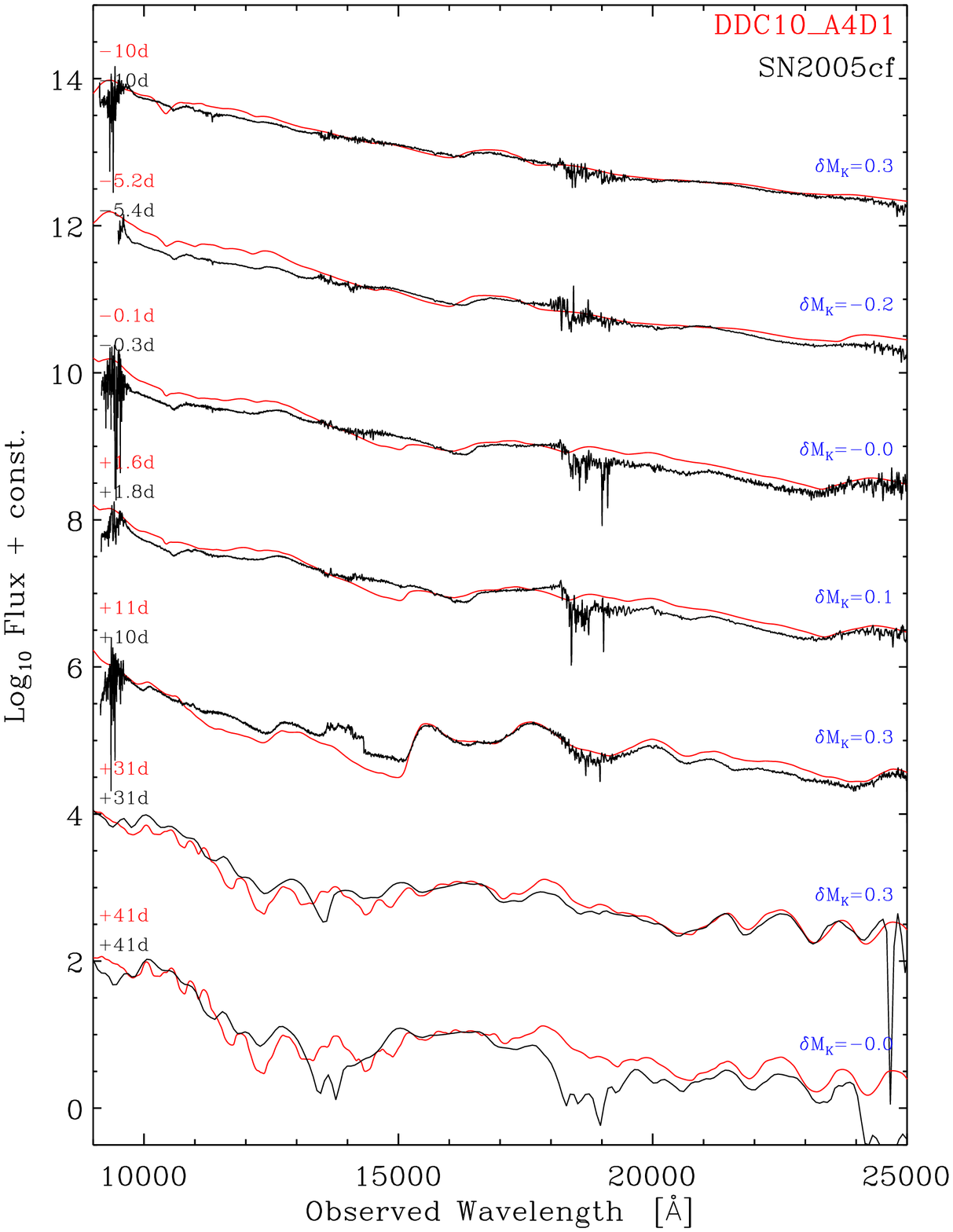,width=17.5cm}
\caption{
Same as Fig.~\ref{fig_spec_A4D1}, but now showing the near-IR range.
We show $\log F_{\lambda}$ for better visibility. For each spectrum, we quote the
true magnitude offset in the $K$ band, while the vertical positioning is adjusted
for optimal visibility.
\label{fig_spec_A4D1_nearir}
}
\end{figure*}

We compare the resulting multi-band light curves of model DDC10\_A4D1 with SN\,2005cf
in Fig.~\ref{fig_mag_a4d1}. Overall, the color agreement in all optical and near-IR bands is good.
We show the comparison with optical and near-IR spectra in Figs.~\ref{fig_spec_A4D1}---\ref{fig_spec_A4D1_nearir}.
The labels appearing on the right indicate the magnitude offset with the contemporaneous $B$-band or $K$-band
magnitude. This offset is typically on the order of 0.1\,mag, with only a few larger offsets of $\lesssim$\,0.3\,mag
at some epochs.
Hence, in model DDC10\_A4D1, the flux offset in the red part of the optical is gone, primarily because
allowance of non-thermal ionization for Co leads to enhanced ionization overall, and in particular for
Co. The general agreement with observations is improved, although we obtain a poorer fit in the
5500\,\AA\ region. In the near-IR, the agreement with SN\,2005cf spectra is satisfactory
(Fig.~\ref{fig_spec_A4D1_nearir}), although some
features do not follow the same evolution as observed (they appear too soon, or are too pronounced etc).
One concern is also the reliability of observed near-IR spectra, in particular for the relative flux.
The remaining mis-matches with SN\,2005cf could be due to the choice of the delayed-detonation
model but might also indicate that further improvement in the adopted atomic models and processes
are needed.

The critical role of forbidden-line transitions in SNe Ia, in particular their influence on the gas and radiation
properties as early as the peak of the light curve does not seem documented in the literature.
Early works on SN Ia radiation modelling did emphasize the critical role of
forbidden-line transitions, but the focus of these studies was on nebular times exclusively
\citep{axelrod_80,pinto_eastman_93,kuchner_etal_94} ---
here we demonstrate their importance as early as bolometric maximum.
\citet{kasen_etal_06} do not include forbidden-line transitions in their
radiative-transfer simulations of the SN Ia near-IR secondary maximum that takes place
$\sim$\,40\,d after explosion, nor
in their subsequent study of the width-luminosity relation \citep{kasen_etal_07}.

Selecting lines based on their oscillator strength is fundamentally inadequate since
the forbidden lines that are so critical for cooling the gas have very low oscillator strengths
and are generally optically thin. They are not critical for trapping photons, but they are key for
cooling the gas, controlling its temperature and ionization state, and thus determining what ions provide
opacity sources. To some extent, this suggests that the SN Ia radiation properties
are not exclusively controlled by opacity and fluorescence/branching, but also
by the way the ejecta cool through
thin lines. These lines may be intrinsically optically thin, but the photons they radiate may
nonetheless be scattered or absorbed if they overlap with other lines.

\section{Local versus non-local energy deposition}
\label{sect_loc_noloc}

\begin{figure*}
\centering
\begin{flushleft}
\begin{minipage}[b]{0.5\linewidth}
\epsfig{file=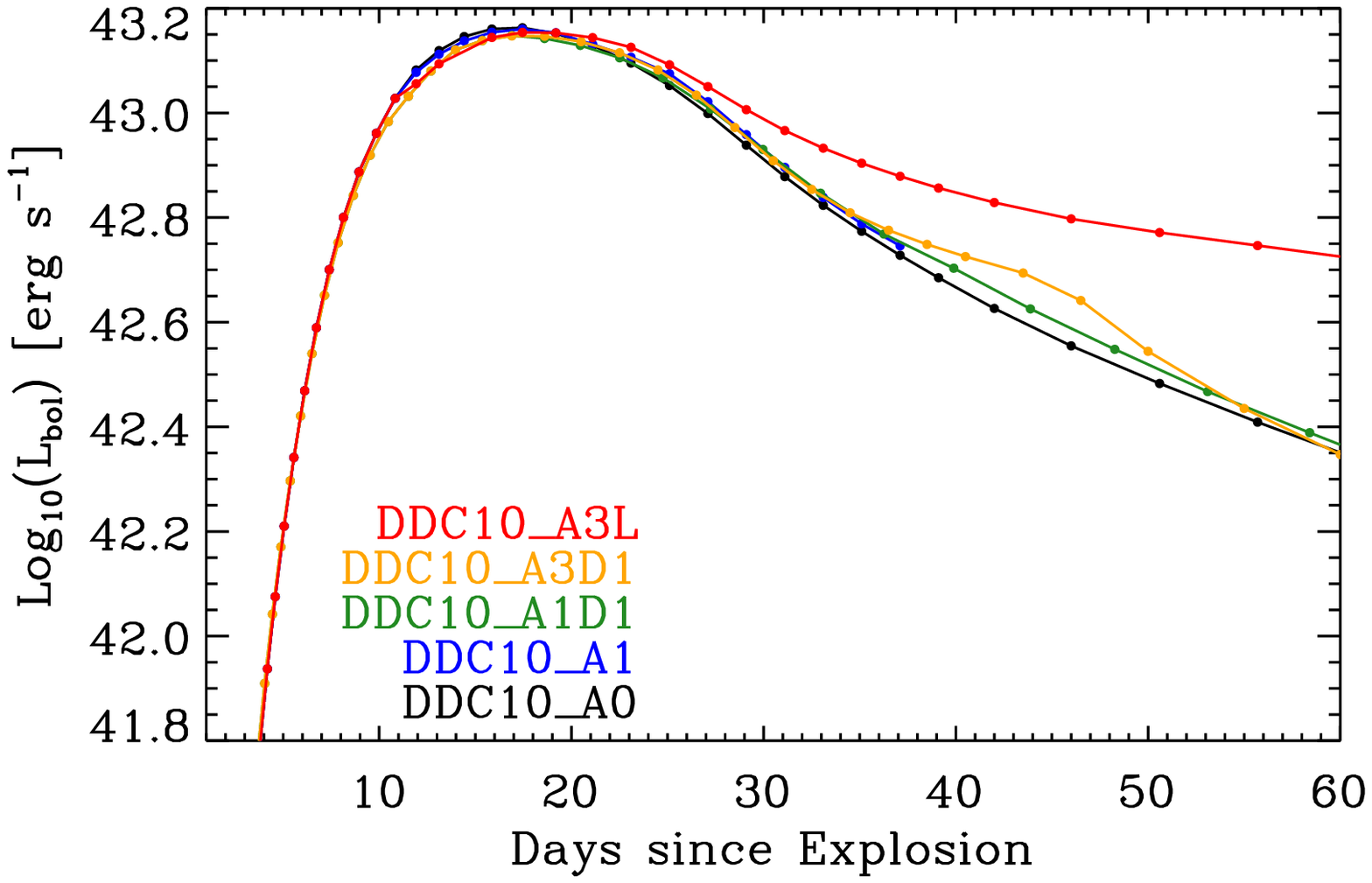,width=8.0cm}
\end{minipage}
\end{flushleft}
\vspace{-5.6cm}
\begin{flushright}
\begin{minipage}[b]{0.5\linewidth}
\epsfig{file=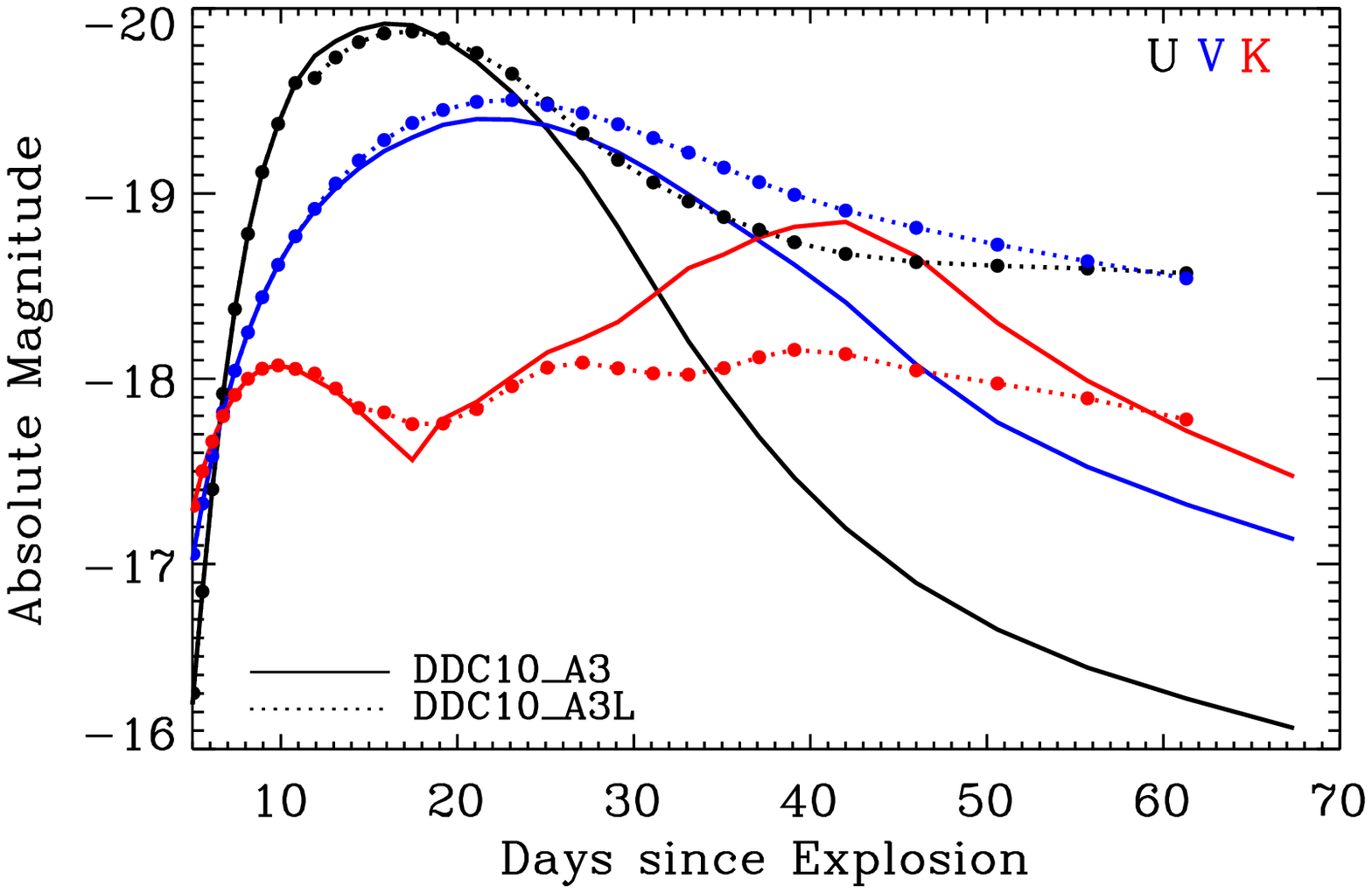,width=8.0cm} \\
\end{minipage}
\end{flushright}
\begin{flushleft}
\begin{minipage}[b]{0.5\linewidth}
\epsfig{file=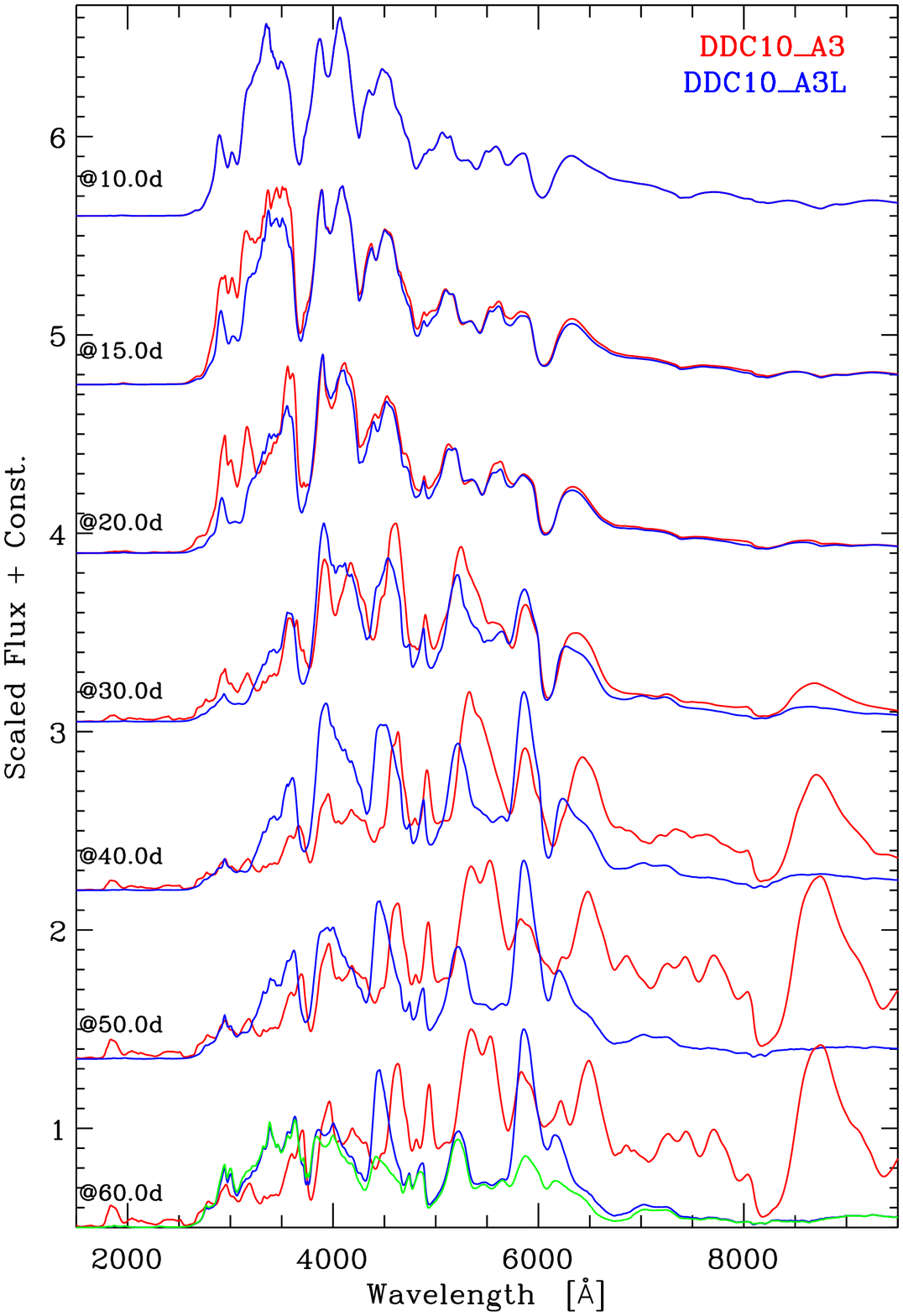,width=8.0cm}
\end{minipage}
\end{flushleft}
\vspace{-12.15cm}
\begin{flushright}
\begin{minipage}[b]{0.5\linewidth}
\epsfig{file=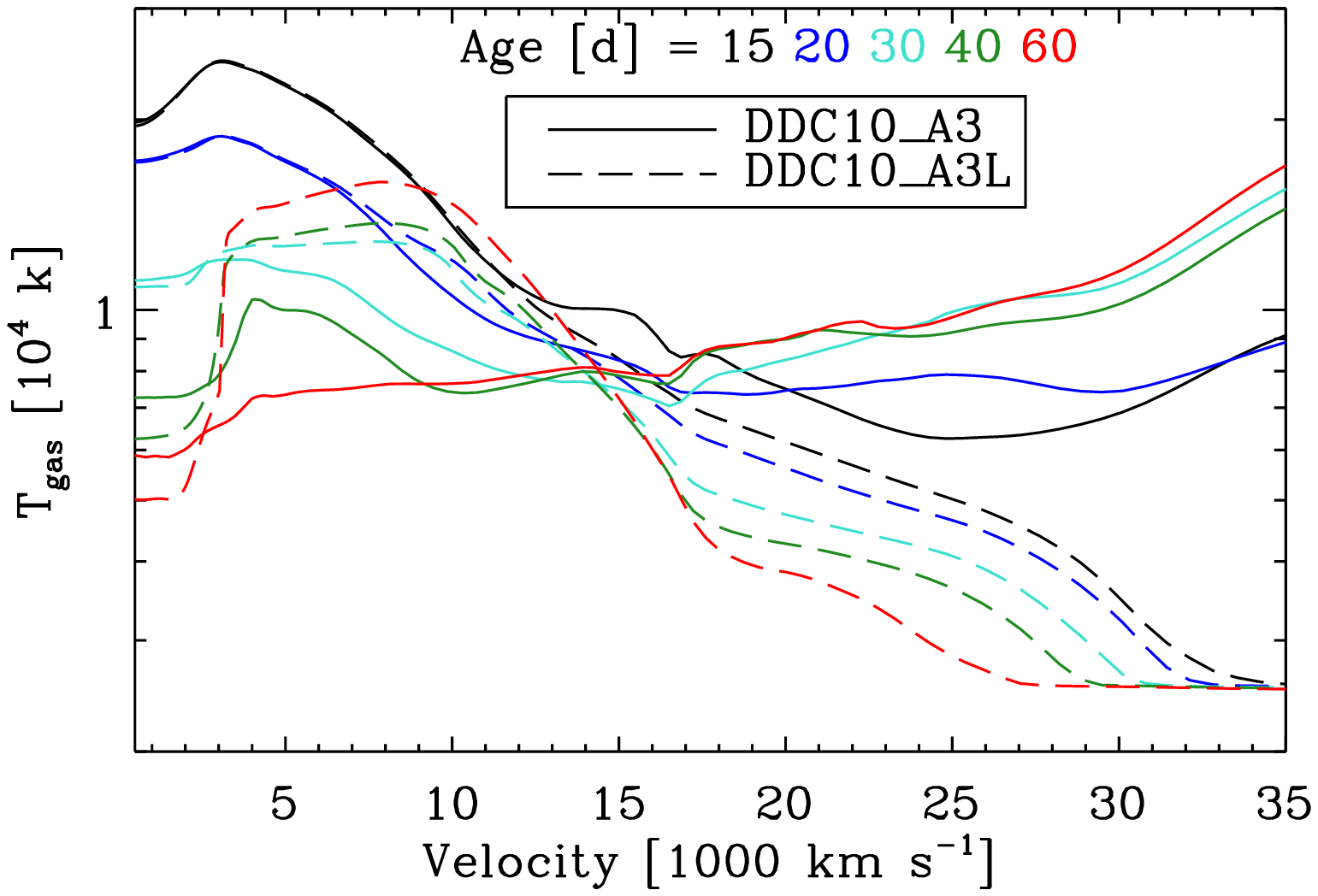,width=8.0cm} \\
\epsfig{file=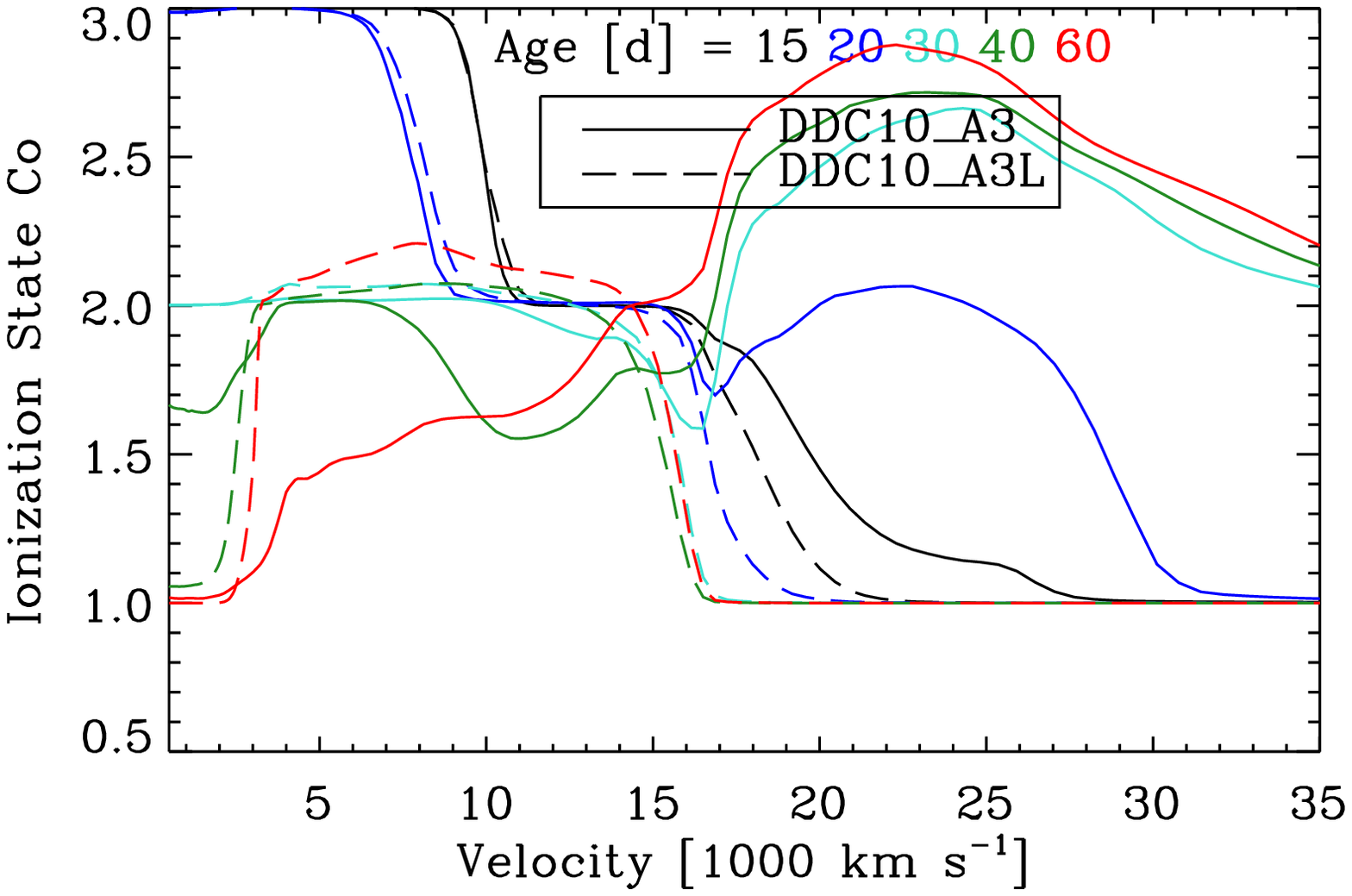,width=8.0cm}
\end{minipage}
\end{flushright}
\vspace{0.75cm}
\caption{
{\bf Left Column, top:}
Bolometric luminosity for models DDC10\_A0, DDC10\_A1, DDC10\_A1D1, DDC10\_A3D1,
and DDC10\_A3L.
Modifications in the model atom, and hence opacity, lead to very modest changes
in the bolometric luminosity. However, if we prevent $\gamma$-ray escape (model DDC10\_A3L),
there is a dramatic change in the bolometric luminosity.
{\bf Left column, bottom:} Spectral comparison between model DDC10\_A3 (red; non-local energy deposition and
$\gamma$-ray escape are allowed for) with model DDC10\_A3L (blue) in which we assume local energy
deposition at all times. The green curve corresponds to the spectrum obtained by taking out
all the forbidden-line transitions of Co\three\ treated in model DDC10\_A3L.
{\bf Right column, top:}  Same as top/left, but now showing the light-curve evolution
in the $U$, $V$, and $K$ bands.
{\bf Right column, middle and bottom:}
Snapshots of the gas temperature and Co ionization state
at selected post-explosion times. Local energy deposition maintains a higher (lower) temperature
in the \iso{56}Ni rich (poor) regions, which directly impacts the gas ionization. For local energy deposition,
Co remains as Co\three\ in the region 3000--15000\,\kms.
\label{fig_A3_A3L}}
\end{figure*}

The low mass and high expansion rate of SN Ia ejecta, combined with the presence
of \iso{56}Ni nuclei at large ejecta velocities (or Lagrangian mass), allows
$\gamma$-ray escape as early as 10-15\,d after explosion. This affects the SN properties
at all times beyond the peak of the light curve (Fig.~\ref{fig_A3_A3L}), by a magnitude that
supersedes any variation in model atoms we have tested in this study. $\gamma$-ray escape
is in fact one of the key ingredients shaping SN Ia bolometric light curves.

   By enforcing local energy deposition in model DDC10\_A3L, we obtain gas temperatures
   and ionization states that are much higher than in model DDC10\_A3. The temperature in
   the \iso{56}Ni rich region stays high. Interestingly, the temperature in the \iso{56}Ni hole
   becomes much lower than in the shells above it at nebular times, suggesting that radiative cooling
   completely inhibits the diffusion of heat to deeper layers.
   Importantly, the gas ionization stays high, despite the treatment of [Co\three] lines.
   For example, Co does not recombine any more after the light curve peak but remains
   as Co$^{2+}$ in the region 3000--15000\,\kms. Although the spectrum reddens because of intense
   blanketing, it remains bluer than in standard SNe Ia. Rather than developing strong Fe\two\ lines,
   e.g.,  at 5169\,\AA, the model shows very strong [Co\three] lines. For the last time displayed,
   we overlay the synthetic spectrum when these forbidden-line transitions are omitted in the calculation
   (green line). As in model DDC10\_A3, [Co\three]
   lines are very strong coolants in SNe Ia, but here because of the much larger energy deposition (i.e.
   no $\gamma$-ray escape) and no recombination (Co\three\ is the dominant Co ion), these lines
   play an even stronger role. Because it will be the subject of a forthcoming paper, we note only
   in passing that the SN Ia feature that forms $\lesssim$\,6000\,\AA\ at $\sim$\,10\,d  after the light curve
   peak is associated with [Co\three] --- the association with Na\one\ is unfounded on numerous
   grounds \citep{dessart_etal_14b}.

   We note that model DDC10\_A3L does not show any bump in the post-maximum bolometric luminosity, nor any bump
   in the near-IR light curves, despite the strong cooling through [Co\three]. Both are in fact intimately related, as we
   discuss in section~\ref{sect_nearir}.

\begin{figure*}
\centering
\begin{flushleft}
\begin{minipage}[b]{0.5\linewidth}
\epsfig{file=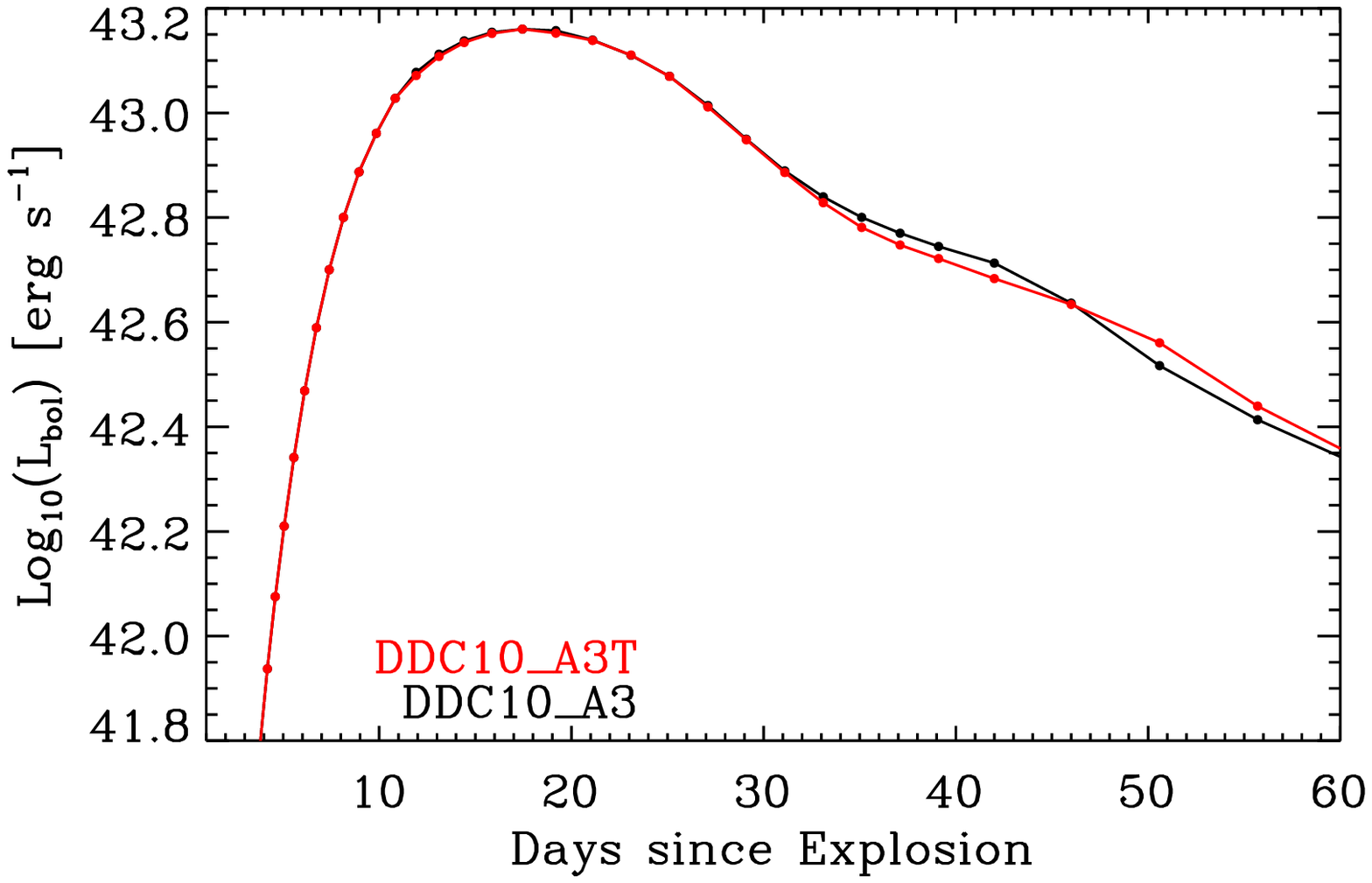,width=8.0cm}
\end{minipage}
\end{flushleft}
\vspace{-5.6cm}
\begin{flushright}
\begin{minipage}[b]{0.5\linewidth}
\epsfig{file=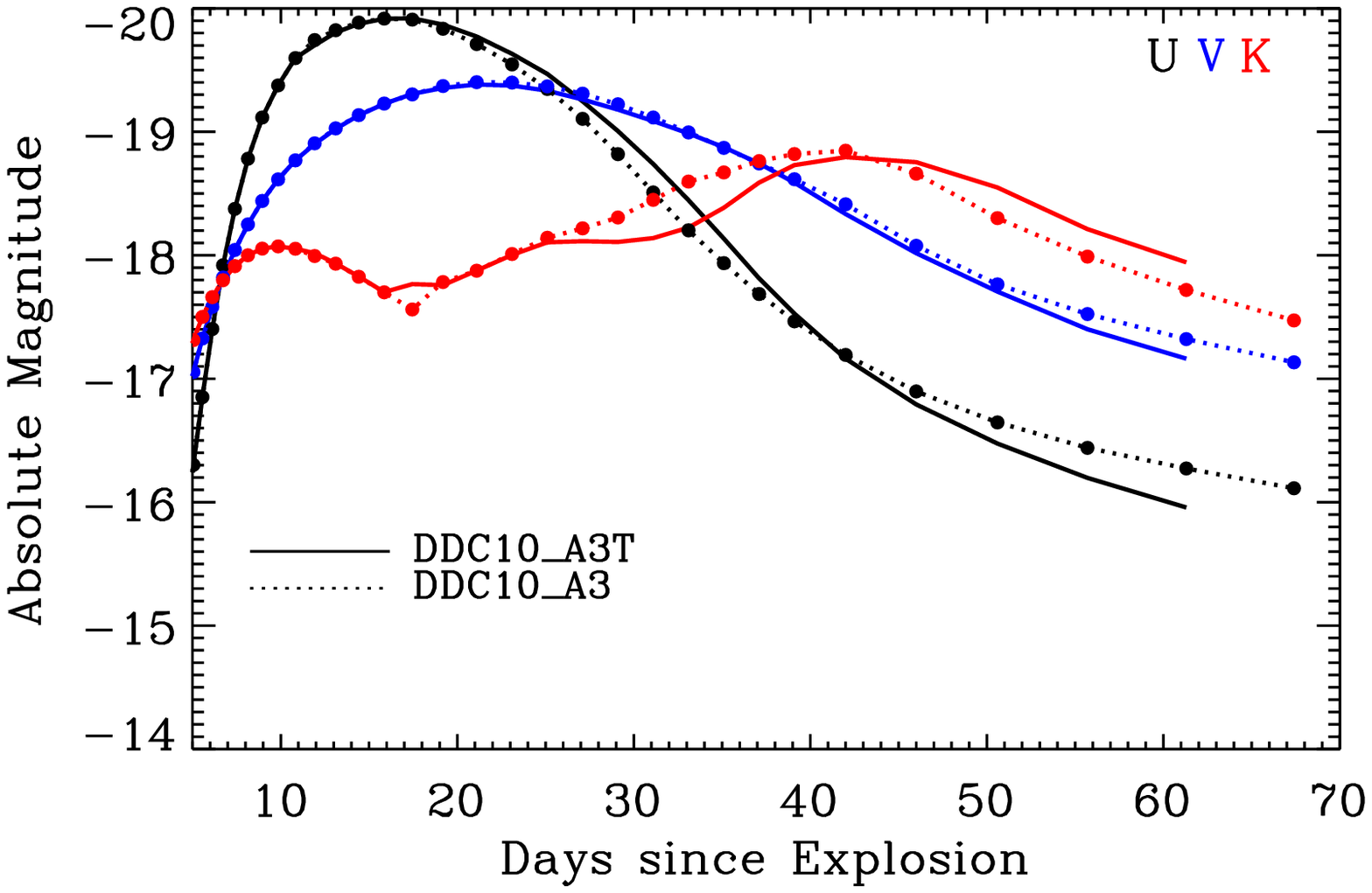,width=8.0cm} \\
\end{minipage}
\end{flushright}
\begin{flushleft}
\begin{minipage}[b]{0.5\linewidth}
\epsfig{file=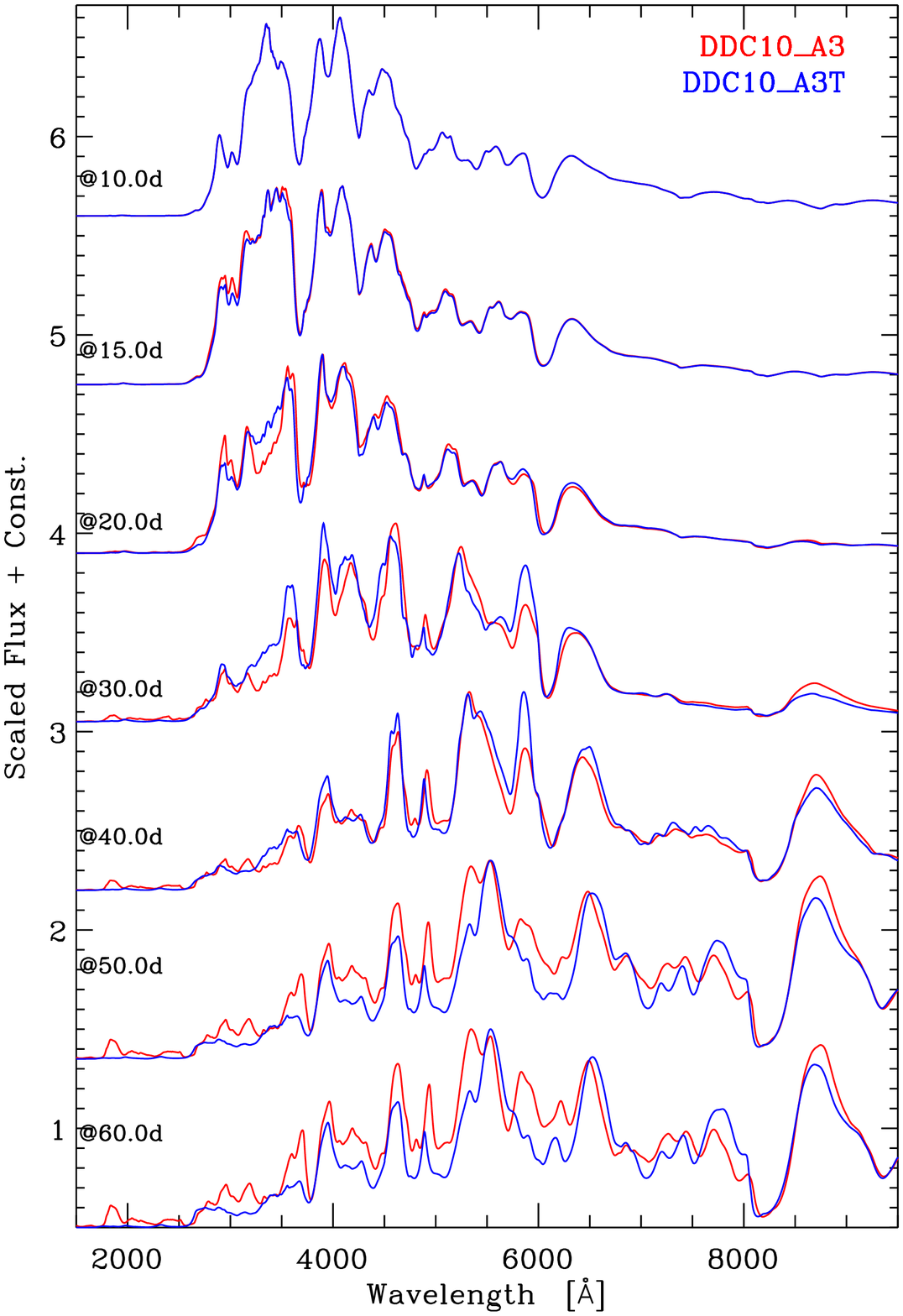,width=8.0cm}
\end{minipage}
\end{flushleft}
\vspace{-12.15cm}
\begin{flushright}
\begin{minipage}[b]{0.5\linewidth}
\epsfig{file=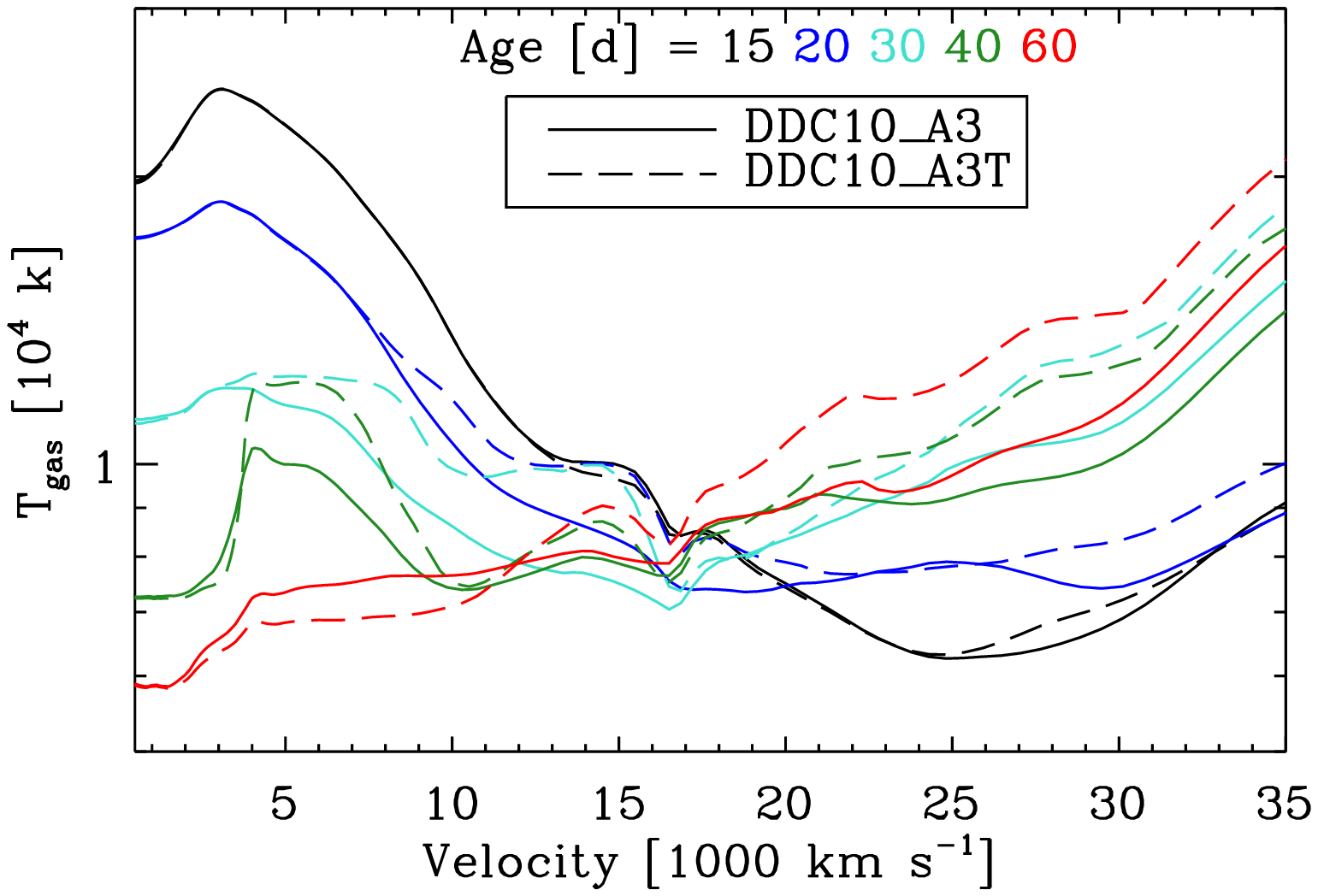,width=8.0cm} \\
\epsfig{file=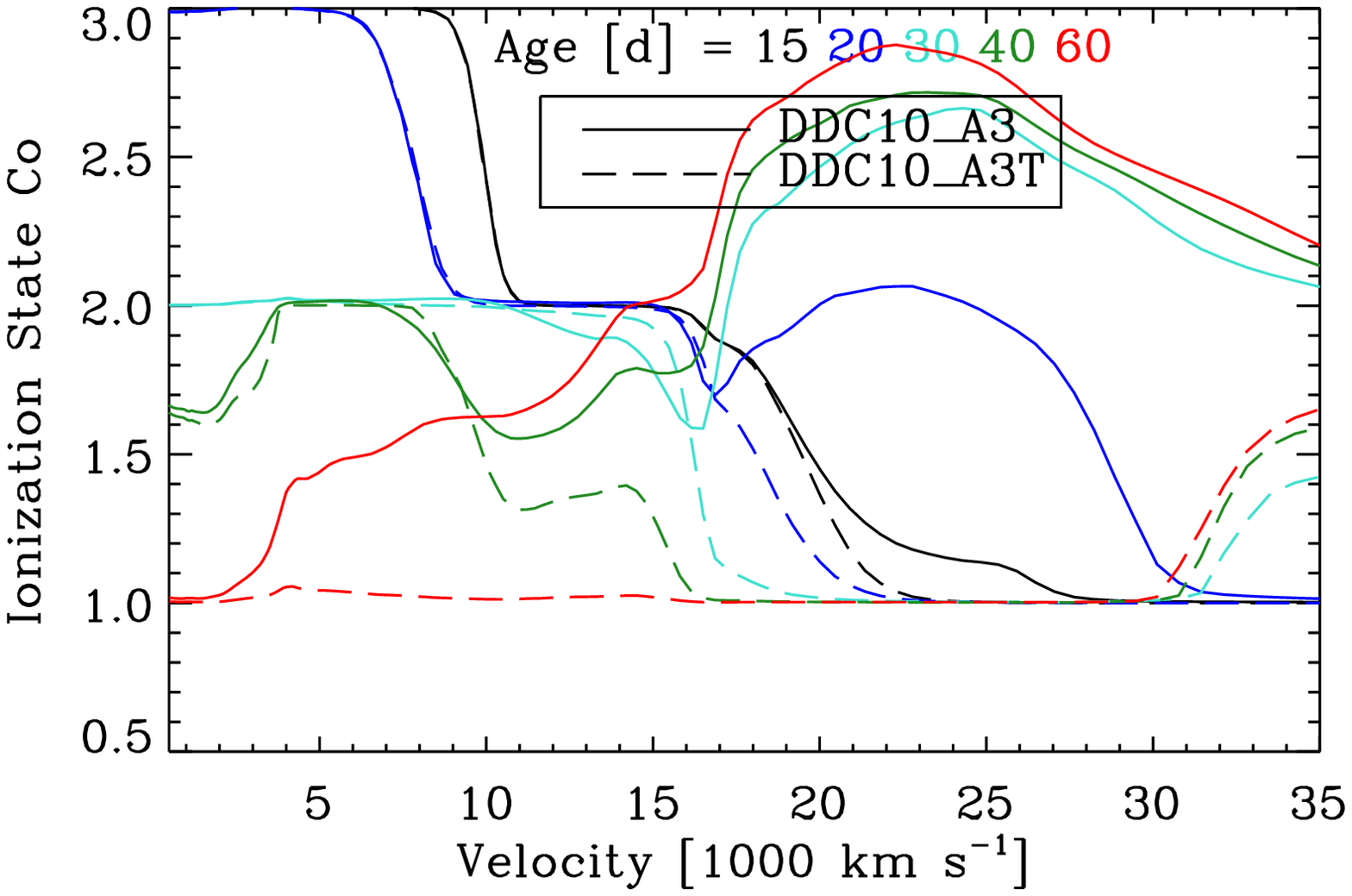,width=8.0cm}
\end{minipage}
\end{flushright}
\vspace{0.75cm}
\caption{Same as Fig.~\ref{fig_A3_A3L}, but this time showing the differences
in bolometric luminosity, colors, and spectra between the model sequence DDC10\_A3,
in which non-thermal processes are treated, and model sequence DDC10\_A3T, in which all decay energy
deposited in the ejecta is treated as heat.
\label{fig_nonte}}
\end{figure*}

\section{Non-thermal processes associated with radioactive decay}
\label{sect_nonte}

   All simulations described so far in this paper include a treatment of non-thermal processes,
following the procedure presented in \citet{li_etal_12}.
However, to assess the effect of non-thermal processes on the gas and radiation
properties, we have run the sequence DDC10\_A3T, identical to DDC10\_A3, but forcing
all decay energy to be deposited as heat, i.e., forcing non-thermal rates associated
with radioactive decay to zero.

   Without detailed non-LTE simulations, it is difficult to guess the importance of such non-thermal
processes in SNe Ia. Indeed, the large abundance of unstable nuclei (in particular \iso{56}Ni and \iso{56}Co),
should favor the strength of non-thermal processes. However, the large ionization of SN Ia ejecta makes
the electron density relatively high, a property that tends to quench non-thermal ionization and excitation
\citep{XM91_87A_energetic,dessart_etal_12}. Furthermore, since \nifs\ is produced primarily at depth
in the ejecta (how deep depends on mixing during the explosion phase),
in layers moving at $\lesssim$\,15000\,\kms, non-thermal processes are irrelevant
at early times when the spectrum formation region is located in the faster moving outer ejecta layers,
i.e. non-thermal processes are confined at such times to layers where
thermalization is secured by the large ejecta optical depth.

   In Fig.~\ref{fig_nonte}, we show the impact of these non-thermal processes when they start becoming
visible around the peak of the bolometric light curve. The thermal model DDC10\_A3T indeed starts
appearing bluer at $\sim$\,20\,d after explosion, but the effect is
rather small. This occurs because
model DDC10\_A3T has a larger ejecta temperature and ionization in the originally \nifs-rich ejecta
layers (i.e., more energy is deposited in the form of heat).
The main effect is to make the forbidden-line transition [Co\three]\,5888\,\AA\
stronger. However, as the ejecta becomes optically thin, the thermal model becomes somewhat
cooler, but more importantly, it becomes significantly less ionized, making Co$^+$ dominate
throughout the ejecta. In contrast, non-thermal
processes maintain an equal share of Co$^+$ and Co$^{2+}$ for Co (we find the same holds for Fe).
At the peak of the light curve, of the total decay energy deposited, we find that the fraction going into
non-thermal ionization is 5-10\% at all depths, and systematically about 1--2 times larger than that going
into non-thermal excitation.
At 60\,d after explosion, this fraction is 5\% for both non-thermal ionization and excitation.
Although small, these fractions correspond to large non-thermal ionization
and excitation contributions, given that modest energies are needed to alter
the thermodynamic state of the gas.

As time progresses further into the nebular phase, non-thermal processes
maintain this high ionization,
although the decay of \iso{56}Co into \iso{56}Fe eventually makes iron dominate over cobalt.
We in fact find that the Co and Fe ionization, as well as the ejecta
temperatures at $\lesssim$\,15000\,\kms, remain roughly constant at $\sim$\,7500\,K from 60\,d
until 200\,d after explosion.
Our delayed detonation models are eventually primarily
cooled through few forbidden-line transitions of Fe\three\ and Fe\two, in particular transitions
like [Fe\three]\,4658\,\AA, which are connected to the ground state (for a discussion,
see \citealt{axelrod_80,kuchner_etal_94,maurer_etal_11}).
A forthcoming study will present these results in detail.

As we mentioned earlier,  the non-thermal solver did not treat non-thermal {\it ionization}
for Ti, Cr, and Co in model DDC10\_A3D1.
When we include the associated rates in model DDC10\_A4D1, the ejecta ionization and temperature
increase a little, exacerbating the contrast with model DDC10\_A3T. However, the enhanced blanketing
we obtain in DDC10\_A4D1, caused by the huge Co\two\ and Co\three\ model atoms employed, leads
only to a modest hardening of the DDC10\_A4D1 synthetic spectra compared to those obtained
for model DDC10\_A3D1 (See Section\,\ref{sect_sol}).

\begin{figure*}
\epsfig{file=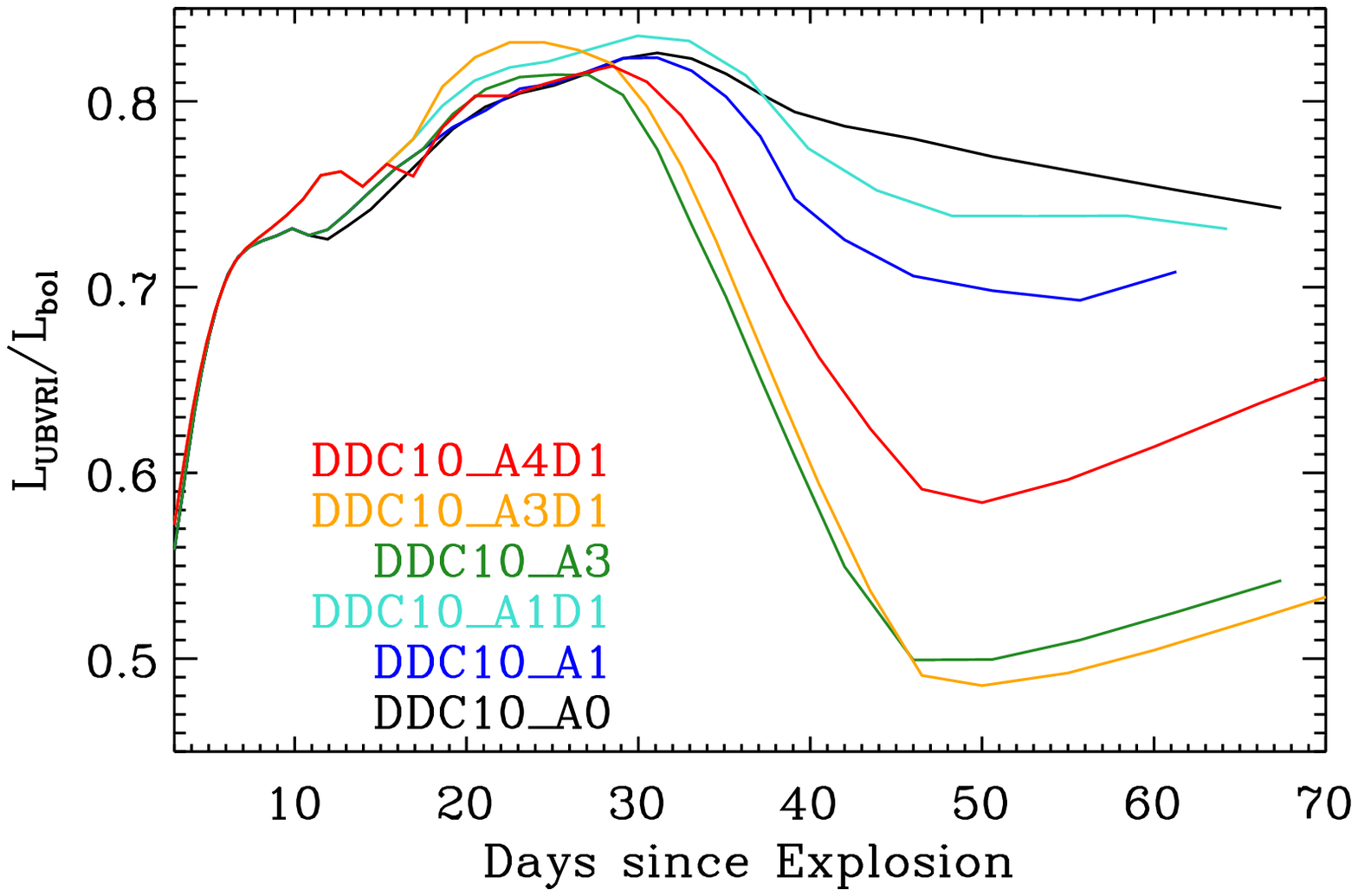,width=8.5cm}
\epsfig{file=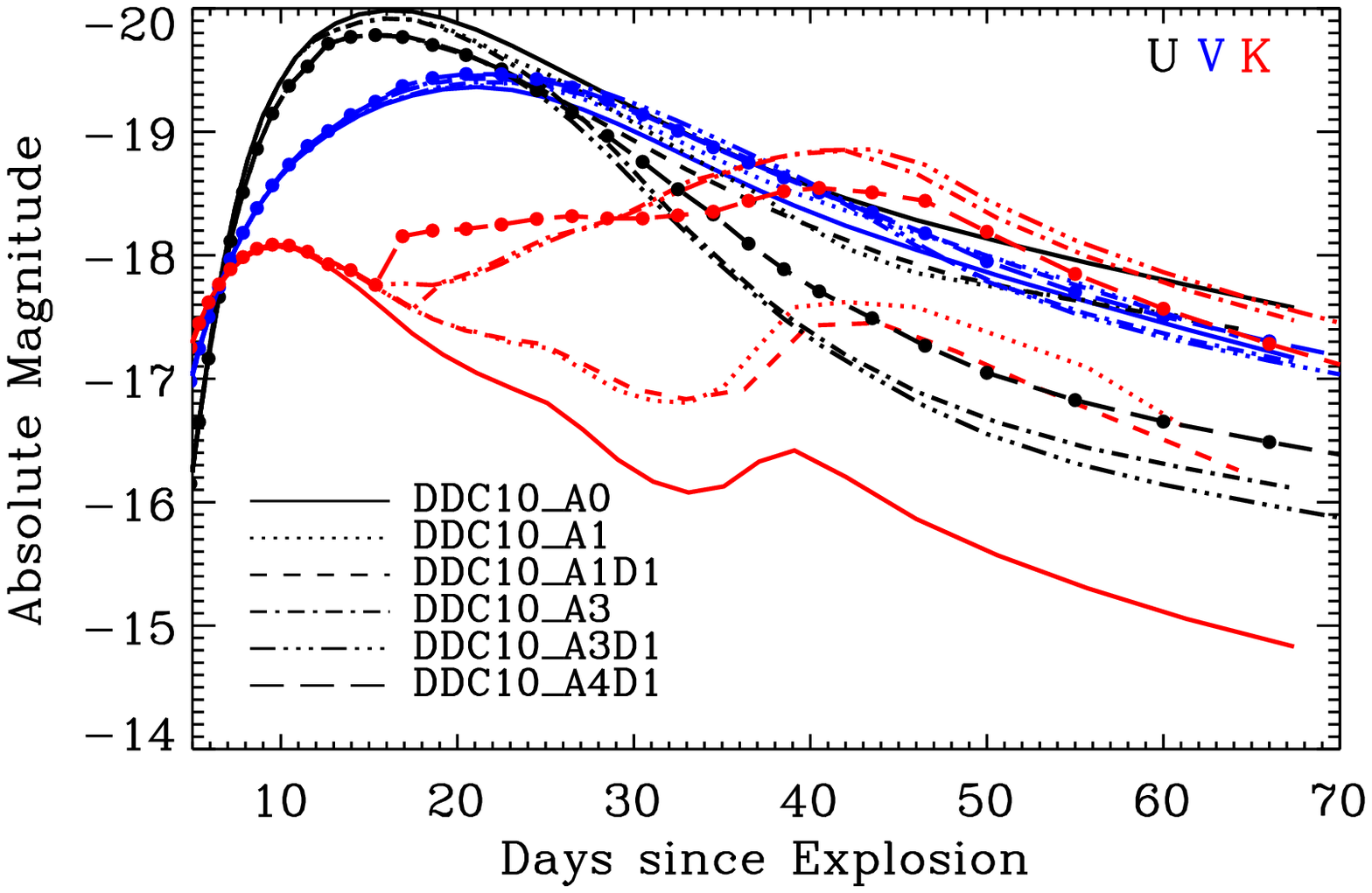,width=8.5cm}
\caption{
{\it Left:} Illustration of the relative fraction of the flux that contributes to the $UBVRI$ brightness
for models DDC10\_A0, DDC10\_A1, DDC10\_A1D1, DDC10\_A3,  DDC10\_A3D1, and
DDC10\_A4D1.
{\it Right:} Same as left, but now showing the  light-curve evolution in the $U$, $V$, and $K$ bands.
Dots correspond to the actual times the \cmfgen\ models are computed.
\label{fig_color_pb}
}
\end{figure*}

\section{Near-IR secondary maximum}
\label{sect_nearir}

  The near-IR bump seen in SNe Ia has been studied in the past by, e.g., \citet{hoeflich_etal_95},
and more recently by \citet{kasen_06} and \citet{jack_etal_12}, who find that it stems from an ionization effect associated
with Cobalt primarily. Here, we investigate the photometric properties of our various DDC10 time
sequences and discuss what controls the behavior, in our simulations, of the near-IR light curves.
This also serves as an additional test for \cmfgen.

   While optical bands show a single pronounced peak around bolometric maximum,
near-IR bands show two peaks, one before the bolometric maximum and one after.
At early times, when a well defined photosphere exists, the brightness in any given spectral
region is controlled primarily by the radius and temperature at the photosphere. On the rise
to light curve peak, the photosphere dramatically expands but heats up
moderately. At such times, the diffusion of heat from greater depth is
strongly degraded by ejecta expansion. Hence, early on, the SN brightens in all bands.
However, as time proceeds, heating is strong enough to raise the temperature in the
spectrum formation region, causing
the SED to shift to the blue, with a peak around 3000\,\AA\ at bolometric
maximum.
The hardening of the radiation is so strong that, despite
the fast expansion of the photosphere, the flux in the red decreases.
The resulting first peak in red bands occurs in
all models (DDC10\_A0, DDC10\_A1, DDC10\_A3 etc.), irrespective of model atoms used,
at about 14.0 ($I$ band), 11.0 ($J$ band), 9.7 ($H$ band), and 9.8\,d ($K$ band). This time differs by a
few 0.1\,d between models at most. The bolometric maximum occurs at 17.8\,d (again,
a difference of $\pm$\,0.2\,d is seen between models).
In our delayed-detonation model DDC10, the optical  colors continuously change prior to peak,
thus making the SN behave quite differently from a ``fireball" \citep{nugent_etal_11}.

All our models discussed above and presented in Table~\ref{tab_modset} show a secondary bump
in the near-IR light curves, but this bump is pronounced and mimics a secondary maximum only
in models DDC10\_A3, DDC10\_A3D1, and DDC10\_A4D1.
As shown in the left panel of Fig.~\ref{fig_color_pb},\footnote{The glitches
in the light curves correspond to times when we vary radiative-transfer ``ingredients'' , i.e.,
when we alter model atom characteristics or introduce non-thermal processes. In the future,
we will need to try to incorporate all the necessary microphysics at the start to avoid such
unwanted variations, although it can be hard to do due to memory or CPU constraints.}
the relative flux emitted outside of the optical range
remains $\lesssim$\,30\% beyond bolometric maximum, leading to a mild re-brightening in the near-IR.
The larger that fraction, the stronger the near-IR ``maximum".
Strictly speaking, and as evidenced in the right panel of Fig.~\ref{fig_color_pb},
only models DDC10\_A3 and DDC10\_A3D1
exhibit a genuine secondary maximum. For model  DDC10\_A3D1, the specific numbers for the
corresponding post-explosion times
are 43.3\,d (-18.95\,mag), 46.4\,d (-18.19\,mag), 41.2\,d (-18.96\,mag), and 42.32\,d (-18.87\,mag)
in the $I$, $J$, $H$, and $Ks$ bands.

Interestingly, our model bolometric light curves sometimes show an excess at about 40\,d after explosion,
although this excess is entirely absent in all models with the over-ionization and spurious temperature jump.
In other words, the only models that exhibit a late bump in bolometric luminosity treat [Co\three] lines
and recombine to Co\two. Recombination energy of 1\,\msun\ of Co from twice to once ionized
liberates about 5$\times$10$^{44}$\,erg,
which, if radiated over a week, produces a meagre 1.6$\times$10$^6$\,\lsun, i.e., too little to cause the bump.
However, SN ejecta are radiation dominated, so that the energy held up in {\it trapped} radiation
completely overwhelms what may be stored in excitation and ionization energy.
Assuming the gas and the trapped radiation are in equilibrium, a decrease by 1000\,K
at 10000\,K in region between [$r_0,r_1$] of [3,5] $\times$\,10$^{15}$\,cm liberates an energy
of  $a\Delta T^4 \Delta V$, where $a$ is the radiation constant, $\Delta V \sim 4 (r_1^3 - r_0^3)$.
Radiated over 10\,d, the corresponding power is 10$^{42.5}$\ergs. In practice, the radiation and the
gas are not in equilibrium, i.e. the mean intensity drops below the Planck source function, but
this suggests that the release of trapped radiation is indeed large enough to produce the bump seen and observed.

\begin{figure}
\epsfig{file=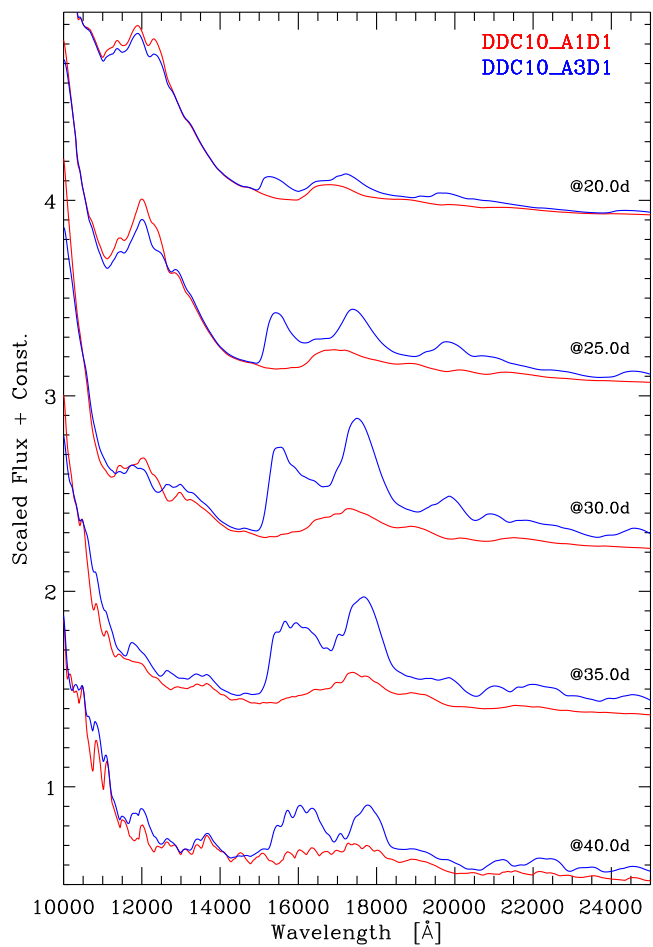,width=8.5cm}
\caption{Comparison between synthetic spectra of models DDC10\_A1D1 and DDC10\_A3D1
from 20 to 40\,d after explosion, i.e. when the near-IR generally presents a secondary maximum.
The same scaling is applied to both sets of synthetic spectra, so the offset between them is genuine.
(see Fig.~\ref{fig_color_pb}).
\label{fig_A1D1_A3D1_nearir_spec}
}
\end{figure}

There is no doubt the bump in bolometric luminosity is associated with a change in ionization
\citep{hoeflich_etal_95}, which comes with a large change in temperature.
This ionization change affects primarily cobalt, which goes from Co\three\ to
Co\two\ soon after the bolometric maximum in our delayed-detonation model.
While the main coolant for Co\three\ are forbidden transitions in the optical (primarily at $\sim$\,5900\,\AA),
where the opacity from overlapping lines is relatively large (especially at $\lesssim$\,5000\,\AA),
the main coolants for Co\two\ are forbidden transitions in the near-IR
associated with the states 4s$^2$--4p$^1$,
As shown in Fig~\ref{fig_A1D1_A3D1_nearir_spec},
besides a general increase in the overall flux, strong lines develop in model DDC10\_A3D1 at
1.6--1.8$\mu$m (model DDC10\_A4D1 has similar near-IR spectral properties),
while they are weak or absent in model DDC10\_A1D1. These lines are primarily
due to permitted transitions of Co\two.
To conclude, both the bump in bolometric luminosity and in near-IR light
curves stem from an ionization shift and the sudden strengthening of Co\two\
emission.

\citet{kasen_06} model the near-IR light curves of SNe Ia and reproduces
the basic morphology, including the secondary maximum. Surprisingly, he
does not account for forbidden line transitions, which we demonstrate here are
key for getting the proper ejecta ionization state (i.e., Co\three\ versus Co\two,
Fe\three\ versus Fe\two), and associated emission features. 

\section{Discussion and conclusions}
\label{sect_conc}

In the present work we have utilized a single delayed-detonation explosion model
of a Chandrasekhar mass WD as input for spectral
calculations of type Ia SNe. Such explosion models, although very attractive
for their ejecta properties and radiative properties, need to be further
studied to understand the growing lack of evidence for companion stars, giving further support
to the notion that many SNe Ia may arise from the coalescence of two WDs.
Despite the limitations of the progenitor model,
it is important to build confidence in the radiative-transfer modeling of SNe Ia, in particular
be able to accurately compute the color and spectral evolution of such models
and understand the dependencies of synthetic observables.
In this and future studies, we wish to investigate how well 1-D delayed-detonation models can
reproduce standard SN Ia spectra and multi-band light curves, i.e., whether their spectral signatures
support the delayed-detonation explosion scenario for SNe Ia.
Can a 1-D treatment, as employed in \cmfgen, be at all successful in that task?

Unfortunately, accurate modeling of SN Ia radiation, in particular of their
spectra, is very difficult. The large abundance of IMEs and IGEs, and
the lack of hydrogen, means that the continuum opacity is small and that the opacity, in stark contrast to
type II SNe, is  dominated by line opacity at most wavelengths. In addition the small ejecta masses, and
low densities, mean that departures from LTE are large, and significantly
influence spectral formation.
Since non-LTE effects are important we need accurate atomic data (not just line opacities)
which for IMEs and IGEs is often lacking, or of insufficient quality. Further, the scale of the problem means
that approximate techniques are often used to simplify the radiative transfer and/or the determination of
the thermodynamic state of the gas.

In the present work we have tried to overcome many of the limitations so that we can accurately model
the spectra of 1D delayed-detonation models. Using \cmfgen\ we have undertaken time-dependent radiative transfer
and statistical equilibrium calculations to model Ia spectra. We make no assumptions about how photons
are thermalized, and we also utilize the same model for comparison of spectra at different epochs.
The latter minimizes the influence of free-parameters. However, these calculations have limitations.
While we use large model atoms, these remain of limited size to match with the current
computer capabilities (each time step takes about 2 days and requires 10\,Gb of RAM).
We also use a moderately large turbulent velocity of 50\,\kms, which enhances intrinsic line overlap,
and we have used super-levels to help facilitate the solution of the rate equations. While we
have made some tests of the influence of these assumptions, such tests have not been exhaustive.

Using our DDC10 model, we were initially unable to match the gross
properties of SN Ia spectra  after bolometric maximum.
Various tests were undertaken to determine the cause of the discrepancy. Surprisingly, the
mismatch in spectra was not due to missing opacity. Increasing the size  of the model atoms
and the number of lines treated did not solve the problem. Rather, the spectral mismatch was due to
the neglect of [Co\three] lines in the Co\three\ model atom. These lines provide crucial cooling, even
at densities well above their critical density, which shifts the Co ionization from Co\three\ towards
Co\two. This, in turn, enhances the opacity in the $U$ band leading to better
agreement with observations.
Using the same hydrodynamical model we are now able to match the spectrum and
multi-band light curves of SN\,2005cf
from pre-maximum (-12\,d) to well beyond maximum (+40\,d). Thus we can reproduce
the basic light curve and fundamental spectral properties of type Ia SNe with the standard
delayed-detonation scenario, even with the assumption of {\it spherical symmetry}.

Of all processes we allow to vary in this work,
the most important one that controls the light curve morphology of SNe Ia
is $\gamma$-ray escape. Indeed, the low mass of SNe Ia ejecta causes a huge leakage of
energy. It starts to be visible about a week after explosion through an
increase in the luminosity;
the non-local energy deposition ``speeds" up the diffusion of radiant energy.
Beyond bolometric maximum, non-local energy deposition is superseded by $\gamma$-ray
escape so that models that treat $\gamma$-ray transport, rather than assuming full trapping,
fade significantly faster. In Nature, this leakage is function of the trapping efficiency and should
thus vary with ejecta mass, expansion rate, and \nifs\ distribution (see, e.g., \citealt{pinto_eastman_00a}).

Despite the ionized conditions in SN Ia ejecta, we find that  non-thermal processes play
an important role, even at bolometric maximum.
Earlier on, decay energy is deposited at high optical depths,
which inhibits non-thermal effects in the spectrum formation region. However, as the ejecta thins out,
non-thermal processes maintain a much higher ionization in simulations that include them, allowing for
the concomitant presence of Co\two\ and Co\three\ instead of Co\two\ alone.
This alters the color evolution after light curve peak. The magnitude of the effect
varies subtly with the optical depth, since thermal excitation/ionization can also occur if the heat deposited
is large enough to cause a significant temperature difference.
From 60 to 200\,d after explosion, the spectrum formation, located at
$\lesssim$\,10000\,\kms, retains a fairly constant temperature of
$\sim$\,7500\,K and a stable ionization with equal fractions for
Co$^+$--Co$^{2+}$ and Fe$^+$--Fe$^{2+}$.
We expect a modulation in the magnitude of such non-thermal effects in SN Ia
ejecta endowed with different initial masses of \nifs. We will explore this issue in
a forthcoming study.

Our radiative-transfer simulations cover from the UV to the far-IR at all
epochs computed. The near-IR light curves for all models, even those
characterized by a small model atom, develop a bump after bolometric
maximum. However, only models that reproduce the basic color evolution
in the optical (and thus capture adequately the shift in ionization)
develop a pronounced secondary maximum in the near-IR, as observed.
In fact, both the bump in bolometric luminosity and near-IR brightness
are tied to an ionization shift, in particular from Co\three\ to Co\two,
which causes enhanced ejecta cooling, and thus an increase in bolometric
luminosity. This escaping radiation appears primarily through [Co\two]
emission, which takes place primarily in the near-IR.

The intense development and testing we have performed  with {\sc cmfgen} for SN Ia calculations
in the last five years suggests that it is in fact possible to reproduce with unprecedented fidelity the
fundamental radiative properties of SNe Ia, both at pre-peak, peak, and post-peak epochs.
This suggests that while multi-dimensionality may play a role, it is not paramount for the
radiative transfer solution. Far more important is to treat explicitly non-LTE, line blanketing,
non-thermal processes, and to include all relevant radiative and collisional processes.
Although generally neglected, forbidden line transitions are found to play an essential role
in controlling the ejecta ionization and temperature as early as the peak of the light curve.
Following these fruitful benchmarking and educational explorations, we are now in a position
to investigate the physics of SN Ia explosions and search for clues about the progenitor systems.

\section*{Acknowledgments}

We acknowledge useful discussions with Carles Badenes, and interaction with Monique Arnaud
and Pasquale Mazzotta on ionization cross sections.  We also thank Daniel Kasen for
his insightful comments regarding the two emission features near 1.7$\mu$m.
LD and SB acknowledge financial support from the European Community through an
International Re-integration Grant, under grant number PIRG04-GA-2008-239184,
and from ``Agence Nationale de la Recherche" grant ANR-2011-Blanc-SIMI-5-6-007-01.
DJH acknowledges support from STScI theory grant HST-AR-12640.01, and NASA theory grant NNX10AC80G.
This work was also supported in part by the National Science Foundation under Grant No. PHYS-1066293 and
benefited from the hospitality of the Aspen Center for Physics.
AK acknowledges the NSF support through the NSF grants AST-0709181 and TG-AST090074.
This work was granted access to the HPC resources of CINES under the
allocation c2013046608 made by GENCI (Grand Equipement
National de Calcul Intensif).

\appendix

\section{Model atoms}
\label{appendix_atom}

The model atoms adopted for all simulations in this work are essentially the same origin as those
used in \citet{DH11}.
The sources of atomic data are varied, and in many cases multiple data sets for a given ion are available.
In some cases these multiple data sets represent an evolution in data quality and/or quantity, while in
other cases they represent different sources and/or computational methods. Comparisons of models
calculated with different data sets and atomic models potentially provide insights into the sensitivity
of our results to the adopted model atoms and hydrodynamical inputs (although such calculations have yet to be
undertaken for SNe).

Oscillator strengths for CO elements were originally taken from
\citet{NS83_LTDR, NS84_CNO_LTDR}. These authors also provide transition probabilities to
states in the ion
continuum. The largest source of oscillator data is from \citet{Kur09_ATD}; its principal advantage
over many other sources (e.g., Opacity Project) is that LS coupling is not assumed. More recently, non-LS
oscillator strengths have become available through the Iron Project \citep{HBE93_IP}, and work done by
the atomic-data group at Ohio State University \citep{Nahar_OSU}. Other important
sources of radiative data for Fe include \citet{BB92_FeV,  BB95_FeVI, BB95_FeIV}, \cite{Nahar95_FeII}.
Atomic data from the opacity project comes from TOPBASE \citep{Topbase93}.
Energy levels have generally been obtained from National Institute of Standards and Technology.
Collisional data is sparse, particularly for
states far from the ground state. The principal source for collisional data among low lying states
for a variety of species is the tabulation by \citet{Men83_col}; other sources include
\citet{BBD85_col}, \citet{LDH85_CII_col}, \citet{LB94_N2}, \citet{SL74},
\citet{T97_SII_col,T97_SIII_col}, Zhang \& Pradhan (\citeyear{ZP95_FeII_col,ZP95_FeIII_col,ZP97_FeIV_col}).
Photoionization data is taken from the Opacity Project \citep{Sea87_OP,Topbase93}, the
Iron Project \citep{HBE93_IP,NP96_FeIII}, and \citet{NP93_SiI}. Unfortunately, Ni and Co
photoionization data is generally unavailable so we have utilized crude approximations,
except for photoionization from the ground state for which we use data from \citet{Vy95_phot}.
Charge exchange cross-sections are from the tabulation
by \citet{KF96_chg}.
Atomic data for C\four\  was obtained from \citet{Lei72_CIV,PSS88_LI_seq},
and for the carbon isoelectronic sequence from \citet{LP89_C_seq}.
Collision strengths for Ar\two\ are from \citet{TH96_ArII_col}.
The LS Ne\,{\sc i} photoionization cross-sections were modified according to \cite{Sea98_NeI_phot}.
The same procedure was applied to using Ar\,{\sc i} mixing coefficients computed at
http://aphysics2.lanl.gov/tempweb/lanl.
Additional data for Ne\,\one\ was obtained from the MCHF/MCDHF web site: http://nlte.nist.gov/MCHF.

For the modelling of SNe Ia, the main issues concern the Cobalt atomic data, and in particular
Co\two\ and Co\three. For these, accurate photo-ionization cross sections and collisional rates
are needed to improve the accuracy of the radiation transfer modelling.

\begin{table}
\begin{center}
\caption[]{
Summary of the model atom A0, refered to as ``Small" in
Table~\ref{tab_modset}. The source of the atomic datasets is given in \citet{DH10} and in Section~\ref{appendix_atom}.
N$_{\rm f}$ (N$_{\rm s}$) refers to the number of full (super) levels, and N$_{\rm trans}$
to the corresponding number of bound-bound transitions. The last column refers to the upper level for each ion
treated. In this configuration, the total number of full (super) levels treated is 8370 (1773), which corresponds to
174\,674 bound-bound transitions.}
\label{tab_atom_A0}
\begin{tabular}{l@{\hspace{3mm}}r@{\hspace{3mm}}r@{\hspace{3mm}}r@{\hspace{3mm}}l}
\hline
 Species        &  N$_{\rm f}$  &  N$_{\rm s}$ & N$_{\rm trans}$ & Upper Level \\
\hline
     C\,{\sc i}\,    &  26  &   14 &    120 & 2s2p$^3$$\,^3$P\opar\                \\
     C\,{\sc ii}\,   &  26  &   14 &     87 & 2s$^2$4d\,$^2$D$_{5/2}$                \\
     C\,{\sc iii}\,  & 112  &   62 &    891 & 2s8f\,$^1$F\opar\                      \\
     C\,{\sc iv}\,   &  64  &   59 &   1446 & n=30                         \\
     O\,{\sc i}\,    &  51  &   19 &    214 & 2s$^2$2p$^3$($^4$S)4f$\,^3$F$_{3}$         \\
     O\,{\sc ii}\,   & 111  &   30 &   1157 & 2s$^2$2p$^2$($^3$P)4d\,$^2$D$_{5/2}$       \\
     O\,{\sc iii}\,  &  86  &   50 &    646 & 2p4f\,$^1$D                       \\
     O\,{\sc iv}\,   &  72  &   53 &    835 & 2p$^2$($^3$P)3p\,$^2$P\opar\                   \\
     Ne\,{\sc i}\,  &  139  &   70 &   1587 & 2s$^2$2p$^5$($^2$P\oparsub{3/2})6d\,$^2$[5/2]\oparsub{3}       \\
     Ne\,{\sc ii}\,  &  91  &   22 &   1106 & 2s$^2$2p$^4$($^3$P)4d\,$^2$P$_{3/2}$       \\
     Ne\,{\sc iii}\, &  71  &   23 &    460 & 2s$^2$2p$^3$($^2$D\opar)3d\,$^3$S$_{1}$         \\
     Na\,{\sc i}\,   &  71  &   22 &   1614 & 30w\,$^2$W                         \\
     Mg\,{\sc ii}\,  &  65  &   22 &   1452 & 30w\,$^2$W                         \\
     Mg\,{\sc iii}\, &  99  &   31 &    775 & 2p$^5$7s$\,^1$P\opar                    \\
     Al\,{\sc ii}\,  &  44  &   26 &    171 & 3s5d\,$^1$D$_{2}$                    \\
     Al\,{\sc iii}\, &  45  &   17 &    362 & 10z\,$^2$Z                         \\
     Si\,{\sc ii}\,  &  59  &   31 &    354 & 3s$^2$($^1$S)7g\,$^2$G$_{7/2}$             \\
     Si\,{\sc iii}\, &  61  &   33 &    310 & 3s5g\,$^1$Ge$_{4}$                    \\
     Si\,{\sc iv}\,  &  48  &   37 &    405 & 10f\,$^2$Fo                        \\
     S\,{\sc ii}\,   & 324  &   56 &   8208 & 3s3p$^3$($^5$S\opar)4p\,$^6$P             \\
     S\,{\sc iii}\,  &  98  &   48 &    837 & 3s3p$^2$($^2$D)3d$\,^3$P             \\
     S\,{\sc iv}\,   &  67  &   27 &    396 & 3s3p($^3$P\opar)4p\,$^2$D$_{5/2}$         \\
     Ar\one\      & 110 & 56  & 1541        & 3s$^2$3p$^5$($^2$P\oparsub{3/2})7p\,$^2$[3/2]$_2$  \\
     Ar\two\       & 415 & 134   & 20197  &     3s$^2$3p$^4$($^3$P$_1$)7i\,$^2$[6]$_{11/2}$  \\
     Ar\,{\sc iii}\, & 346  &   32 &   6898 & 3s$^2$3p$^3$($^2$Do)8s\,$^1$Do            \\
     Ca\,{\sc ii}\,  &  77  &   21 &   1736 & 3p$^6$30w\,$^2$W                     \\
     Ca\,{\sc iii}\, &  40  &   16 &    108 & 3s$^2$3p$^5$5s$\,^1$P\opar               \\
     Ca\,{\sc iv}\,  &  69  &   18 &    335 & 3s3p$^5$($^3$P\opar)3d\,$^4$D\oparsub{1/2}        \\
     Sc\,{\sc ii}\,  &  85  &   38 &    979 & 3p$^6$3d4f$\,^1$P\oparsub{1}             \\
     Sc\,{\sc iii}\, &  45  &   25 &    235 & 7h\,$^2$H\oparsub{11/2}                  \\
     Ti\,{\sc ii}\,  & 152  &   37 &   3134 & 3d$^2$($^3$F)5p\,$^4$D\oparsub{7/2}            \\
     Ti\,{\sc iii}\, & 206  &   33 &   4735 & 3d6f\,$^3$H\oparsub{6}                   \\
     Cr\,{\sc ii}\,  & 196  &   28 &   3629 & 3d$^4$($^3$G)4p\,x$^4$G\oparsub{11/2}          \\
     Cr\,{\sc iii}\, & 145  &   30 &   2359 & 3d$^3$($^2$D2)4p\,$^3$D\oparsub{3}            \\
     Cr\,{\sc iv}\,  & 234  &   29 &   6354 & 3d$^2$($^3$P)5p$^4$P\oparsub{5/2}            \\
     Mn\,{\sc ii}\,  &  97  &   25 &    236 & 3d$^4$($^5$D)4s$^2\,$c$^5$D$_{4}$            \\
     Mn\,{\sc iii}\, & 175  &   30 &   3173 & 3d$^4$($^3$G)4p\,y$^4$H\oparsub{13/2}         \\
       Fe\,{\sc i}\,  &   136 &     44 &     1900 & 3d$^6$($^5$D)4s4p\,x$^5$F\oparsub{3}           \\
       Fe\,{\sc ii}\, &   115 &     50 &     1437 & 3d$^6$($^1$G1)4s\,d$^2$G$_{7/2}$           \\
     Fe\,{\sc iii}\,  &   477 &     61 &     6496 & 3d$^5$($^4$F)5s\,$^5$F$_{1}$               \\
      Fe\,{\sc iv}\,  &   294 &     51 &     8068 & 3d$^4$($^5$D)4d\,$^4$G$_{5/2}$            \\
       Fe\,{\sc v}\,  &   191 &     47 &     3977 & 3d$^3$($^4$F)4d\,$^5$F$_{3}$              \\
     Fe\,{\sc vi}\,   &   433 &     44 &    14\,103 & 3p5($^2$P)3d$^4$($^1$S)\,$^2$Pc\oparsub{3/2}      \\
     Fe\,{\sc vii}\,  &   153 &     29 &     1753 & 3p5($^2$P)3d$^3$\,(b$^2$D)$\,^1$P$_{1}$        \\
       Co\,{\sc ii}\, &   144 &     34 &     2088 & 3d$^6$($^5$D)4s4p$\,^7$D\oparsub{1}            \\
     Co\,{\sc iii}\,  &   361 &     37 &    10\,937 & 3d$^6$($^5$D)5p\,$^4$P$_{3/2}$            \\
      Co\,{\sc iv}\,  &   314 &     37 &     8684 & 3d$^5$($^2$P)4p$\,^3$P\oparsub{1}              \\
       Co\,{\sc v}\,  &   387 &     32 &    13\,605 & 3d$^4$($^3$F)4d\,$^2$H$_{9/2}$            \\
     Co\,{\sc vi}\,   &   323 &     28 &     9608 & 3d$^3$($^2$D)4d$\,^1$S$_{0}$              \\
     Co\,{\sc vii}\,  &   319 &     31 &     9096 & 3p5($^2$P)3d$^4$($^3$F)\,$^2$D$_{3/2}$        \\
       Ni\,{\sc ii}\, &    93 &     19 &      842 & 3d$^7$($^4$F)4s4p\,$^6$D\oparsub{1/2}          \\
     Ni\,{\sc iii}\,  &    67 &     15 &      379 & 3d$^7$ $^4$F 4p\,$^3$D\oparsub{1}              \\
      Ni\,{\sc iv}\,  &   200 &     36 &     4085 & 3d$^6$($^3$D)4p\,$^2$D\oparsub{5/2}            \\
       Ni\,{\sc v}\,  &   183 &     46 &     3065 & 3d$^5$($^2$D3)4p$\,^3$F\oparsub{3}             \\
     Ni\,{\sc vi}\,   &   314 &     37 &     9569 & 3d$^4$($^5$D)4d\,$^4$F$_{9/2}$            \\
     Ni\,{\sc vii}\,  &   308 &     37 &     9225 & 3d$^3$($^2$D)4d$\,^3$P$_{2}$              \\
\hline
\end{tabular}
\end{center}
\end{table}

\begin{table}
\begin{center}
\caption[]{
Same as for Table~\ref{tab_modset}, but now showing the bigger Fe/Co/Ni atoms used
for the model sequence DDC10\_A1 (the model atom for other species is kept the same and the details
about these are not repeated here). Because the
larger model atom only starts at later times when the ejecta has cooled, some high ionization
stages for IGEs are excluded.
The total number of full (super) levels treated is 13\,959 (2149), which corresponds to
629\,396 bound-bound transitions.
Model atom A3 is identical to A1 except that it includes forbidden-line transitions for Co\three.}
\label{tab_atom_A1}
\begin{tabular}{l@{\hspace{3mm}}r@{\hspace{3mm}}r@{\hspace{3mm}}r@{\hspace{3mm}}l}
\hline
 Species        &  N$_{\rm f}$  &  N$_{\rm s}$ & N$_{\rm trans}$ & Upper Level \\
\hline
     Fe\,{\sc i}\,   & 136  &   44 &   1900 & 3d$^6$($^5$D)4s4p\,x$^5$F\oparsub{3}          \\
     Fe\,{\sc ii}\,  & 827  &  275 &  44\,831 & 3d$^5$($^6$S)4p$^2$($^3$P)\,$^4$P$_{1/2}$        \\
     Fe\,{\sc iii}\, & 607  &   69 &   9794 & 3d$^5$($^4$D)6s\,$^3$D$_{2}$               \\
     Fe\,{\sc iv}\,  &1000  &  100 &  72\,223 & 3d$^4$($^3$G)4f\,$^4$P\oparsub{5/2}            \\
     Fe\,{\sc v}\,   & 191  &   47 &   3977 & 3d$^3$($^4$F)4d\,$^5$F$_{3}$              \\
     Fe\,{\sc vi}\,  & 433  &   44 &  14\,103 & 3p5($^2$P\opar)3d$^4$($^1$S)\,$^2$Pc\oparsub{3/2}      \\
     Co\,{\sc ii}\,  &1000  &   81 &  61\,986 & 3d$^7$($^4$P)4f\,$^5$F\oparsub{4}              \\
     Co\,{\sc iii}\, &1000  &   72 &  68\,462 & 3d$^6$($^5$D)5f\,$^4$F\oparsub{9/2}            \\
     Co\,{\sc iv}\,  &1000  &   56 &  69\,425 & 3d$^5$($^2$D)5s\,$^1$D$_{2}$              \\
     Co\,{\sc v}\,   & 387  &   32 &  13\,605 & 3d$^4$($^3$F)4d\,$^2$H$_{9/2}$            \\
     Co\,{\sc vi}\,  & 323  &   28 &   9608 & 3d$^3$($^2$D)4d$\,^1$S$_{0}$              \\
     Ni\,{\sc ii}\,  &1000  &   59 &  51\,707 & 3d$^8$($^3$F)7f\,$^4$I\oparsub{9/2}      \\
     Ni\,{\sc iii}\, &1000  &   47 &  66\,486 & 3d$^7$($^2$D)4d\,$^3$Sb$_{1}$             \\
     Ni\,{\sc iv}\,  &1000  &   54 &  72\,898 & 3d$^6$($^5$D)6p\,$^6$F$_{11/2}$           \\
     Ni\,{\sc v}\,   & 183  &   46 &   3065 & 3d$^5$($^2$D3)4p$\,^3$F$_{3}$             \\
     Ni\,{\sc vi}\,  & 314  &   37 &   9569 & 3d$^4$($^5$D)4d\,$^4$F$_{9/2}$            \\
\hline
\end{tabular}
\end{center}
\end{table}

\begin{table}
\begin{center}
\caption[]{Same as for Table~\ref{tab_modset}, but now showing the huge Co\,\two\ and Co\,\three\
model atom A2 used in model DDC10\_A2 --- all other ions have the same characteristics as in DDC10\_A1
with the exception of Fe\,\six, Co\,\six, and Ni\,\six, which are excluded in DDC10\_A2 because of the lower
ionization of the ejecta at the times we perform our tests.
Model atom A4 is identical to A2 except that it also treats all important forbidden-line
transitions of metal ions --- this model atom is used in the sequence DDC10\_A4D1.
With this configuration, the total number of full (super) levels treated is now only 17\,553 (2338),
which corresponds to 1\,738\,088 bound-bound transitions.
}
\label{tab_atom_A2}
\begin{tabular}{l@{\hspace{3mm}}r@{\hspace{3mm}}r@{\hspace{3mm}}r@{\hspace{3mm}}l}
\hline
 Species        &  N$_{\rm f}$  &  N$_{\rm s}$ & N$_{\rm trans}$ & Upper Level \\
     \hline
       Co\,{\sc ii}\, &   2747 &    136 &    593\,140 &      3d$^7$($^2$D)6p$\,^3$P\oparsub{1}         \\
     Co\,{\sc iii}\,  &   3917 &     315&    679\,280 &      3d$^6$($^3$D)6d$\,^4$P$_{3/2}$    \\
\hline
\end{tabular}
\end{center}
\end{table}

\begin{figure}
\epsfig{file=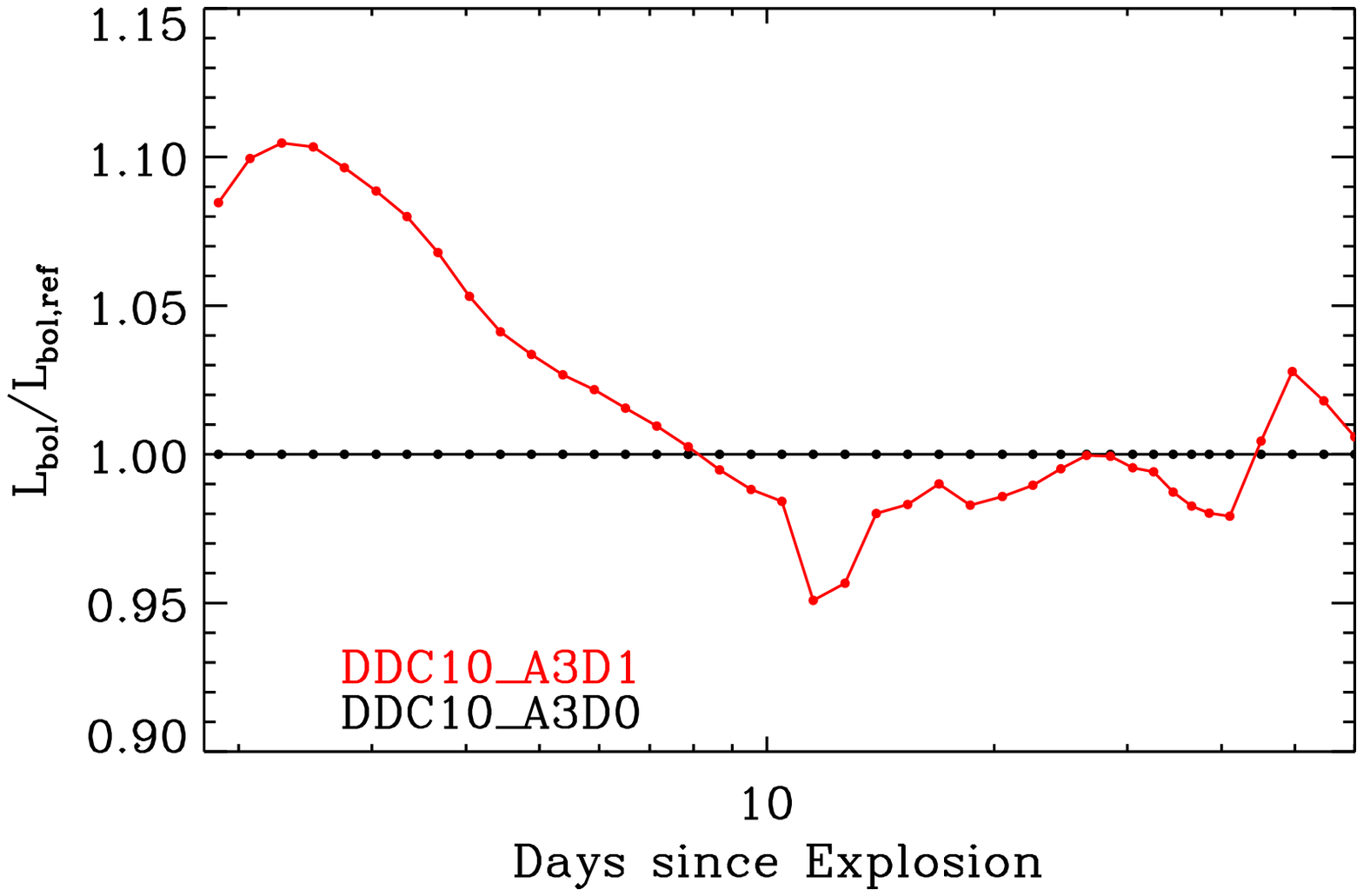,width=8.5cm}
\caption{
Illustration of the bolometric light curve for model with (DDC10\_A3D1)
and without (DDC10\_A3D0) the additional two-step decay chains.
To better reveal the small contrast, we show the ratio with the
DDC10\_A3D0 used as reference.}
\label{fig_lc_nuc}
\end{figure}

\section{Decay routes}
\label{app_decay}

 Energy wise, the  \iso{56}Ni decay chain is the most important for SNe Ia calculation.
 However, other chains may need to be considered if they are associated with unstable
 isotopes present in regions where \iso{56}Ni is absent (for example in the outer ejecta),
 or if these chains produce isotopes with a potentially strong line blanketing power. This
 is the case of Ti\two, whose mass fraction can be increased by two orders of magnitude
 through the decay of \iso{48}Cr and \iso{48}V.

   We have modified  \cmfgen\ to handle  multiple decay chains, either 2-step like
\iso{56}Ni $\rightarrow$ \iso{56}Co $\rightarrow$ \iso{56}Fe, or 1-step. In this work,
we generally treat only the \iso{56}Ni decay chain, but investigate at times the impact
of having all 2-step decay chains presented in Tables~\ref{tab_nuc1}--\ref{tab_nuc2}.
The role of 1-step decay chains is discussed in \citet{dessart_etal_14a}.
The Monte Carlo transport code that calculates the non-local
$\gamma$-ray energy deposition was modified to handle the same chains.

  Nuclear masses are taken from \citet{audi_etal_03}, while decay products ($\gamma$-ray lines,
  electrons/positrons, and neutrinos) and energies are taken from http://www.nndc.bnl.gov/chart.

  For an illustration, we show in Fig.~\ref{fig_lc_nuc} the effect on the bolometric light curve
  of introducing all 2-step decay chains rather than \iso{56}Ni decay chain only. The effect
  is at most of $\lesssim$\,10\%, and limited to early, i.e. pre-peak, times.

\begin{table}
\begin{center}
\caption{Summary of the 2-step decay chains used.
For all chains, we give the characteristics for each
of the two steps, starting with the half-life, the total
energy radiated in $\gamma$-rays $Q_{\nu}$, and  the total
energy liberated in the form of particles $Q_{\rm th}$.
We then list the main $\gamma$-ray lines emitted in each decay
together with their probability.
\label{tab_nuc1}}
\vspace{0.3cm}
\begin{tabular}{cccc}
\hline
 \multicolumn{4}{c}{  $\iso{56}Ni \rightarrow \iso{56}Co \rightarrow \iso{56}Fe$ } \\
   \multicolumn{2}{c}{$\iso{56}Ni \rightarrow \iso{56}Co$  }  &   \multicolumn{2}{c}{$\iso{56}Co \rightarrow \iso{56}Fe$  }  \\
\hline
  \multicolumn{2}{c}{$t_{1/2}=$          6.075      d} &  \multicolumn{2}{c}{$t_{1/2}=$         77.233     d} \\
        \multicolumn{2}{c}{$Q_{\gamma}=$          1.718    MeV} &        \multicolumn{2}{c}{$Q_{\gamma}=$          3.633   MeV} \\
        \multicolumn{2}{c}{$Q_{\rm th}=$          0.000    MeV} &        \multicolumn{2}{c}{$Q_{\rm th}=$          0.116   MeV} \\
\hline
       $E_\gamma$ &             Prob. &        $E_\gamma$ &             Prob. \\
     0.158  &      98.8  &     0.511  &      38.0   \\
     0.270  &      36.5  &     0.847  &     100.0   \\
     0.480  &      36.5  &     0.977  &       1.4   \\
     0.750  &      49.5  &     1.038  &      14.0   \\
     0.812  &      86.0  &     1.175  &       2.3   \\
     1.562  &      14.0  &     1.238  &      67.6   \\
            &            &     1.360  &       4.3   \\
            &            &     1.771  &      15.7   \\
            &            &     2.015  &       3.1   \\
            &            &     2.035  &       7.9   \\
            &            &     2.598  &      17.3   \\
            &            &     3.010  &       1.0   \\
            &            &     3.202  &       3.2   \\
            &            &     3.253  &       7.9   \\
            &            &     3.273  &       1.9   \\
\hline
\hline
 \multicolumn{4}{c}{  $\iso{57}Ni \rightarrow \iso{57}Co \rightarrow \iso{57}Fe$ } \\
     \multicolumn{2}{c}{$\iso{57}Ni \rightarrow \iso{57}Co$}  &     \multicolumn{2}{c}{$\iso{57}Co \rightarrow \iso{57}Fe$}  \\
\hline
  \multicolumn{2}{c}{$t_{1/2}=$          1.483      d} &  \multicolumn{2}{c}{$t_{1/2}=$        271.740     d} \\
        \multicolumn{2}{c}{$Q_{\gamma}=$          1.937    MeV} &        \multicolumn{2}{c}{$Q_{\gamma}=$          0.122   MeV} \\
        \multicolumn{2}{c}{$Q_{\rm th}=$          0.154    MeV} &        \multicolumn{2}{c}{$Q_{\rm th}=$          0.000   MeV} \\
\hline
       $E_\gamma$ &             Prob. &        $E_\gamma$ &             Prob. \\
     0.127  &      16.7  &     0.014  &       9.2   \\
     0.511  &      87.0  &     0.122  &      85.6   \\
     1.378  &      81.7  &     0.137  &      10.7   \\
     1.758  &       5.8  &            &             \\
     1.919  &      12.3  &            &             \\
\hline
\hline
 \multicolumn{4}{c}{   $\iso{48}Cr \rightarrow \iso{48}V \rightarrow \iso{48}Ti$ } \\
      \multicolumn{2}{c}{$\iso{48}Cr \rightarrow \iso{48}V$}  &      \multicolumn{2}{c}{$\iso{48}V \rightarrow \iso{48}Ti$}  \\
\hline
  \multicolumn{2}{c}{$t_{1/2}=$          0.898      d} &  \multicolumn{2}{c}{$t_{1/2}=$         15.973     d} \\
        \multicolumn{2}{c}{$Q_{\gamma}=$          0.432    MeV} &        \multicolumn{2}{c}{$Q_{\gamma}=$          2.910   MeV} \\
        \multicolumn{2}{c}{$Q_{\rm th}=$          0.002    MeV} &        \multicolumn{2}{c}{$Q_{\rm th}=$          0.145   MeV} \\
\hline
       $E_\gamma$ &             Prob. &        $E_\gamma$ &             Prob. \\
     0.112  &      96.0  &     0.511  &      99.8   \\
     0.308  &     100.0  &     0.944  &       7.8   \\
     0.511  &       3.2  &     0.984  &     100.0   \\
            &            &     1.312  &      97.5   \\
            &            &     2.240  &       2.4   \\
\hline
\hline
 \multicolumn{4}{c}{   $\iso{49}Cr \rightarrow \iso{49}V \rightarrow \iso{49}Ti$ } \\
      \multicolumn{2}{c}{$\iso{49}Cr \rightarrow \iso{49}V$}  &      \multicolumn{2}{c}{$\iso{49}V \rightarrow \iso{49}Ti$}  \\
\hline
  \multicolumn{2}{c}{$t_{1/2}=$          0.029      d} &  \multicolumn{2}{c}{$t_{1/2}=$        330.000     d} \\
        \multicolumn{2}{c}{$Q_{\gamma}=$          1.055    MeV} &        \multicolumn{2}{c}{$Q_{\gamma}=$          0.000   MeV} \\
        \multicolumn{2}{c}{$Q_{\rm th}=$          0.598    MeV} &        \multicolumn{2}{c}{$Q_{\rm th}=$          0.000   MeV} \\
\hline
       $E_\gamma$ &             Prob. &        $E_\gamma$ &             Prob. \\
     0.062  &      16.4  &     0.511  &       0.0   \\
     0.091  &      53.2  &            &             \\
     0.153  &      30.3  &            &             \\
     0.511  &     186.0  &            &             \\
\hline
\end{tabular}
\end{center}
\end{table}

\begin{table}
\begin{center}
\caption{Cont.
\label{tab_nuc2}}
\vspace{0.3cm}
\begin{tabular}{cccc}
\hline
 \multicolumn{4}{c}{   $\iso{51}Mn \rightarrow \iso{51}Cr \rightarrow \iso{51}V$ } \\
     \multicolumn{2}{c}{$\iso{51}Mn \rightarrow \iso{51}Cr$}  &      \multicolumn{2}{c}{$\iso{51}Cr \rightarrow \iso{51}V$}  \\
\hline
  \multicolumn{2}{c}{$t_{1/2}=$          0.032      d} &  \multicolumn{2}{c}{$t_{1/2}=$         27.700     d} \\
        \multicolumn{2}{c}{$Q_{\gamma}=$          0.992    MeV} &        \multicolumn{2}{c}{$Q_{\gamma}=$          0.032   MeV} \\
        \multicolumn{2}{c}{$Q_{\rm th}=$          0.933    MeV} &        \multicolumn{2}{c}{$Q_{\rm th}=$          0.000   MeV} \\
\hline
       $E_\gamma$ &             Prob. &        $E_\gamma$ &             Prob. \\
     0.511  &     194.2  &     0.320  &       9.9   \\
\hline
\hline
 \multicolumn{4}{c}{  $\iso{55}Co \rightarrow \iso{55}Fe \rightarrow \iso{55}Mn$ } \\
     \multicolumn{2}{c}{$\iso{55}Co \rightarrow \iso{55}Fe$}  &     \multicolumn{2}{c}{$\iso{55}Fe \rightarrow \iso{55}Mn$}  \\
\hline
  \multicolumn{2}{c}{$t_{1/2}=$          0.730      d} &  \multicolumn{2}{c}{$t_{1/2}=$       1002.200     d} \\
        \multicolumn{2}{c}{$Q_{\gamma}=$          1.943    MeV} &        \multicolumn{2}{c}{$Q_{\gamma}=$          0.000   MeV} \\
        \multicolumn{2}{c}{$Q_{\rm th}=$          0.430    MeV} &        \multicolumn{2}{c}{$Q_{\rm th}=$          0.000   MeV} \\
\hline
       $E_\gamma$ &             Prob. &        $E_\gamma$ &             Prob. \\
     0.477  &      20.2  &     0.511  &       0.0   \\
     0.511  &     152.0  &            &             \\
     0.931  &      75.0  &            &             \\
     1.317  &       7.1  &            &             \\
     1.370  &       2.9  &            &             \\
     1.408  &      16.9  &            &             \\
\hline
\hline
 \multicolumn{4}{c}{   $\iso{37}K \rightarrow \iso{37}Ar \rightarrow \iso{37}Cl$ } \\
      \multicolumn{2}{c}{$\iso{37}K \rightarrow \iso{37}Ar$}  &     \multicolumn{2}{c}{$\iso{37}Ar \rightarrow \iso{37}Cl$}  \\
\hline
  \multicolumn{2}{c}{$t_{1/2}=$          1.226      s} &  \multicolumn{2}{c}{$t_{1/2}=$         35.040     d} \\
        \multicolumn{2}{c}{$Q_{\gamma}=$          1.072    MeV} &        \multicolumn{2}{c}{$Q_{\gamma}=$          0.000   MeV} \\
        \multicolumn{2}{c}{$Q_{\rm th}=$          2.347    MeV} &        \multicolumn{2}{c}{$Q_{\rm th}=$          0.000   MeV} \\
\hline
       $E_\gamma$ &             Prob. &        $E_\gamma$ &             Prob. \\
     0.511  &     199.8  &     0.003  &       5.5   \\
     2.796  &       1.8  &            &             \\
     3.601  &       0.0  &            &             \\
\hline
\hline
 \multicolumn{4}{c}{  $\iso{52}Fe \rightarrow \iso{52}Mn \rightarrow \iso{52}Cr$ } \\
     \multicolumn{2}{c}{$\iso{52}Fe \rightarrow \iso{52}Mn$}  &     \multicolumn{2}{c}{$\iso{52}Mn \rightarrow \iso{52}Cr$}  \\
\hline
  \multicolumn{2}{c}{$t_{1/2}=$          0.345      d} &  \multicolumn{2}{c}{$t_{1/2}=$          0.015     d} \\
        \multicolumn{2}{c}{$Q_{\gamma}=$          0.751    MeV} &        \multicolumn{2}{c}{$Q_{\gamma}=$          2.447   MeV} \\
        \multicolumn{2}{c}{$Q_{\rm th}=$          0.191    MeV} &        \multicolumn{2}{c}{$Q_{\rm th}=$          1.113   MeV} \\
\hline
       $E_\gamma$ &             Prob. &        $E_\gamma$ &             Prob. \\
     0.169  &      99.2  &     0.511  &     190.0   \\
     0.378  &       1.6  &     1.434  &      98.3   \\
     0.511  &     112.0  &     2.965  &       1.0   \\
            &            &     3.129  &       1.0   \\
\hline
\hline
 \multicolumn{4}{c}{  $\iso{44}Ti \rightarrow \iso{44}Sc \rightarrow \iso{44}Ca$ } \\
     \multicolumn{2}{c}{$\iso{44}Ti \rightarrow \iso{44}Sc$}  &     \multicolumn{2}{c}{$\iso{44}Sc \rightarrow \iso{44}Ca$}  \\
\hline
  \multicolumn{2}{c}{$t_{1/2}=$      21915.000      d} &  \multicolumn{2}{c}{$t_{1/2}=$          0.165     d} \\
        \multicolumn{2}{c}{$Q_{\gamma}=$          0.000    MeV} &        \multicolumn{2}{c}{$Q_{\gamma}=$          2.136   MeV} \\
        \multicolumn{2}{c}{$Q_{\rm th}=$          0.000    MeV} &        \multicolumn{2}{c}{$Q_{\rm th}=$          0.596   MeV} \\
\hline
       $E_\gamma$ &             Prob. &        $E_\gamma$ &             Prob. \\
     0.511  &       0.0  &     0.511  &     188.5   \\
            &            &     1.157  &      99.9   \\
\hline
\end{tabular}
\end{center}
\end{table}

\label{lastpage}


\begin{thebibliography}{103}
\expandafter\ifx\csname natexlab\endcsname\relax\def\natexlab#1{#1}\fi

\bibitem[{{Arnaud} \& {Rothenflug}(1985)}]{AR85}
{Arnaud}, M. \& {Rothenflug}, R. 1985, \aaps, 60, 425

\bibitem[{{Arnett}(1982)}]{arnett_82}
{Arnett}, W.~D. 1982, \apj, 253, 785

\bibitem[{{Audi} {et~al.}(2003){Audi}, {Wapstra}, \& {Thibault}}]{audi_etal_03}
{Audi}, G., {Wapstra}, A., \& {Thibault}, C. 2003, Nuclear Physics A729, 337

\bibitem[{{Axelrod}(1980)}]{axelrod_80}
{Axelrod}, T.~S. 1980, PhD thesis, California Univ., Santa Cruz.

\bibitem[{{Baron} {et~al.}(1996){Baron}, {Hauschildt}, {Nugent}, \&
  {Branch}}]{baron_etal_96}
{Baron}, E., {Hauschildt}, P.~H., {Nugent}, P., \& {Branch}, D. 1996, \mnras,
  283, 297

\bibitem[{{Becker} \& {Butler}(1992)}]{BB92_FeV}
{Becker}, S.~R. \& {Butler}, K. 1992, \aap, 265, 647

\bibitem[{{Becker} \& {Butler}(1995{\natexlab{a}})}]{BB95_FeVI}
---. 1995{\natexlab{a}}, \aap, 294, 215

\bibitem[{{Becker} \& {Butler}(1995{\natexlab{b}})}]{BB95_FeIV}
---. 1995{\natexlab{b}}, \aap, 301, 187

\bibitem[{{Berrington} {et~al.}(1985){Berrington}, {Burke}, {Dufton}, \&
  {Kingston}}]{BBD85_col}
{Berrington}, K.~A., {Burke}, P.~G., {Dufton}, P.~L., \& {Kingston}, A.~E.
  1985, Atomic Data and Nuclear Data Tables, 33, 195

\bibitem[{{Blinnikov} {et~al.}(1998){Blinnikov}, {Eastman}, {Bartunov},
  {Popolitov}, \& {Woosley}}]{blinnikov_etal_98}
{Blinnikov}, S.~I., {Eastman}, R., {Bartunov}, O.~S., {Popolitov}, V.~A., \&
  {Woosley}, S.~E. 1998, \apj, 496, 454

\bibitem[{{Blinnikov} {et~al.}(2006){Blinnikov}, {R{\"o}pke}, {Sorokina},
  {Gieseler}, {Reinecke}, {Travaglio}, {Hillebrandt}, \&
  {Stritzinger}}]{blinnikov_etal_06}
{Blinnikov}, S.~I., {R{\"o}pke}, F.~K., {Sorokina}, E.~I., {Gieseler}, M.,
  {Reinecke}, M., {Travaglio}, C., {Hillebrandt}, W., \& {Stritzinger}, M.
  2006, \aap, 453, 229

\bibitem[{{Blondin} {et~al.}(2013){Blondin}, {Dessart}, {Hillier}, \&
  {Khokhlov}}]{blondin_etal_13}
{Blondin}, S., {Dessart}, L., {Hillier}, D.~J., \& {Khokhlov}, A.~M. 2013,
  \mnras, 429, 2127

\bibitem[{{Blondin} {et~al.}(2006){Blondin}, {Dessart}, {Leibundgut}, {Branch},
  {H{\"o}flich}, {Tonry}, {Matheson}, {Foley}, {Chornock}, {Filippenko},
  {Sollerman}, {Spyromilio}, {Kirshner}, {Wood-Vasey}, {Clocchiatti},
  {Aguilera}, {Barris}, {Becker}, {Challis}, {Covarrubias}, {Davis},
  {Garnavich}, {Hicken}, {Jha}, {Krisciunas}, {Li}, {Miceli}, {Miknaitis},
  {Pignata}, {Prieto}, {Rest}, {Riess}, {Salvo}, {Schmidt}, {Smith}, {Stubbs},
  \& {Suntzeff}}]{blondin_etal_06}
{Blondin}, S., {Dessart}, L., {Leibundgut}, B., {Branch}, D., {H{\"o}flich},
  P., {Tonry}, J.~L., {Matheson}, T., {Foley}, R.~J., {Chornock}, R.,
  {Filippenko}, A.~V., {Sollerman}, J., {Spyromilio}, J., {Kirshner}, R.~P.,
  {Wood-Vasey}, W.~M., {Clocchiatti}, A., {Aguilera}, C., {Barris}, B.,
  {Becker}, A.~C., {Challis}, P., {Covarrubias}, R., {Davis}, T.~M.,
  {Garnavich}, P., {Hicken}, M., {Jha}, S., {Krisciunas}, K., {Li}, W.,
  {Miceli}, A., {Miknaitis}, G., {Pignata}, G., {Prieto}, J.~L., {Rest}, A.,
  {Riess}, A.~G., {Salvo}, M.~E., {Schmidt}, B.~P., {Smith}, R.~C., {Stubbs},
  C.~W., \& {Suntzeff}, N.~B. 2006, \aj, 131, 1648

\bibitem[{{Branch} {et~al.}(2006){Branch}, {Dang}, {Hall}, {Ketchum},
  {Melakayil}, {Parrent}, {Troxel}, {Casebeer}, {Jeffery}, \&
  {Baron}}]{branch_etal_06b}
{Branch}, D., {Dang}, L.~C., {Hall}, N., {Ketchum}, W., {Melakayil}, M.,
  {Parrent}, J., {Troxel}, M.~A., {Casebeer}, D., {Jeffery}, D.~J., \& {Baron},
  E. 2006, \pasp, 118, 560

\bibitem[{{Bufano} {et~al.}(2009){Bufano}, {Immler}, {Turatto}, {Landsman},
  {Brown}, {Benetti}, {Cappellaro}, {Holland}, {Mazzali}, {Milne}, {Panagia},
  {Pian}, {Roming}, {Zampieri}, {Breeveld}, \& {Gehrels}}]{bufano_etal_09}
{Bufano}, F., {Immler}, S., {Turatto}, M., {Landsman}, W., {Brown}, P.,
  {Benetti}, S., {Cappellaro}, E., {Holland}, S.~T., {Mazzali}, P., {Milne},
  P., {Panagia}, N., {Pian}, E., {Roming}, P., {Zampieri}, L., {Breeveld},
  A.~A., \& {Gehrels}, N. 2009, \apj, 700, 1456

\bibitem[{{Cardelli} {et~al.}(1989){Cardelli}, {Clayton}, \&
  {Mathis}}]{cardelli_etal_89}
{Cardelli}, J.~A., {Clayton}, G.~C., \& {Mathis}, J.~S. 1989, \apj, 345, 245

\bibitem[{{Cunto} {et~al.}(1993){Cunto}, {Mendoza}, {Ochsenbein}, \&
  {Zeippen}}]{Topbase93}
{Cunto}, W., {Mendoza}, C., {Ochsenbein}, F., \& {Zeippen}, C.~J. 1993, \aap,
  275, L5

\bibitem[{{Dessart} {et~al.}(2013{\natexlab{a}}){Dessart}, {Blondin},
  {Hillier}, \& {Khokhlov}}]{dessart_etal_14a}
{Dessart}, L., {Blondin}, S., {Hillier}, D.~J., \& {Khokhlov}, A.
  2013{\natexlab{a}}, ArXiv:1310.7747

\bibitem[{{Dessart} \& {Hillier}(2005{\natexlab{a}})}]{DH05b}
{Dessart}, L. \& {Hillier}, D.~J. 2005{\natexlab{a}}, \aap, 439, 671

\bibitem[{{Dessart} \& {Hillier}(2005{\natexlab{b}})}]{DH05a}
---. 2005{\natexlab{b}}, \aap, 437, 667

\bibitem[{{Dessart} \& {Hillier}(2008)}]{DH08}
---. 2008, \mnras, 383, 57

\bibitem[{{Dessart} \& {Hillier}(2010)}]{DH10}
---. 2010, \mnras, 405, 2141

\bibitem[{{Dessart} \& {Hillier}(2011)}]{DH11}
---. 2011, \mnras, 410, 1739

\bibitem[{{Dessart} {et~al.}(2014){Dessart}, {Hillier}, {Blondin}, \&
  {Khokhlov}}]{dessart_etal_14b}
{Dessart}, L., {Hillier}, D.~J., {Blondin}, S., \& {Khokhlov}, A. 2014, \mnras,
  439, 3114

\bibitem[{{Dessart} {et~al.}(2012){Dessart}, {Hillier}, {Li}, \&
  {Woosley}}]{dessart_etal_12}
{Dessart}, L., {Hillier}, D.~J., {Li}, C., \& {Woosley}, S. 2012, \mnras, 424,
  2139

\bibitem[{{Dessart} {et~al.}(2011){Dessart}, {Hillier}, {Livne}, {Yoon},
  {Woosley}, {Waldman}, \& {Langer}}]{dessart_etal_11}
{Dessart}, L., {Hillier}, D.~J., {Livne}, E., {Yoon}, S.-C., {Woosley}, S.,
  {Waldman}, R., \& {Langer}, N. 2011, \mnras, 414, 2985

\bibitem[{{Dessart} {et~al.}(2013{\natexlab{b}}){Dessart}, {Hillier},
  {Waldman}, \& {Livne}}]{dessart_etal_13b}
{Dessart}, L., {Hillier}, D.~J., {Waldman}, R., \& {Livne}, E.
  2013{\natexlab{b}}, \mnras, 433, 1745

\bibitem[{{Dessart} {et~al.}(2013{\natexlab{c}}){Dessart}, {Waldman}, {Livne},
  {Hillier}, \& {Blondin}}]{dessart_etal_13}
{Dessart}, L., {Waldman}, R., {Livne}, E., {Hillier}, D.~J., \& {Blondin}, S.
  2013{\natexlab{c}}, \mnras, 428, 3227

\bibitem[{{Eastman} \& {Pinto}(1993)}]{pinto_eastman_93}
{Eastman}, R.~G. \& {Pinto}, P.~A. 1993, \apj, 412, 731

\bibitem[{{Filippenko} {et~al.}(1992){Filippenko}, {Richmond}, {Branch},
  {Gaskell}, {Herbst}, {Ford}, {Treffers}, {Matheson}, {Ho}, {Dey}, {Sargent},
  {Small}, \& {van Breugel}}]{filippenko_etal_92}
{Filippenko}, A.~V., {Richmond}, M.~W., {Branch}, D., {Gaskell}, M., {Herbst},
  W., {Ford}, C.~H., {Treffers}, R.~R., {Matheson}, T., {Ho}, L.~C., {Dey}, A.,
  {Sargent}, W.~L.~W., {Small}, T.~A., \& {van Breugel}, W.~J.~M. 1992, \aj,
  104, 1543

\bibitem[{{Gall} {et~al.}(2012){Gall}, {Taubenberger}, {Kromer}, {Sim},
  {Benetti}, {Blanc}, {Elias-Rosa}, \& {Hillebrandt}}]{gall_etal_12}
{Gall}, E.~E.~E., {Taubenberger}, S., {Kromer}, M., {Sim}, S.~A., {Benetti},
  S., {Blanc}, G., {Elias-Rosa}, N., \& {Hillebrandt}, W. 2012, \mnras, 427,
  994

\bibitem[{{Garavini} {et~al.}(2007){Garavini}, {Nobili}, {Taubenberger},
  {Pastorello}, {Elias-Rosa}, {Stanishev}, {Blanc}, {Benetti}, {Goobar},
  {Mazzali}, {Sanchez}, {Salvo}, {Schmidt}, \&
  {Hillebrandt}}]{garavini_etal_07}
{Garavini}, G., {Nobili}, S., {Taubenberger}, S., {Pastorello}, A.,
  {Elias-Rosa}, N., {Stanishev}, V., {Blanc}, G., {Benetti}, S., {Goobar}, A.,
  {Mazzali}, P.~A., {Sanchez}, S.~F., {Salvo}, M., {Schmidt}, B.~P., \&
  {Hillebrandt}, W. 2007, \aap, 471, 527

\bibitem[{{Hansen} {et~al.}(1984){Hansen}, {Raassen}, \&
  {Uylings}}]{hansen_etal_84}
{Hansen}, J.~E., {Raassen}, A.~J.~J., \& {Uylings}, P.~H.~M. 1984, \apj, 277,
  435

\bibitem[{{Hillier} \& {Dessart}(2012)}]{HD12}
{Hillier}, D.~J. \& {Dessart}, L. 2012, \mnras, 424, 252

\bibitem[{{Hillier} \& {Miller}(1998)}]{HM98_lb}
{Hillier}, D.~J. \& {Miller}, D.~L. 1998, \apj, 496, 407

\bibitem[{{Hoeflich} \& {Khokhlov}(1996)}]{hoeflich_khokhlov_96}
{Hoeflich}, P. \& {Khokhlov}, A. 1996, \apj, 457, 500

\bibitem[{{Hoeflich} {et~al.}(1992){Hoeflich}, {Khokhlov}, \&
  {Mueller}}]{hoeflich_etal_92}
{Hoeflich}, P., {Khokhlov}, A., \& {Mueller}, E. 1992, \aap, 259, 549

\bibitem[{{Hoeflich} {et~al.}(1996){Hoeflich}, {Khokhlov}, {Wheeler},
  {Phillips}, {Suntzeff}, \& {Hamuy}}]{hoeflich_etal_96}
{Hoeflich}, P., {Khokhlov}, A., {Wheeler}, J.~C., {Phillips}, M.~M.,
  {Suntzeff}, N.~B., \& {Hamuy}, M. 1996, \apjl, 472, L81

\bibitem[{{Hoeflich} {et~al.}(1993){Hoeflich}, {Mueller}, \&
  {Khokhlov}}]{hoeflich_etal_93}
{Hoeflich}, P., {Mueller}, E., \& {Khokhlov}, A. 1993, \aap, 268, 570

\bibitem[{{Hoflich}(1995)}]{hoeflich_95}
{Hoflich}, P. 1995, \apj, 443, 89

\bibitem[{{H{\"o}flich} {et~al.}(2002){H{\"o}flich}, {Gerardy}, {Fesen}, \&
  {Sakai}}]{hoeflich_etal_02}
{H{\"o}flich}, P., {Gerardy}, C.~L., {Fesen}, R.~A., \& {Sakai}, S. 2002, \apj,
  568, 791

\bibitem[{{Hoflich} {et~al.}(1995){Hoflich}, {Khokhlov}, \&
  {Wheeler}}]{hoeflich_etal_95}
{Hoflich}, P., {Khokhlov}, A.~M., \& {Wheeler}, J.~C. 1995, \apj, 444, 831

\bibitem[{{Hoyle} \& {Fowler}(1960)}]{hoyle_fowler_60}
{Hoyle}, F. \& {Fowler}, W.~A. 1960, \apj, 132, 565

\bibitem[{{Hummer} {et~al.}(1993){Hummer}, {Berrington}, {Eissner}, {Pradhan},
  {Saraph}, \& {Tully}}]{HBE93_IP}
{Hummer}, D.~G., {Berrington}, K.~A., {Eissner}, W., {Pradhan}, A.~K.,
  {Saraph}, H.~E., \& {Tully}, J.~A. 1993, \aap, 279, 298

\bibitem[{{Iben} \& {Tutukov}(1984)}]{iben_tutukov_84}
{Iben}, Jr., I. \& {Tutukov}, A.~V. 1984, \apjs, 54, 335

\bibitem[{{Jack} {et~al.}(2011){Jack}, {Hauschildt}, \& {Baron}}]{jack_etal_11}
{Jack}, D., {Hauschildt}, P.~H., \& {Baron}, E. 2011, \aap, 528, A141

\bibitem[{{Jack} {et~al.}(2012){Jack}, {Hauschildt}, \& {Baron}}]{jack_etal_12}
---. 2012, \aap, 538, A132

\bibitem[{{Karp} {et~al.}(1977){Karp}, {Lasher}, {Chan}, \&
  {Salpeter}}]{karp_etal_77}
{Karp}, A.~H., {Lasher}, G., {Chan}, K.~L., \& {Salpeter}, E.~E. 1977, \apj,
  214, 161

\bibitem[{{Kasen}(2006)}]{kasen_06}
{Kasen}, D. 2006, \apj, 649, 939

\bibitem[{{Kasen} {et~al.}(2006){Kasen}, {Thomas}, \& {Nugent}}]{kasen_etal_06}
{Kasen}, D., {Thomas}, R.~C., \& {Nugent}, P. 2006, \apj, 651, 366

\bibitem[{{Kasen} {et~al.}(2008){Kasen}, {Thomas}, {R{\"o}pke}, \&
  {Woosley}}]{kasen_etal_08}
{Kasen}, D., {Thomas}, R.~C., {R{\"o}pke}, F., \& {Woosley}, S.~E. 2008,
  Journal of Physics Conference Series, 125, 012007

\bibitem[{{Kasen} \& {Woosley}(2007)}]{kasen_etal_07}
{Kasen}, D. \& {Woosley}, S.~E. 2007, \apj, 656, 661

\bibitem[{{Khokhlov} {et~al.}(1993){Khokhlov}, {Mueller}, \&
  {Hoeflich}}]{khokhlov_etal_93}
{Khokhlov}, A., {Mueller}, E., \& {Hoeflich}, P. 1993, \aap, 270, 223

\bibitem[{{Kingdon} \& {Ferland}(1996)}]{KF96_chg}
{Kingdon}, J.~B. \& {Ferland}, G.~J. 1996, \apjs, 106, 205

\bibitem[{{Kromer} \& {Sim}(2009)}]{kromer_sim_09}
{Kromer}, M. \& {Sim}, S.~A. 2009, \mnras, 398, 1809

\bibitem[{{Kuchner} {et~al.}(1994){Kuchner}, {Kirshner}, {Pinto}, \&
  {Leibundgut}}]{kuchner_etal_94}
{Kuchner}, M.~J., {Kirshner}, R.~P., {Pinto}, P.~A., \& {Leibundgut}, B. 1994,
  \apjl, 426, L89

\bibitem[{{Kurucz}(2009)}]{Kur09_ATD}
{Kurucz}, R.~L. 2009, in American Institute of Physics Conference Series, Vol.
  1171, American Institute of Physics Conference Series, ed. {I.~Hubeny,
  J.~M.~Stone, K.~MacGregor, \& K.~Werner}, 43--51

\bibitem[{{Leibowitz}(1972)}]{Lei72_CIV}
{Leibowitz}, E.~M. 1972, Journal of Quantitative Spectroscopy and Radiative
  Transfer, 12, 299

\bibitem[{{Lennon} \& {Burke}(1994)}]{LB94_N2}
{Lennon}, D.~J. \& {Burke}, V.~M. 1994, \aaps, 103, 273

\bibitem[{{Lennon} {et~al.}(1985){Lennon}, {Dufton}, {Hibbert}, \&
  {Kingston}}]{LDH85_CII_col}
{Lennon}, D.~J., {Dufton}, P.~L., {Hibbert}, A., \& {Kingston}, A.~E. 1985,
  \apj, 294, 200

\bibitem[{{Li} {et~al.}(2012){Li}, {Hillier}, \& {Dessart}}]{li_etal_12}
{Li}, C., {Hillier}, D.~J., \& {Dessart}, L. 2012, \mnras, 426, 1671

\bibitem[{{Lucy}(2005)}]{lucy_05}
{Lucy}, L.~B. 2005, \aap, 429, 19

\bibitem[{{Luo} \& {Pradhan}(1989)}]{LP89_C_seq}
{Luo}, D. \& {Pradhan}, A.~K. 1989, Journal of Physics B Atomic Molecular
  Physics, 22, 3377

\bibitem[{{Maurer} {et~al.}(2011){Maurer}, {Jerkstrand}, {Mazzali},
  {Taubenberger}, {Hachinger}, {Kromer}, {Sim}, \&
  {Hillebrandt}}]{maurer_etal_11}
{Maurer}, I., {Jerkstrand}, A., {Mazzali}, P.~A., {Taubenberger}, S.,
  {Hachinger}, S., {Kromer}, M., {Sim}, S., \& {Hillebrandt}, W. 2011, \mnras,
  418, 1517

\bibitem[{{Mazzali} \& {Lucy}(1993)}]{mazzali_etal_93}
{Mazzali}, P.~A. \& {Lucy}, L.~B. 1993, \aap, 279, 447

\bibitem[{{Mazzotta} {et~al.}(1998){Mazzotta}, {Mazzitelli}, {Colafrancesco},
  \& {Vittorio}}]{mazzotta_etal_98}
{Mazzotta}, P., {Mazzitelli}, G., {Colafrancesco}, S., \& {Vittorio}, N. 1998,
  \aaps, 133, 403

\bibitem[{{Mendoza}(1983)}]{Men83_col}
{Mendoza}, C. 1983, in IAU Symposium, Vol. 103, Planetary Nebulae, ed.
  {D.~R.~Flower}, 143--172

\bibitem[{{Mihalas}(1978)}]{mihalas_78}
{Mihalas}, D. 1978, {Stellar atmospheres /2nd edition/} (San Francisco,
  W.~H.~Freeman and Co., 1978.~650 p.)

\bibitem[{{Nahar}(1995)}]{Nahar95_FeII}
{Nahar}, S.~N. 1995, \aap, 293, 967

\bibitem[{{Nahar}(2010)}]{Nahar_OSU}
---. 2010, NORAD-Atomic-Data

\bibitem[{{Nahar} \& {Pradhan}(1993)}]{NP93_SiI}
{Nahar}, S.~N. \& {Pradhan}, A.~K. 1993, Journal of Physics B Atomic Molecular
  Physics, 26, 1109

\bibitem[{{Nahar} \& {Pradhan}(1996)}]{NP96_FeIII}
---. 1996, \aaps, 119, 509

\bibitem[{{Nomoto}(1982)}]{nomoto_82}
{Nomoto}, K. 1982, \apj, 253, 798

\bibitem[{{Nugent} {et~al.}(1995){Nugent}, {Phillips}, {Baron}, {Branch}, \&
  {Hauschildt}}]{nugent_etal_95}
{Nugent}, P., {Phillips}, M., {Baron}, E., {Branch}, D., \& {Hauschildt}, P.
  1995, \apjl, 455, L147

\bibitem[{{Nugent} {et~al.}(2011){Nugent}, {Sullivan}, {Cenko}, {Thomas},
  {Kasen}, {Howell}, {Bersier}, {Bloom}, {Kulkarni}, {Kandrashoff},
  {Filippenko}, {Silverman}, {Marcy}, {Howard}, {Isaacson}, {Maguire},
  {Suzuki}, {Tarlton}, {Pan}, {Bildsten}, {Fulton}, {Parrent}, {Sand},
  {Podsiadlowski}, {Bianco}, {Dilday}, {Graham}, {Lyman}, {James}, {Kasliwal},
  {Law}, {Quimby}, {Hook}, {Walker}, {Mazzali}, {Pian}, {Ofek}, {Gal-Yam}, \&
  {Poznanski}}]{nugent_etal_11}
{Nugent}, P.~E., {Sullivan}, M., {Cenko}, S.~B., {Thomas}, R.~C., {Kasen}, D.,
  {Howell}, D.~A., {Bersier}, D., {Bloom}, J.~S., {Kulkarni}, S.~R.,
  {Kandrashoff}, M.~T., {Filippenko}, A.~V., {Silverman}, J.~M., {Marcy},
  G.~W., {Howard}, A.~W., {Isaacson}, H.~T., {Maguire}, K., {Suzuki}, N.,
  {Tarlton}, J.~E., {Pan}, Y.-C., {Bildsten}, L., {Fulton}, B.~J., {Parrent},
  J.~T., {Sand}, D., {Podsiadlowski}, P., {Bianco}, F.~B., {Dilday}, B.,
  {Graham}, M.~L., {Lyman}, J., {James}, P., {Kasliwal}, M.~M., {Law}, N.~M.,
  {Quimby}, R.~M., {Hook}, I.~M., {Walker}, E.~S., {Mazzali}, P., {Pian}, E.,
  {Ofek}, E.~O., {Gal-Yam}, A., \& {Poznanski}, D. 2011, \nat, 480, 344

\bibitem[{{Nussbaumer} \& {Storey}(1983)}]{NS83_LTDR}
{Nussbaumer}, H. \& {Storey}, P.~J. 1983, \aap, 126, 75

\bibitem[{{Nussbaumer} \& {Storey}(1984)}]{NS84_CNO_LTDR}
---. 1984, \aaps, 56, 293

\bibitem[{{Pastorello} {et~al.}(2007){Pastorello}, {Taubenberger},
  {Elias-Rosa}, {Mazzali}, {Pignata}, {Cappellaro}, {Garavini}, {Nobili},
  {Anupama}, {Bayliss}, {Benetti}, {Bufano}, {Chakradhari}, {Kotak}, {Goobar},
  {Navasardyan}, {Patat}, {Sahu}, {Salvo}, {Schmidt}, {Stanishev}, {Turatto},
  \& {Hillebrandt}}]{pastorello_etal_07}
{Pastorello}, A., {Taubenberger}, S., {Elias-Rosa}, N., {Mazzali}, P.~A.,
  {Pignata}, G., {Cappellaro}, E., {Garavini}, G., {Nobili}, S., {Anupama},
  G.~C., {Bayliss}, D.~D.~R., {Benetti}, S., {Bufano}, F., {Chakradhari},
  N.~K., {Kotak}, R., {Goobar}, A., {Navasardyan}, H., {Patat}, F., {Sahu},
  D.~K., {Salvo}, M., {Schmidt}, B.~P., {Stanishev}, V., {Turatto}, M., \&
  {Hillebrandt}, W. 2007, \mnras, 376, 1301

\bibitem[{{Pauldrach} {et~al.}(1996){Pauldrach}, {Duschinger}, {Mazzali},
  {Puls}, {Lennon}, \& {Miller}}]{pauldrach_etal_96}
{Pauldrach}, A.~W.~A., {Duschinger}, M., {Mazzali}, P.~A., {Puls}, J.,
  {Lennon}, M., \& {Miller}, D.~L. 1996, \aap, 312, 525

\bibitem[{{Pauldrach} {et~al.}(2013){Pauldrach}, {Hoffmann}, \&
  {Hultzsch}}]{pauldrach_etal_13}
{Pauldrach}, A.~W.~A., {Hoffmann}, T.~L., \& {Hultzsch}, P.~J.~N. 2013, ArXiv
  e-prints

\bibitem[{{Peach} {et~al.}(1988){Peach}, {Saraph}, \& {Seaton}}]{PSS88_LI_seq}
{Peach}, G., {Saraph}, H.~E., \& {Seaton}, M.~J. 1988, Journal of Physics B
  Atomic Molecular Physics, 21, 3669

\bibitem[{{Perlmutter} {et~al.}(1999){Perlmutter}, {Aldering}, {Goldhaber},
  {Knop}, {Nugent}, {Castro}, {Deustua}, {Fabbro}, {Goobar}, {Groom}, {Hook},
  {Kim}, {Kim}, {Lee}, {Nunes}, {Pain}, {Pennypacker}, {Quimby}, {Lidman},
  {Ellis}, {Irwin}, {McMahon}, {Ruiz-Lapuente}, {Walton}, {Schaefer}, {Boyle},
  {Filippenko}, {Matheson}, {Fruchter}, {Panagia}, {Newberg}, {Couch}, \&
  {Supernova Cosmology Project}}]{perlmutter_etal_99}
{Perlmutter}, S., {Aldering}, G., {Goldhaber}, G., {Knop}, R.~A., {Nugent}, P.,
  {Castro}, P.~G., {Deustua}, S., {Fabbro}, S., {Goobar}, A., {Groom}, D.~E.,
  {Hook}, I.~M., {Kim}, A.~G., {Kim}, M.~Y., {Lee}, J.~C., {Nunes}, N.~J.,
  {Pain}, R., {Pennypacker}, C.~R., {Quimby}, R., {Lidman}, C., {Ellis}, R.~S.,
  {Irwin}, M., {McMahon}, R.~G., {Ruiz-Lapuente}, P., {Walton}, N., {Schaefer},
  B., {Boyle}, B.~J., {Filippenko}, A.~V., {Matheson}, T., {Fruchter}, A.~S.,
  {Panagia}, N., {Newberg}, H.~J.~M., {Couch}, W.~J., \& {Supernova Cosmology
  Project}. 1999, \apj, 517, 565

\bibitem[{{Pinto} \& {Eastman}(2000{\natexlab{a}})}]{pinto_eastman_00a}
{Pinto}, P.~A. \& {Eastman}, R.~G. 2000{\natexlab{a}}, \apj, 530, 744

\bibitem[{{Pinto} \& {Eastman}(2000{\natexlab{b}})}]{pinto_eastman_00b}
---. 2000{\natexlab{b}}, \apj, 530, 757

\bibitem[{{Quinet}(1998)}]{quinet_98}
{Quinet}, P. 1998, \aaps, 129, 147

\bibitem[{{Riess} {et~al.}(1998){Riess}, {Filippenko}, {Challis},
  {Clocchiatti}, {Diercks}, {Garnavich}, {Gilliland}, {Hogan}, {Jha},
  {Kirshner}, {Leibundgut}, {Phillips}, {Reiss}, {Schmidt}, {Schommer},
  {Smith}, {Spyromilio}, {Stubbs}, {Suntzeff}, \& {Tonry}}]{riess_etal_98}
{Riess}, A.~G., {Filippenko}, A.~V., {Challis}, P., {Clocchiatti}, A.,
  {Diercks}, A., {Garnavich}, P.~M., {Gilliland}, R.~L., {Hogan}, C.~J., {Jha},
  S., {Kirshner}, R.~P., {Leibundgut}, B., {Phillips}, M.~M., {Reiss}, D.,
  {Schmidt}, B.~P., {Schommer}, R.~A., {Smith}, R.~C., {Spyromilio}, J.,
  {Stubbs}, C., {Suntzeff}, N.~B., \& {Tonry}, J. 1998, \aj, 116, 1009

\bibitem[{{Seaton}(1987)}]{Sea87_OP}
{Seaton}, M.~J. 1987, Journal of Physics B Atomic Molecular Physics, 20, 6363

\bibitem[{{Seaton}(1998)}]{Sea98_NeI_phot}
---. 1998, \mnras, 300, L1

\bibitem[{{Seitenzahl} {et~al.}(2013){Seitenzahl}, {Ciaraldi-Schoolmann},
  {R{\"o}pke}, {Fink}, {Hillebrandt}, {Kromer}, {Pakmor}, {Ruiter}, {Sim}, \&
  {Taubenberger}}]{seitenzahl_etal_13}
{Seitenzahl}, I.~R., {Ciaraldi-Schoolmann}, F., {R{\"o}pke}, F.~K., {Fink}, M.,
  {Hillebrandt}, W., {Kromer}, M., {Pakmor}, R., {Ruiter}, A.~J., {Sim}, S.~A.,
  \& {Taubenberger}, S. 2013, \mnras, 429, 1156

\bibitem[{{Shine} \& {Linsky}(1974)}]{SL74}
{Shine}, R.~A. \& {Linsky}, J.~L. 1974, \solphys, 39, 49

\bibitem[{{Sim}(2007)}]{sim_07}
{Sim}, S.~A. 2007, \mnras, 375, 154

\bibitem[{{Sim} {et~al.}(2013){Sim}, {Seitenzahl}, {Kromer},
  {Ciaraldi-Schoolmann}, {R{\"o}pke}, {Fink}, {Hillebrandt}, {Pakmor},
  {Ruiter}, \& {Taubenberger}}]{sim_etal_13}
{Sim}, S.~A., {Seitenzahl}, I.~R., {Kromer}, M., {Ciaraldi-Schoolmann}, F.,
  {R{\"o}pke}, F.~K., {Fink}, M., {Hillebrandt}, W., {Pakmor}, R., {Ruiter},
  A.~J., \& {Taubenberger}, S. 2013, \mnras, 436, 333

\bibitem[{{Tayal}(1997{\natexlab{a}})}]{T97_SII_col}
{Tayal}, S.~S. 1997{\natexlab{a}}, \apjs, 111, 459

\bibitem[{{Tayal}(1997{\natexlab{b}})}]{T97_SIII_col}
---. 1997{\natexlab{b}}, \apj, 481, 550

\bibitem[{{Tayal} \& {Henry}(1996)}]{TH96_ArII_col}
{Tayal}, S.~S. \& {Henry}, R.~J.~W. 1996, Journal of Physics B Atomic Molecular
  Physics, 29, 3443

\bibitem[{{Verner} \& {Yakovlev}(1995)}]{Vy95_phot}
{Verner}, D.~A. \& {Yakovlev}, D.~G. 1995, \aaps, 109, 125

\bibitem[{{Wang} {et~al.}(2009){Wang}, {Filippenko}, {Ganeshalingam}, {Li},
  {Silverman}, {Wang}, {Chornock}, {Foley}, {Gates}, {Macomber}, {Serduke},
  {Steele}, \& {Wong}}]{wang_etal_09}
{Wang}, X., {Filippenko}, A.~V., {Ganeshalingam}, M., {Li}, W., {Silverman},
  J.~M., {Wang}, L., {Chornock}, R., {Foley}, R.~J., {Gates}, E.~L.,
  {Macomber}, B., {Serduke}, F.~J.~D., {Steele}, T.~N., \& {Wong}, D.~S. 2009,
  \apjl, 699, L139

\bibitem[{{Webbink}(1984)}]{webbink_84}
{Webbink}, R.~F. 1984, \apj, 277, 355

\bibitem[{{Whelan} \& {Iben}(1973)}]{whelan_iben_73}
{Whelan}, J. \& {Iben}, Jr., I. 1973, \apj, 186, 1007

\bibitem[{{Xu} \& {McCray}(1991)}]{XM91_87A_energetic}
{Xu}, Y. \& {McCray}, R. 1991, \apj, 375, 190

\bibitem[{{Zhang} \& {Pradhan}(1995{\natexlab{a}})}]{ZP95_FeII_col}
{Zhang}, H.~L. \& {Pradhan}, A.~K. 1995{\natexlab{a}}, \aap, 293, 953

\bibitem[{{Zhang} \& {Pradhan}(1995{\natexlab{b}})}]{ZP95_FeIII_col}
---. 1995{\natexlab{b}}, Journal of Physics B Atomic Molecular Physics, 28,
  3403

\bibitem[{{Zhang} \& {Pradhan}(1997)}]{ZP97_FeIV_col}
---. 1997, \aaps, 126, 373

\end{thebibliography}
\end{document}